\documentclass[galaxies,review,accept,pdftex,moreauthors]{Definitions/mdpi} 

\usepackage{color}

\firstpage{1} 
\pubvolume{1}
\issuenum{1}
\articlenumber{0}
\pubyear{2023}
\copyrightyear{2023}
\datereceived{ } 
\daterevised{ } 
\dateaccepted{ } 
\datepublished{ } 
\hreflink{https://doi.org/} 

\Title{Spin evolution of neutron stars}

\TitleCitation{Spin evolution of neutron stars}

\newcommand{\Msun}{\ensuremath{M_{\odot}}}
\newcommand{\diff}{\ensuremath{{\rm d}}}

\Author{Pavel Abolmasov $^{1}$\orcidA{}, Anton Biryukov $^{2,3}$\orcidB{} and Sergei B. Popov $^{2,4}$\orcidC{}*}

\AuthorNames{Pavel Abolmasov, Anton Biryukov and Sergei B. Popov}

\AuthorCitation{Abolmasov, P.; Biryukov, A.; Popov, S.B.}

\address{%
$^{1}$ \quad The Raymond and Beverly Sackler School of Physics and Astronomy, Tel Aviv University, Tel Aviv 69978, Israel; pavel.abolmasov@gmail.com\\
$^{2}$ \quad Lomonosov Moscow State University, Sternberg Astronomical Institute; Universitetsky pr. 13, 119234 Moscow, Russia; polar@sai.msu.ru\\
$^{3}$ \quad Faculty of Physics, HSE University, 21/4 Staraya Basmannaya str., Moscow, 105066, Russia; avbiryukov@hse.ru\\
$^{4}$ \quad ICTP, Strada Costiera 11, I-34151, Trieste, Italy; spopov@ictp.it}

\corres{Correspondence: spopov@ictp.it; Tel.: +39-040-2240-397 (S.B.P.) }

\abstract{In this paper we review the basics of magneto-rotational properties of neutron stars focusing on spin-up/spin-down behavior at different evolutionary stages. The main goal is to provide equations for the spin frequency changes in various regimes (radio pulsar, propeller, accretor, etc.). Since presently spin behavior of neutron stars at all stages remains a subject of many uncertainties, we review different suggestions made over the years in the literature.  }


\begin{document}




\section{Introduction}\label{introduction}

 In many respects, observational appearances of neutron stars (NSs) are determined by their spin period and magnetic field.  Thus, on the one hand, an understanding of magneto-rotational evolution is necessary to explain observed phenomena and to predict properties of yet undiscovered types of sources. 
On the other hand, it allows us to derive parameters of NSs (e.g., magnetic field) from observations and understand the formation and evolution of various sources. 

\begin{figure}
\includegraphics[width=0.8\columnwidth]{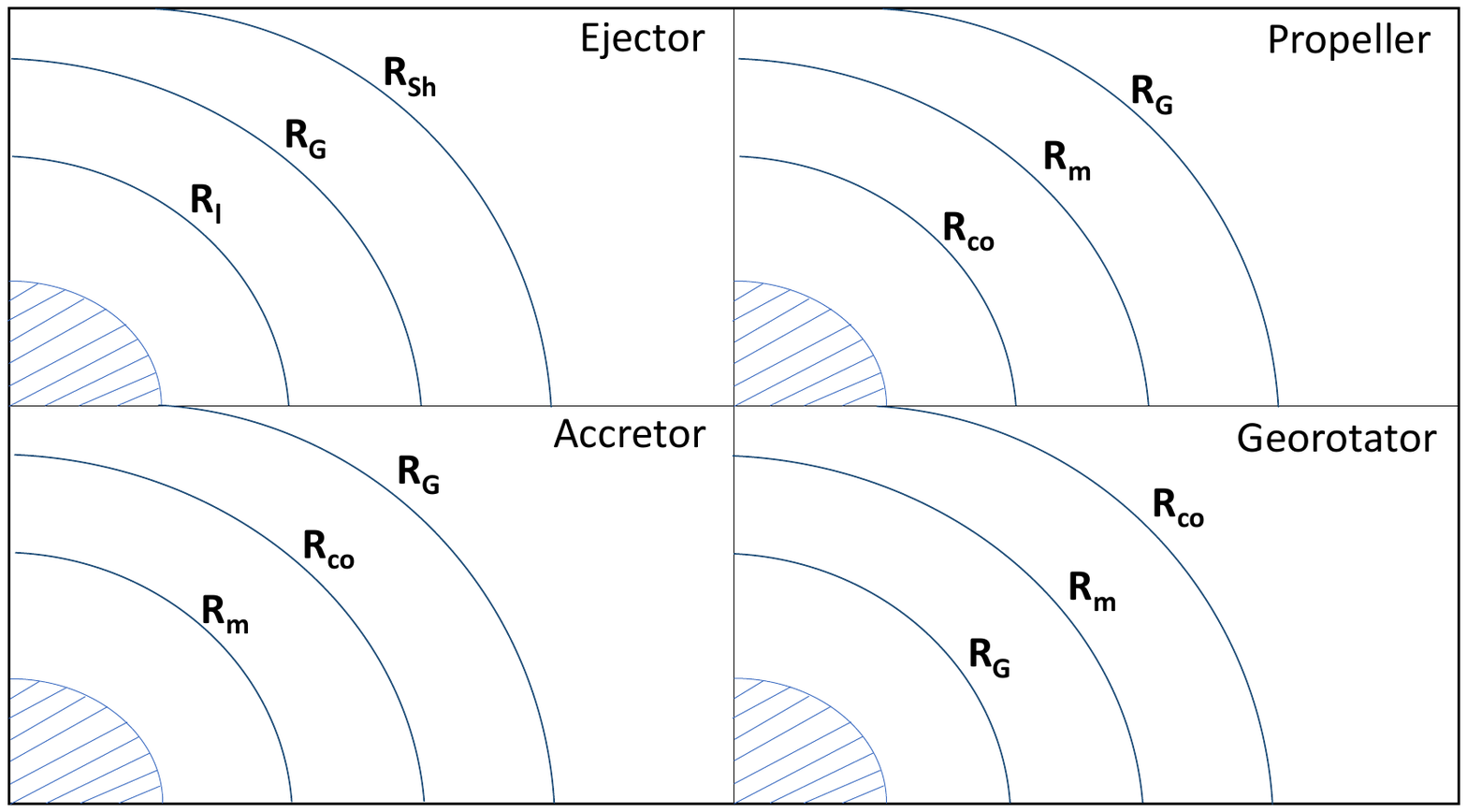}
 \caption{ A schematic representation of relation between critical radii at different stages of magneto-rotational evolution. }\label{fig:stages}
\end{figure}

 Magneto-rotational evolution is partly driven by the internal physics of NSs (e.g., field decay), and partly by the interaction of an NS with the surrounding medium. Typically, NS activity (particles wind, electromagnetic radiation, magnetic field) works against a tendency of external matter to penetrate down to the NS surface either due to gravity or just ram pressure. This counterbalance of external and internal forces determines the stage of evolution of an NS. Here we use the following classification (see e.g., \cite{1992ans..book.....L}): ejector, propeller, accretor, and georotator. Transitions from stage to stage often can be specified in terms of the equality of some specific radii, see Fig. \ref{fig:stages}. Let us define the most important of them.

The equality of kinetic energy and the absolute value of  potential energy of the matter surrounding a compact object defines the gravitational capture radius (aka Bondi radius):

 \begin{equation}
     R_\mathrm{G}= \frac{2GM}{v^2}.
\label{rg}     
 \end{equation}
 Here $G$ is the Newton constant, $M$ is the compact object mass, and $v$ is velocity relative to the medium accounting for the sound velocity $c_\mathrm{s}$: $v^2=v_\infty^2 +c_\mathrm{s}^2$ ($v_\infty$ is the NS spatial velocity).

 Magnetic field lines corotate with the NS. Thus, the linear velocity of a field line grows (in the equatorial plane) as $\Omega R$, where $R$ is the distance from the star's center and $\Omega=2\pi/P$ is spin frequency. As this velocity is limited by the speed of light, there is a critical distance called the
 light cylinder radius:

 \begin{equation}
     R_\mathrm{l}=c/\Omega,
     \label{rl}
 \end{equation}
 where $c$ is the velocity of light. 

Plasma frozen in the magnetosphere corotates with it with the maximum velocity $\Omega R$. 
 If this value is larger than the local Keplerian velocity $\sqrt{GM/R}$ then a centrifugal barrier prevents accretion down to the NS surface. Equality of the linear and Keplerian velocity defines the
 corotation radius:

 \begin{equation}
     R_\mathrm{co}=(GM/\Omega^2)^{1/3}.
     \label{rco}
 \end{equation}

Recently, Lyutikov \cite{2023MNRAS.520.4315L} (see also \cite{2020MNRAS.496...13A}, Sec. 2.3) reconsidered the penetration of matter in the cases of spherical and disc accretion regarding centrifugal force. He derived a value of the centrifugal barrier, $R_\mathrm{br}$, that is smaller than $R_\mathrm{co}$. In particular, for the spherical accretion for an aligned dipole in the equatorial plane: $R_\mathrm{cb}=(2GM/3\Omega^2)^{1/3}$.

Magnetic pressure (energy density) of a dipolar magnetosphere behaves as $P_\mathrm{mag}\propto R^{-6}$. Thus, it rapidly grows toward the compact object. The pressure of gravitationally captured free-falling matter grows as $P_\mathrm{ext}\propto R^{-5/2}$ under the standard assumption that the sound velocity is about the free-fall velocity, i.e. $T\propto r^{-1}$, and density is $\propto r^{-3/2}$ from the continuity relation. Then at some distance $R_\mathrm{m}$ the matter can be stopped by the magnetic field (for low fields it can happen that $R_\mathrm{m}$ is smaller than the NS radius $R_\mathrm{NS}$, then the influence of the field can be neglected). 
 The magnetospheric radius $R_\mathrm{m}$ might be calculated differently in the case of disc and spherical accretion. Also, the radius is calculated differently depending on the relative value of $R_\mathrm{m}$ and $R_\mathrm{G}$. For spherical accretion and $R_\mathrm{m}<R_\mathrm{G}$ we have:

 \begin{equation}
     R_\mathrm{m}=\left( \frac{\mu^2}{2 \dot M\sqrt{2GM}} \right)^{2/7}.
     \label{ra}
 \end{equation}
 This critical distance is also called the Alfven radius, $R_\mathrm{A}$.
 Here $\mu=B R_\mathrm{NS}^3$ is a magnetic moment, where $B$ is the equatorial surface dipolar field. As it is an approximate estimate, the numerical coefficient in the denominator can be different for different assumptions. The accretion rate can be estimated as:

 \begin{equation}
     \dot M=\xi_\mathrm{acc} R_\mathrm{G}^2 \rho v, 
     \label{mdot}
 \end{equation}
 where $\rho=m_\mathrm{p} \, n$ is the density of the accreted medium at $R_\mathrm{G}$ ($m_\mathrm{p}$ is proton mass). 
 The coefficient $\xi_\mathrm{acc}$ depends on details of accretion (for example, eq.~\ref{ra} is obtained for $\xi_\mathrm{acc}=4 \pi$), and below we discuss it in every particular case. Note, that $\dot M$ can appear in equations even if accretion is not possible at that stage; then this value just characterizes the properties of the external medium.

 Shvartsman radius $R_\mathrm{Sh}$ is determined by the balance between the pressure of relativistic particles wind and external medium. For $R_\mathrm{Sh}>R_\mathrm{G}$ and standard magneto-dipole rate of losses we have: 

  \begin{equation}
     R_\mathrm{Sh}=\left(\frac{2\mu^2(GM)^2\Omega^4}{3\dot M v^5 c^4}\right)^{1/2}.
     \label{rsh}
 \end{equation}
However, the coefficient in the numerator can differ for other models of rotation energy losses (see Sec.~\ref{sec:pulsars} below). 

$R_\mathrm{Sh}$ must be larger than the light cylinder radius as it is assumed that the wind is generated in the vicinity of the light cylinder boundary. 
 If $R_\mathrm{Sh}<R_\mathrm{G}$ then the boundary between the wind and external plasma is rapidly shrinking as the wind pressure $P_\mathrm{w} \propto R^{-2}$ and the pressure of the gravitationally captured matter grows as $P_\mathrm{ext}\propto R^{-5/2}$. Thus, at the ejector stage, we have $R_\mathrm{Sh}> \max{(R_\mathrm{G}, R_\mathrm{l})}$. 

 The ejector stage is over when the relativistic particle wind is not produced anymore and the Poynting flux becomes sufficiently low. Usually, it is assumed that this happens when external plasma can penetrate the light cylinder. This can happen due to spin-down, magnetic field decay, or modifications of external conditions. Inside the light cylinder, matter starts to interact with the magnetosphere. Unless some very exotic situation is realized, immediately after the transition from the ejector stage, $R_\mathrm{m}$ is just slightly smaller than $R_\mathrm{l}$. This means that $R_\mathrm{m}>R_\mathrm{co}$ as the Keplerian velocity cannot be larger than $c$. This corresponds to the propeller regime. 
 Finally, when $R_\mathrm{m}<R_\mathrm{co}$ and $R_\mathrm{m}<R_\mathrm{G}$ matter can reach the surface of the NS. So, we have accretion. 
 Some other more exotic situations can be realized. 
 We discuss them in Sec. \ref{exotic}.

\begin{figure}
\includegraphics[width=0.8\columnwidth]{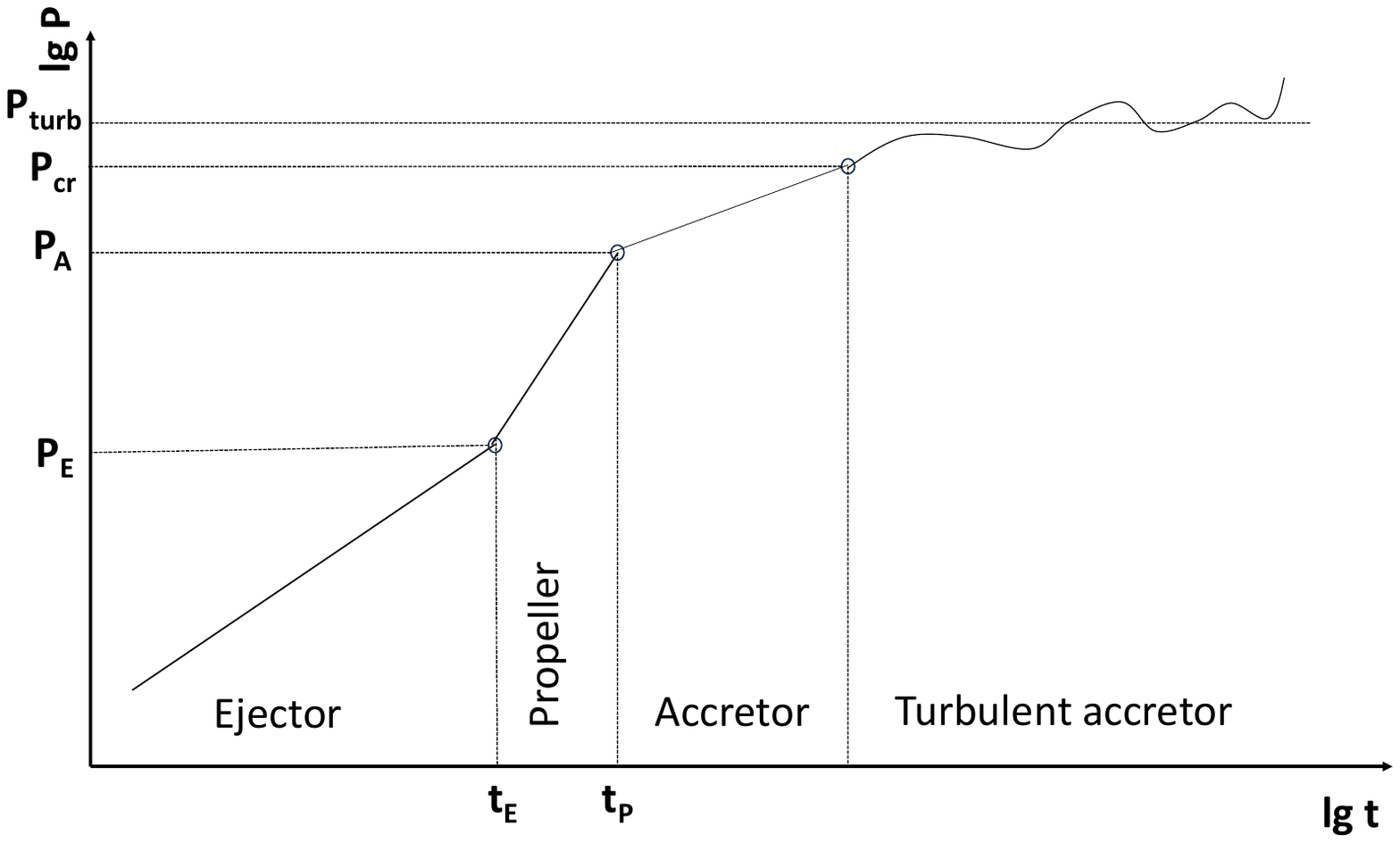}
 \caption{ A schematic representation of spin evolution of an isolated NS. The compact object consequently passes the stages of ejection, propeller, and accretor. Finally, it can reach the stage when the spin behavior is determined by the accretion of turbulent angular momentum. }\label{fig:evol}
\end{figure}

 The basic picture of magneto-rotational evolution looks like this (see Fig. \ref{fig:evol}). 
An NS is born rapidly rotating ($P\lesssim 1$~s) with a sufficient magnetic field ($B\sim 10^{12}$~G) at the ejector stage (often as a radio pulsar). Then it slows down (and, maybe, the magnetic field decays) coming to the propeller stage at some $t=t_\mathrm{E}$. The transition happens at a critical period $P_\mathrm{E}$. Consequent spin-down can bring the NS to the stage of accretion, especially in the case of an isolated object,  when another critical period, $P_\mathrm{A}$, is reached at $t=t_\mathrm{P}$  This route can be followed by an isolated  NS as well as by an NS in a binary system. At the accretion stage, the spin period can change due to changes in the angular momentum of the incoming matter. For example, in the case of an isolated NS, this can be due to the interstellar turbulence (see Sec. 5.2). Turbulence becomes important at the period $P_\mathrm{cr}$. At the stage of turbulent accretion, the spin period fluctuates around the value $P_\mathrm{turb}$.

This simple picture can be perplexed by an initial stage of fallback, non-trivial field evolution, 
variations of properties of the surrounding medium, etc. 
Mainly, in this review we focus on the spin evolution. Magnetic field evolution was recently reviewed, e.g. in \cite{2021Univ....7..351I}. 

  In the following sections, we discuss the main evolutionary stages and transitions between them in different astrophysical scenarios. We describe the spin evolution of an NS in `chronological' order, starting from the early stage of fallback of a part of supernova ejecta onto a newborn NS (Sec. 2). Then we describe the ejector stage paying special attention to radio pulsars (sec. 3). This stage is followed by the propeller (Sec. 4) and accretor stages (Sec. 5). In Sec. 6 we discuss some hypothetical regimes of interaction of an NS with surrounding medium. Our conclusions are presented in Sec. 7. 

 In the text we use convention that $X_\mathrm{x}=X/10^x$.

\section{Fallback}
\label{fallback}

In this section, we discuss the so-called fallback -- an episode of very intense accretion of the ejecta  onto a newborn compact object. The total amount of accreted matter can reach $\lesssim 0.1 M_\odot$ and the accretion rate at early stages can be from a $\sim$few up to $\sim 10^4 M_\odot$~yr$^{-1}$. 
For the first time this process was discussed more than half a century ago in
\cite{1971ApJ...163..221C, 1972SvA....16..209Z}. 
 Modern approaches to modeling fallback are considerably influenced by the seminal paper by Roger Chevalier
\cite{1989ApJ...346..847C}. 

 In \cite{1989ApJ...346..847C} the author analyzed in detail different regimes of fallback accounting for heating due to radioactive decay and changes in opacity (in time and at different distances from the compact object). Asymmetries due to the kick velocity, rotation, and pulsar emission were also briefly discussed. The author used mainly a semianalytical approach so that all involved physical processes could be clearly visible and understood. 
 Strong initial fallback is due to a reverse shock and interaction of the ejecta  with the envelope. 
 At the initial stages, the accretion rate dependence on time roughly follows the law $\dot M \propto t^{-3/2}$. Later, when the ballistic regime is established, the standard law of fallback $\dot M \propto t^{-5/3}$ is realized. Duration of a quasi-spherical fallback depends on the properties of the progenitor and for red supergiants can reach $\sim 1$~year.
 
 Fallback properties at early stages significantly depend on the structure of the pre-SN. Thus, the total amount of the accreted matter also depends on the parameters of the progenitor. Emission properties are dependent on the trapping of photons by the envelope. Only at a low accretion rate, do they start to diffuse out. Before this moment, the effects of the central compact object are hidden. 

Fallback can be important also to explain some properties of gamma-ray bursts and superluminous supernovae, see e.g. \cite{2018ApJ...857...95M} and references therein. However, here we will not discuss these issues focusing on consequences related to the magneto-rotational evolution of NSs as the fallback can drastically modify the initial properties of these compact objects. For example, it can reduce the surface magnetic field at the early phases of evolution, can significantly change the spin period, and finally, can even change the evolutionary stage of a young NS (e.g., preventing the appearance of a radio pulsar). We discuss all these possibilities below. 

 Significant fallback (here the accretion rate $\dot M$ is more important than the total accreted mass $\Delta M$, however, these two parameters are interlinked as $\Delta M=\int \dot M(t) \diff t $) can strongly influence the observational appearance of a young NS. In particular, the radio pulsar activity can be completely switched off. 
This possibility was suggested and analyzed in \cite{1995ApJ...440L..77M}. 
If the pressure of the external medium falling onto the compact object is larger than both pressure of the relativistic wind and magnetic field pressure for all radii larger than $R_\mathrm{NS}$ then the field is submerged at depth $\sim$~a few tens or hundred meters below the surface. Thus, the external field that determines the radio pulsar activity and many other manifestations of an NS is orders of magnitude lower than the initial value. 
 For a range of parameters used in \cite{1995ApJ...440L..77M} it was found that the field diffuses out on a time scale $\sim 10^3$~yrs. This value mostly depends on the depth of submergence which is determined by the amount of accreted matter. Later this process was studied in several papers in more realistic frameworks. 

In \cite{1999A&A...345..847G}   
 the authors modeled the field diffusion in more detail. 
 In particular, it was demonstrated that due to the field submergence, the characteristic ages of young pulsars, defined as $P/2\dot P$, can be significantly different (larger) in comparison with their actual ages. 
The authors notice that due to compression the field in the crust is amplified by $\sim 2$ orders of magnitude. And as the spatial scale for this field is relatively small it is a subject of relatively rapid Ohmic decay. 
 With the parameters used in this paper, the field diffuses out in $\sim10^3-10^4$~yrs. Also, the authors notice that submergence can explain the absence of radio pulsars in some supernova remnants (SNRs). 
  
A new interest in the fallback field submergence appeared in the 2010s. This was related to the necessity to explain the properties of central compact objects in SN remnants (CCOs). 
Wynn Ho was probably the first who used the field submergence to explain the observational appearance of CCOs 
\cite{2011MNRAS.414.2567H}. 
The author shows that accretion of $\Delta M  \lesssim 10^{-4} M_\odot$ is enough for significant submergence with the time of re-emergence $\lesssim 10^4$~yrs. This is sufficient to explain the properties of the most studied CCOs. 

 A 2D model of submergence and re-emergence coupled with detailed calculations of field decay and thermal evolution was presented in
\cite{2012MNRAS.425.2487V}. 
The authors obtained results qualitatively in correspondence with the study \cite{2011MNRAS.414.2567H}. 
Corresponding tracks in the $P - \dot P$ diagram can be found in \cite{2012A&A...547A...9P}. 
In 2D it was possible to analyze changes in the topology of the field during re-emergence.
In more detail, this process was studied in
\cite{2016MNRAS.462.3689I}. 
The authors presented a 2D model of field re-emergence after a fallback episode with $\Delta M = 10^{-3}\, M_\odot$. 
It is shown that higher multipoles, which are strongly suppressed during fallback, re-emerge faster than the dipole. Thus, they can dominate at ages $\sim 10^4-10^5$yrs. This is important for the pulsar switch-on. 

Details of field submergence were calculated numerically in several papers by Bernal et al.:
\cite{2010RMxAA..46..309B},  
\cite{2013ApJ...770..106B}, 
\cite{2018PhRvD..98h3012F}.   
 In \cite{2013ApJ...770..106B} the authors considered the interaction of the fallback matter with a magnetic field loop anchored in the crust. 
They demonstrated with 2D and 3D modeling that for sufficiently strong accretion on a timescale $\sim100$~ms after the reverse shock starts to interact with the loop, the field (which is always below the accretion shock) is submerged. Convection in the accretion envelope is crucially important for the fate of the field. If the envelope remains convective then the field is not submerged completely. If the reverse shock is not developed at all the field is not modified significantly. 


Field submergence can modify the evolutionary status of a newborn NS. In a 1D approach (i.e., for spherical symmetry) it was analyzed in \cite{2018PASJ...70..107S} and 
\cite{2021ApJ...917...71Z}.   
The authors aim to account for the back reaction of the NS (due to a relativistic particle wind and magnetic pressure) on the fallback flow. This allows them to explain different types of young neutron stars (radio pulsars, CCOs, magnetars, etc.) by different outputs of such interaction.
 In \cite{2018PASJ...70..107S} the authors presented self-similar analytical considerations and in \cite{2021ApJ...917...71Z} -- numerical modeling. 

 In \cite{2021ApJ...917...71Z} the model is mostly determined by two parameters: characteristic fallback time, $t_\mathrm{fb}$, and the fallback mass, $M_\mathrm{fb}$. The latter parameter can be calculated from the dependence of the fallback rate on time (notice, that it is not the rate of fallback on the NS surface {as accretion can be prevented by relativistic wind or the magnetosphere}):

 \begin{equation}
     \dot M_\mathrm{fb}=\dot M_\mathrm{fb,ini} \times
     \begin{cases} 
    1 & t \leq t_\mathrm{fb} \\
    (t/t_\mathrm{fb})^{-l} & t > t_\mathrm{fb}
    \end{cases}.
 \end{equation}

 The fallback time was set to $t_\mathrm{fb,1}=({t_\mathrm{fb}}/{10\, \mathrm{s}})=1$. Among other parameters, it determines the initial encounter radius $R_\mathrm{enc}$ at which the relativistic wind from the NS collides with the fallback matter:

 \begin{equation}
     R_\mathrm{enc}\approx R_\mathrm{fb}=(GM t_\mathrm{fb}^2)^{1/3}= 2.7\times 10^9 \mathrm{cm} \, t_\mathrm{fb,1}^{2/3}.
 \end{equation}

 The luminosity of the wind, $L_\mathrm{sd}$, is calculated either with the usual magneto-dipole formula, or it can be enhanced due to accretion of some fraction of the fallback matter through a disc as it was suggested by \cite{2016ApJ...822...33P}. 

 The authors determine three possible regimes: significant field submergence, disturbed magnetosphere, and matter expulsion. 
 If $ ({R_\mathrm{enc}L_\mathrm{sd}})/ ({GM \dot M_\mathrm{fb,ini}}) <1$ then the fallback matter reaches the surface and submerges the magnetic field or significantly disturbs the magnetosphere (in the latter case the authors speculate about a magnetar formation due to field enhancement). 

 Matter expulsion can happen due to the usual wind produced by the magnetic dipole. This corresponds to $ M_\mathrm{fb} < 7.7 \times 10^{-8} M_\odot \, B_{13}^2 P_{-2}^{-4} t_\mathrm{fb,1}^{5/3} $. If the fallback mass is larger but $< 5.2 \times 10^{-3} M_\odot \, B_{13}^2 P_{-2}^{-14/3} t_\mathrm{fb,1}^{23/9}$ then the fallback matter is repelled by the enhanced spin-down power. 

Fallback can result in the formation of a disc around the NS if the surrounding matter has enough angular momentum. These can be viscous accretion discs or passive discs like the one observed around the anomalous X-ray pulsar 4U 0142+61 \cite{2006Natur.440..772W}. In this century, such discs have been extensively studied by many authors, starting with the paper by Chatterjee et al. \cite{2000ApJ...534..373C}, especially as an alternative to explain the properties of anomalous X-ray pulsars. See, e.g. \cite{2007MNRAS.382..871J, 2016MNRAS.457.4114B} and references therein. Here we do not pretend to present an extensive review of fallback disc studies. Instead, we focus on a few papers mostly relevant to the subject of the present paper.  

Ideally, fallback properties might be obtained from realistic calculations of SN explosions. At the moment, numerical models of SN are far from being complete. Still, new important results provide some optimism. 
In particular, recent calculations presented in \cite{2022ApJ...926....9J}  
provide a realistic setup for analysis of the fallback influence on magneto-rotational evolution of NSs.

In this model, the newborn NS is not fixed in the center of the grid. The movement of the compact object allows accounting for new effects. In particular, the authors demonstrate that an NS can be spun up by the fallback and the obtained spin is correlated with the kick velocity. Vorticity in the fallback matter is not related to the spin of the progenitor. Instead, it is generated by non-radial hydrodynamic instabilities. Late fallback is associated with the highest angular momentum. Accretion of just $\sim 10^{-2} M_\odot$ can result in the spin period $\sim 0.05$~s which is even slightly shorter than a usual value for young NSs \cite{2012Ap&SS.341..457P}. Typically in this scenario, the angle between the spin axis and velocity vector is $\lesssim 30^\circ$. Alignment is more pronounced for large kicks as in this case, the NS can move farther from the center of the explosion, so accretion is less symmetric. For low kicks, an NS accretes from all directions and this results in averaging of the accreted angular momentum. According to the proposed model, fallback discs might be widely spread among young NSs with not very low spatial velocities. 

Joint effects of fallback disc formation and field emergence on the appearance of radio pulsars are considered in 
\cite{2013ApJ...775..124F}. The authors perform population synthesis modeling accounting for spin up/down due to the NS interaction with the disc. In their model, the magnetic field is not submerged by the initial strong fallback. Instead, the authors use the equation identical to the one used for field behavior in an accreting binary system:

\begin{equation}
    B=\frac{B_0}{1+\Delta M/10^{-5} M_\odot}.
\end{equation}
Here $B_0$ is the field before accretion starts. In the population synthesis it is assumed to have a log-normal distribution with $\langle \log \, B_0/\mathrm{G}\rangle =12.35$ and $\sigma_{\mathrm{log B_0}}=0.4$. And $\Delta M$ is the total amount of the accreted matter. After accretion stops, the field re-emerges with the rate:

\begin{equation}
    \dot B=0.01 \left( \frac{\Delta M}{M_\odot}\right)^{-3} \left( 1- \frac{B}{B_0}\right)^2 \, \rm{G \, yr}^{-1}.
\end{equation}

An NS can pass stages of accretion, propeller, and finally ejector. At the ejector stage, the NS can appear as a radio pulsar. 
The authors of \cite{2013ApJ...775..124F}, in the first place, intended to explain statistics of the braking indices distribution. By construction, in their model braking indices might be $\leq 3$. Thus, the results are more or less compatible with the data on pulsars in SNRs. Still, it is very important that the authors used in their population model complicated early spin evolution due to NS-disc interaction and sub/re-emergence of the magnetic field.  

Discovery of several long-period radio sources attracted new interest to the model of fallback discs as the interaction between an NS and the disc can result in a drastic spin-down of the compact object. 
The first new source with a peculiar period was the 76-second pulsar PSR J0901-4046 \cite{2022NatAs...6..828C}. In addition, two radio sources with periods $\sim1000$~s were found: GLEAM-X J1627-52 \cite{2022Natur.601..526H} and GPM J1839-10 \cite{2023Natur.619..487H}. As their properties point towards magnetar-scale magnetic field, all three objects might be young. So, it is impossible to explain their spin periods simply by a magneto-dipole-like spin-down. 
Ronchi et al. \cite{2022ApJ...934..184R} 
 successfully explain long spin periods of freshly discovered sources in the framework of a fallback disc around a young strongly magnetized NS. 

 As specific angular momentum of the fallback matter is $\sim 10^{16}-10^{17}$cm$^2$~s$^{-1}$, see
 \cite{2022ApJ...926....9J}, the circularization radius where the accreted matter settles at a circular orbit forming a disc is $R_\mathrm{c}\sim 10^6-10^8$~cm. The viscous time scale in the disc is $t_\nu\approx 2.1\times 10^3 T_\mathrm{c,6}^{-1} R_\mathrm{c,8}^{1/2}  $~s. Here $T_\mathrm{c,6}$ is the disc's central temperature  in units $10^6$~K and $R_\mathrm{c,8}=R_\mathrm{c}/10^8$~cm. 
 The mass flow rate in the outer parts of the disc is:

 \begin{equation}
\dot M_\mathrm{d}(t)=\dot M_\mathrm{d,0} \left( 1+ \frac{t}{t_\nu} \right)^{-\alpha}.
 \end{equation}
 The coefficient $\alpha$ depends on the opacity and in \cite{2022ApJ...934..184R} the authors assume it to be equal to 1.2. 
 The accretion rate at the inner edge of the disc is limited by the Eddington rate. 

 The authors do not account for the field submergence, but allow for the decay as they are interested in times scales up to $\gtrsim 10^5$~yrs. The field decay is calculated according to the analytical expression from \cite{2008A&A...486..255A}:

 \begin{equation}\label{E:fallback:B}
     B(t)=B_0\frac{\exp{(-t/\tau_\mathrm{Ohm})}}{1+\frac{\tau_\mathrm{Ohm}}{\tau_\mathrm{Hall,0}}(1-\exp{(-t/\tau_\mathrm{Ohm})})}. 
 \end{equation}
Here $\tau_\mathrm{Ohm}$ is the Ohmic dissipation time scale which is assumed to be $4.4\times 10^6$~yrs. 
And $\tau_\mathrm{Hall,0}=6.4\times 10^4 (10^{14} \rm{G}/B_0)$ is the initial Hall time scale which depends on the initial magnetic field $B_0$. 

 With such a setup and realistic fallback rates, it is possible to obtain spin periods $\sim 100$~s for $B_0\sim10^{13}$~G within $10^5$~yrs and $\sim 1000 $~s for $B_0\sim10^{14}-10^{15}$~G.

\section{Ejector stage and radio pulsar activity}
\label{sec:pulsars}
With more than 3000 known rotation-powered pulsars\footnote{See the ATNF Pulsar Catalogue \cite{2005AJ....129.1993M},\url{https://www.atnf.csiro.au/research/pulsar/psrcat/expert.html}, v.1.73}, ejectors are the most common type of Galactic NSs observed.
The rotational evolution of these sources is not affected by the surrounding interstellar medium but is driven by the complex 
electrodynamical processes within their magnetospheres. 

\begin{figure}
\includegraphics[width=1\columnwidth]{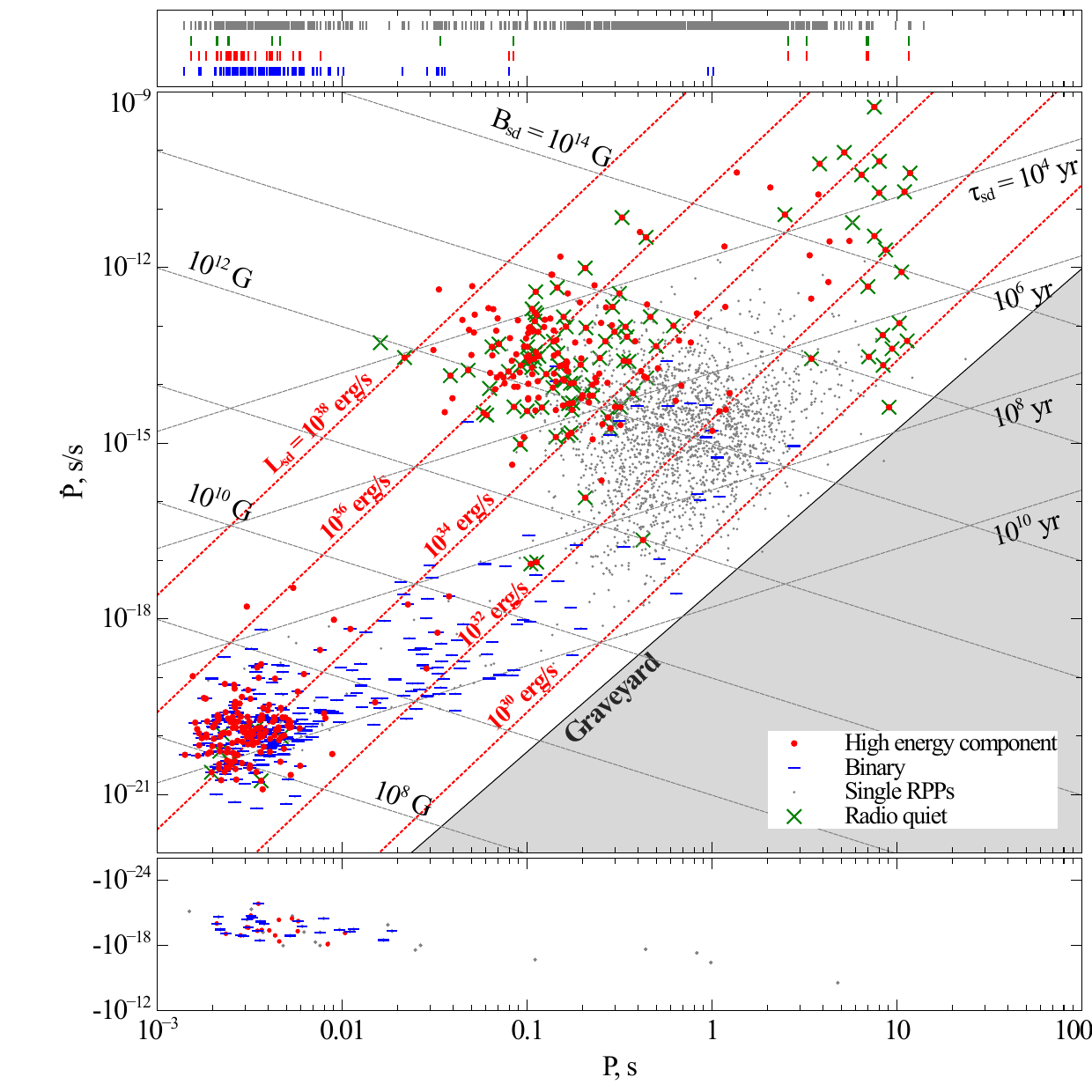}
 \caption{Spin periods and their derivatives for 3473 pulsating neutron stars collected in the ATNF Pulsar Catalogue. \textit{Middle panel}: Classical single rotation-powered pulsars are indicated by grey dots; red circles mark sources with high-energy pulsed emission; green crosses indicate radio quietness, while horizontal blue stripes are for pulsars in binary systems. Lines of constant surface magnetic fields $B_\mathrm{sd} \propto \sqrt{P \dot P}$ Gs, ages $\tau_\mathrm{sd} = P/2\dot P$ and luminosities $L_\mathrm{sd} \propto \dot P P^{-3}$ are also drawn assuming $R_\mathrm{NS} = 10$~km and $I = 10^{45}$~g~cm$^2$. \textit{Bottom panel}: Sources with negative $\dot P$ are shown in the same coordinates. Most of these pulsars are in globular clusters and their ``negative spin-down'' is due to the Shklovskii effect \cite{1970SvA....13..562S}. The pulsar graveyard border is plotted according to \cite{2022MNRAS.516.5084B}. \textit{Upper panel}: Vertical lines show the spin period values for pulsars for which $\dot P$ has not yet been estimated. Grey lines are for all such objects, green ones mark the values for radio-quiet sources, red ones are for those exhibiting high-energy emission, while blue lines are for pulsars in binary systems.}\label{fig:ppdot}
\end{figure}

For now, it is fairly clear, that most of the rotational energy of the ejectors is carried away by the wind of energetic charged particles (electrons and positrons) accelerated by strong electric fields within the magnetosphere of the star. At the same time, the physics of this process is still not well understood (there 
are detailed reviews on this topic \cite{2015SSRv..191..207B, 2018PhyU...61..353B}). Typical observed ejectors have spin periods in the range $P \sim 10^{-3}-10$ s while their surface magnetic fields are of the order $\sim 10^9 - 10^{13}$ G. This is consistent with observed spin-down rates of ejectors $\dot P \sim 10^{-20}-10^{-9}$ s/s (e.g. \cite{2012hpa..book.....L}). Therefore, their spin-down luminocities $L_\mathrm{sd} = 4\pi^2 I \dot P P^{-3}$ are of order $10^{30} - 10^{38}$ erg/s (see Fig.~\ref{fig:ppdot}).

In classical theory, single neutron stars are viewed as rotating, highly magnetized, conducting bodies with a primarily dipolar magnetic field. Indeed, higher-order multipoles decrease very rapidly with the distance from the star. However, multipolar magnetic fields have been also considered in the literature (see e.g. \cite{1991ApJ...373L..69K, 2004ARep...48.1029K, 2015MNRAS.450..714P, 2017MNRAS.472.3304P}). 

The magnetosphere of the star is created by its magnetic field, electric 
fields, and electric currents. If there is an electric field 
$\mathbf{E_\perp}$ perpendicular to the local magnetic field $\mathbf{B}$, 
a non-zero Poynting flux $\mathbf{S}\propto(\mathbf{E}\times\mathbf{B})$ 
exists, which carries away the rotational energy of the star. This is true 
for both vacuum and plasma-filled rotating magnetospheres, which are two 
fundamental models for the physics of the ejector. 

In the vacuum approach, the spin-down of a neutron star is due to the surface electric currents induced by the corresponding electric fields. So the deceleration of the star is driven by the Amp{\`e}re's force. On the other hand, the environment of real neutron stars is not vacuum, but rather filled with plasma \cite{1971ApJ...164..529S, 1975ApJ...196...51R}. 
In general, the magnetosphere of a neutron star consists of two parts, containing closed and open magnetic field lines respectively. The plasma can freely escape through the region of open lines, giving rise to the so-called `pulsar wind'. Its back-reaction will also be responsible for the loss of the rotational energy \cite{2004IAUS..218..357S}. Notice, that magnetospheric currents define local electric and magnetic fields in a self-consistent manner. 

So, finally, the spin-down torque acting on the neutron star can be described as the sum of two components: the vacuum and the wind. Hereafter we assume that the spin-down torque is \textit{positive} and that the neutron star is spherical with a moment of inertia I. This means that for the ejector stage: 
\begin{equation}
    I{\mathbf{\dot \Omega}} = - \mathbf{K}_\mathrm{sd,vac} - \mathbf{K}_\mathrm{sd,w}.
\end{equation}

For the vacuum component, the surface quadrupolar electric field $E \sim (\Omega R/c) B$ is induced by the rotation, and so the corresponding spin-down torque is
\begin{equation}
    K_\mathrm{sd,vac} = f_\perp \dfrac{\mu_{\perp}^2}{R^{3}_\mathrm{l}},
    \label{eq:pulsars:vacuum_torque}
\end{equation}
where $\mu_\perp = \mu\sin\chi$ is the orthogonal component of the magnetic moment, $\chi$ is the angle 
between the magnetic and spin axes, while $f_\perp$ is a structure factor. The latter is of the order of unity to a first approximation, although its details depend on certain assumptions. The vacuum losses are inefficient for a perfectly aligned rotator: $K_\mathrm{sd,vac} = 0$ when $\chi = 0$. This is 
similar to the behavior of the emission from a rotating magnetic dipole. Furthermore, if $f_{\perp} = 2/3$, then the equation 
(\ref{eq:pulsars:vacuum_torque}) coincides with the classical one for the electromagnetic loss of a point-sized dipole. However, it is important to point out that there is no real magneto-dipolar emission from a neutron star (see the discussion in Section 2.3 of \cite{2018PhyU...61..353B}). 
So the above equation is nothing more than a law for the spin-down of a magnetized, conducting, rotating sphere in a vacuum.

As for the wind component of the ejector's spin down, its power $L_\mathrm{sd,w}$ is based on the electric potential drop $\Delta\varphi$ near the star's surface, so that
\begin{equation}
    L_\mathrm{sd,w} = \mathbf{K}_\mathrm{sd,w} \cdot \mathbf{\Omega} = -2\pi r_p^2 c \kappa 
    \rho_\mathrm{GJ} \Delta \varphi,
\end{equation}
where $r_p = R_\mathrm{NS}\sqrt{\Omega R_\mathrm{NS}/c}$ is the star's polar cap radius, $\rho_\mathrm{GJ} 
\approx \omega B_\parallel/c$ is the Goldreich-Julian charge density \cite{1969ApJ...157..869G}, while 
$\kappa \sim 10^3$ is the coefficient describing the primary particle density \cite{2015MNRAS.450.1990K}. 
Taking the maximum potential drop for a rotating dipole $\Delta \Phi = \mu_\parallel/R_\mathrm{l}^2$ 
\cite{1975ApJ...196...51R} as an essential scale factor, the wind braking torque can be formally expressed 
in a manner similar to (\ref{eq:pulsars:vacuum_torque}) as
\begin{equation}
    K_\mathrm{sd,w} = f_\parallel \dfrac{\mu^2_\parallel}{R^{3}_\mathrm{l}},
\end{equation}
where $\mu_\parallel = \mu\cos\chi$ and $f_\parallel = 2\kappa \Delta\varphi/\Delta \Phi$ is another structure 
factor that depends ultimately on the microphysics of the acceleration gap. The wind torque calculated as 
shown above disappears for a perfectly orthogonal rotator ($\chi = 90^\circ$ ) but is present for a perfectly 
aligned one.

At the turn of the millennium Harding et al. \cite{1999ApJ...525L.125H} and then Xu \& Qiao 
\cite{2001ApJ...561L..85X} suggested that in fact, both spin-down mechanisms have a right to exist and that 
the total spin-down is driven by their combination. However, as the pulsar spins down, the potential drop 
may not sustain the need to accelerate particles \cite{1975ApJ...196...51R}. Thus, the maximum period 
$P_\mathrm{death}$ exists where wind losses are efficient. This is the so-called `death line' (or `death 
period') of radiopulsars, beyond which these objects cannot be observed as radio-loud sources (since 
plasma acceleration and secondary particle production are thought to be related to pulsar radio emission). 
After crossing the death line, the wind component become negligible, so the pulsar's spin-down is expected to follow the vacuum approximation \cite{2006ApJ...643.1139C}.

In this way, a complete torque that is responsible for the loss of rotational energy of a single ejector can 
be written as follows:
\begin{equation}
K_\mathrm{sd} = 
\left\{
    \begin{array}{lr}
        K_\mathrm{sd,vac} + K_\mathrm{sd, w} = \dfrac{\mu^2}{R_\mathrm{l}^3} (f_\perp\sin^2\chi + 
        f_\parallel\cos^2\chi), & \text{if } P < P_\mathrm{death}\\
        K_\mathrm{sd,vac} =  f_\perp\dfrac{\mu^2}{R_\mathrm{l}^3}\sin^2\chi , & \text{if } P \ge 
        P_\mathrm{death}
    \end{array}
    \right .
    \label{eq:pulsars:general_spindown}
\end{equation}
The values of the these factors and their dependence on the parameters of the star are
of particular interest for an accurate understanding of the rotation history of ejectors.
Both quantities depend on the details of the physics of 
neutron star magnetospheres -- the topology of the field and currents, the interaction of the currents with 
the stellar surface, and the physical structure and composition of the surface itself. And, in principle, there 
are two approaches to unravelling these factors theoretically: analytical and numerical ones. The purely analytical approach works well when considering a neutron star with a vacuum magnetosphere (and hence in determining $f_\perp$, see the section \ref{sect:pulsars:vacuum}), whereas the complex structure 
of the filled magnetosphere (taking into account both $f_\perp$ and $f_\parallel$) can only be revealed numerically as described in the section \ref{sect:pulsars:plasma}. But first, a brief review of the pulsar `death condition' must be given.

\subsection{Pulsar `death line'}
\label{sect:pulsars:death}

There has been much discussion regarding the `death period' $P_\mathrm{death}$ since the discovery of active radio pulsars. It is widely believed that pulsar radio emission is generated by the secondary plasma near the polar regions of neutron stars \cite{1971ApJ...164..529S, 1975ApJ...196...51R}. The `death' of pulsars is therefore closely linked to the possibility of the formation of electron-positron pairs. As shown by Ruderman and Sutherland \cite{1975ApJ...196...51R}, the critical condition for this process is the equality of the height of the vacuum gap $H \propto B^{-4/7} R_\mathrm{l}^{3/7} \rho_\mathrm{c}^{2/7}$ (where $\rho_\mathrm{c}$ is the radius of curvature of the local magnetic fields) to the polar cap radius $r_\mathrm{p}$. Corresponding `death' condition can be written as
\begin{equation}
    P_\mathrm{death} = a\dot P_{-15}^{b}
    \label{eq:pulsars:death}
\end{equation}
where $a \approx 0.6$ s, $b = 4/11$ and $\dot P_{-15}=\dot P/(10^{-15}\mbox{ s/s})$. This equation represents a monotonous curve (in particular, a straight line) on the $P-\dot P$ diagram for neutron stars when its plotted in logarithmic scales. This is why the pulsar `death condition' is often referred to as the `death line'.

However, equation (\ref{eq:pulsars:death}) only partially explains the properties of the observed population of radio pulsars 
since there are many sources with $P > P_\mathrm{death}$ if the `death period' is calculated using eq.~(\ref{eq:pulsars:death}).
Then Chen and Ruderman \cite{1993ApJ...402..264C} considered several variants of the magnetic field structure near the polar cap and obtained 
refined parameters for the death condition. In particular, they found $(a, b)=$(3.2 s, 4/9) and (11 s, 1/2) as possible solutions. Accounting for relativistic frame dragging affects the condition of the electron-positron pairs production due to additional induced electric fields \cite{1992MNRAS.255...61M}. Thus, Zhang et al. \cite{2000ApJ...531L.135Z} have considered two polar cap acceleration models and found $P_\mathrm{death}\approx0.67\dot P^{2/5}_{-15}$\,s for a realistic death line in the case of dipolar and $P_\mathrm{death}\approx5.75\dot P^{1/2}\rho_{c,6}^{-1/2}$\,s for multipolar field configurations, respectively.

In the influential paper by Faucher-Gigu{\`e}re and Kaspi \cite{2006ApJ...643..332F} describing the population synthesis of isolated radiopulsars, the death line condition was used in the form
\begin{equation}
    \dfrac{B}{P^2} = 0.17\times 10^{12}\mbox{ G/s}^2
\end{equation}
which corresponds to $P_\mathrm{death}\approx3.28\dot P^{1/3}_{-15}$ s. This relation was earlier established empirically by Rawley et al. \cite{1986Natur.319..383R} from observations of a newly discovered long-period pulsar. In spite of its purely empirical nature, it was successful in reproducing the observed population of active pulsars. However, many authors (like Gull{\'o}n et al. \cite{2014MNRAS.443.1891G} and Graber et al. \cite{2023arXiv231214848G}) have not assumed the existence of a death line at all, since pulsars become undetectable earlier due to small luminosities and beaming fractions.

At the same time, Beskin et al. recently presented a more sophisticated analysis of the processes near the polar cap of active pulsars \cite{2022MNRAS.510.2572B, 2022MNRAS.516.5084B} and found that the classical equation (\ref{eq:pulsars:death}) with $b = 4/11$ is still a good approximation to the death line condition but either $a = 8.27$ or $a = 4.6$\,s depending on the spin-down model. In fact, they have considered both a numerical MHD model based on Spitkovsky's work \cite{2006ApJ...648L..51S} and their own analytical model \cite{1993ppm..book.....B}. They argued, that the analytical model (with $a = 8.27$ s) fits the observations much better since it allows many long-period pulsars to exist. However, the pulsars with the longest known periods (J0250+5854 with $P \approx 23.5$ s, J0901-4046 with $P \approx 76$ s) are still contradict this model. So the physics of pulsar `death' is far from complete.

In addition, it is worth mentioning that the `death line' is likely to depend not only on its magnetic field (or spin-down rate), but also on the magnetic angle (see e.g. \cite{2018PhyU...61..353B}). This fact is often ignored in the analysis. A reasonable form for the `death' condition can then be approximately expressed as
\begin{equation}
    \cos\chi < A P^{15/7} B_{12}^{-8/7},
\end{equation}
where $A \approx 0.4-0.5$ from the analysis of more than 150 pulsars with known $\chi$ \cite{2020MNRAS.494.3899N} (see also \cite{2004ARep...48.1029K}), $B_{12} = B/(10^{12}\ G)$ -- is the equatorial magnetic field of a pulsar and period $P$ is in seconds. This means that nearly orthogonal pulsars shut down much earlier than aligned ones, which affects their statistics.

\subsection{Spin-down of a star with vacuum magnetosphere}
\label{sect:pulsars:vacuum}

The braking torque resulting from the non-zero Poynting flux of a magnetized star rotating in a vacuum can be calculated analytically and in great detail.

As early as November 1967 (a few months before the official discovery of 
radiopulsars by Hewish and Bell \cite{1968Natur.217..709H}) Pacini 
published a note briefly discussing the energy budget of a rotating neutron 
star with a dipolar magnetic field \cite{1967Natur.216..567P}. He suggested that such a star could lose its rotational energy with luminosity 
$L_\mathrm{sd} \sim \mu^2\Omega/R^3_\mathrm{l}$.
But more than a decade earlier, Deutsch had derived 
\cite{1955AnAp...18....1D} the first vacuum solution for 
the electromagnetic field of a perfectly conducting, rigidly rotating spherical star.\footnote{See also an 
interesting pedagogical note by Satherley and Gordon  \cite{2022PASA...39...38S} regarding Deutsch's classic 
work.} In general, the braking torque vector $\mathbf{K}_\mathrm{sd,vac}$ acting on such a star consists of 
three orthogonal terms. If $\mathbf{e}_\Omega$ and $\mathbf{e}_\mu$ are unit vectors collinear with the spin 
and magnetic axes of the star, respectively, then 
\begin{equation}
    \dfrac{\mathbf{K}_\mathrm{sd,vac}}{K_\mathrm{sd,0}} = 
    \eta_1 (r_\mathrm{NS}) \cos\chi  (\mathbf{e}_\mu \times \mathbf{e}_\omega) + 
    \eta_2(r_\mathrm{NS}) \cos\chi (\mathbf{e}_\mu \times \mathbf{e}_\omega)\times \mathbf{e}_\omega + 
    \eta_3(r_\mathrm{NS}) \sin^2\chi \mathbf{e}_\omega.
    \label{eq:pulsars:vacuum3Dtorque}
\end{equation}
Here $K_\mathrm{sd,0}~=~\mu^2/R^3_\mathrm{l}$, while $\eta_{1,2,3}(r_\mathrm{NS})$ are dimensionless functions, depending on the assumptions made about the structure of the fields inside and outside the star
\cite{1969ApJ...157.1395O, 1970ApJ...159L..81D, 1985ApJ...299..706G, 2000MNRAS.313..217M}. These functions depend on the small parameter  $r_\mathrm{NS}~=~R_\mathrm{NS}/R_l$. 

In terms of the spin-down equation (\ref{eq:pulsars:general_spindown}) one 
then has $f_\perp~\equiv~-\eta_3$.
As the third term of 
(\ref{eq:pulsars:vacuum3Dtorque}), the second term represents the 
braking torque arising due to radiation in the far (wave) zone. It is responsible for the evolution of the magnetic angle $\chi$. In the vacuum solution $\eta_2~>~0$, therefore the magnetic angle decreases systematically ($\dot\chi~<~0$) on the timescale 
of the pulsar spin-down. Moreover, typically $\eta_3~=~-\eta_2$. 

The Euler equations describing the evolution of a spherical star with moment of inertia $I$, are therefore as follows:
\begin{equation}
    P \dot P = 4\pi^2 \eta_3(r_\mathrm{NS}) \dfrac{\mu^2}{Ic^3}\sin^2\chi
    \label{eq:pulsars:Euler_P}
\end{equation}
for the evolution of the spin period and 
\begin{equation}
    P^{-2}\dot\chi = -4\pi^2 \eta_2(r_\mathrm{NS}) \dfrac{\mu^2}{Ic^3}\sin\chi\cos\chi
    \label{eq:pulsars:Euler_Pdot}
\end{equation}
for the evolution of the magnetic angle. As for the first term in 
(\ref{eq:pulsars:vacuum3Dtorque}), proportional to $\eta_1(r_\mathrm{NS})$, 
it does not directly affect the observed spin evolution of the star. Since it 
is always perpendicular to both the rotational and magnetic axes, it causes 
a precession around the magnetic axis with angular velocity
\begin{equation}
    w_1 = \eta_1(r_\mathrm{NS}) \dfrac{\mu^2}{R_\mathrm{l}^2} \dfrac{\cos\chi}{Ic}, 
\end{equation}
or, in terms of period, $T_1~\approx~6\times10^4PB^{-2}_{12}~\cos^{-1}\chi$ years. Here we assumed $I = 10^{45}$~g~cm$^2$, $R_\mathrm{NS} = 10^6$~cm, and the spin period $P$ is taken in seconds \cite{1985ApJ...299..706G, 
2012MNRAS.420..103B}. The value of $T_1$ is so short relative to the spin-down time scale at the ejector stage 
\begin{equation}
t_\mathrm{ej} \sim \dfrac{I\Omega R_\mathrm{l}^3}{\eta_3(r_\mathrm{NS}) \mu^2} \sim 3\times 10^7 P^2 B_{12}^{-2}\mbox{ years,} 
\end{equation}
because $\eta_3(r_\mathrm{NS}) \sim 1$ to a first approximation, while $\eta_1 \propto 1/r_\mathrm{NS}$.

Indeed, the first term in (\ref{eq:pulsars:vacuum3Dtorque}) can be interpreted either as a result of the 
inertia of the magnetic dipole itself \cite{1970ApJ...160L..11G}, or as a near-field radiation torque 
\cite{2000MNRAS.313..217M}, or as the appearance of an internal field momentum \cite{2018PhyU...61..353B}.  
This torque is often referred to as the `anomalous braking torque' (see e.g. \cite{2014PhyU...57..799B}). 
Melatos \cite{2000MNRAS.313..217M} has given an accurate calculation of $\eta_{1,2,3}$, assuming the 
internal magnetic field to be a point dipole field:
\begin{equation}
    \eta_1(r_\mathrm{NS}) = \dfrac{6(r_\mathrm{NS}^2 + 6)}{5r_\mathrm{NS} (r_\mathrm{NS}^6 -3 
    r_\mathrm{NS}^4 + 36)} + \dfrac{6 - 4 r_\mathrm{NS}^2}{15 r_\mathrm{NS} (r_\mathrm{NS}^2 + 1)} \approx 
    \dfrac{3}{5}r_\mathrm{NS}^{-1} - \dfrac{38}{17}r_\mathrm{NS} + ...
    \label{eq:pulsars:eta1}
\end{equation}
and
\begin{equation}
    \eta_2(r_\mathrm{NS}) = -\eta_3(r_\mathrm{NS}) = \dfrac{2r_\mathrm{NS}^4}{5(r_\mathrm{NS}^6 - 3 
    r_\mathrm{NS}^4 + 36)} + \dfrac{2}{3(r_\mathrm{NS}^2 + 1)} \approx \dfrac{2}{3} - \dfrac{2}
    {3}r_\mathrm{NS}^2 + ...
\end{equation}
respectively. Thus it is clear that if only leading order terms are taken the account, then 
$\eta_{2,3}~=~2/3$ is constant but $\eta_1~\propto~r_\mathrm{NS}^{-1}$. As a result, unlike the 
`conventional torques' which are proportional to $1/R_\mathrm{l}^3$, the `anomalous' one scales as $\propto 
1/R_\mathrm{l}^2$, making it up to four orders of magnitude larger, since $r_\mathrm{NS} \sim 10^{-4} - 
10^{-2}$ for most real pulsars. Extensive analysis made in \cite{2000MNRAS.313..217M} has shown that this 
torque could be of high importance for a non-spherical neutron star, causing significant quasi-periodic 
variations in the spin-down rate and magnetic angle of a star with a typical timescale $T_1$. The first term ($\propto r_\mathrm{NS}^{-1}$) of the equation (\ref{eq:pulsars:eta1}) has also been considered by Beskin \& Zheltoukhov in \cite{2014PhyU...57..799B}. They have shown, that if the angular momentum of the internal electric field is also taken into account, then the anomalous torque depends on the structure of the magnetic field inside the star. Thus, if there is a uniform magnetic field up to radius $R_\mathrm{in} < R_\mathrm{NS}$, while the dipole configuration is relevant for $R > R_\mathrm{in}$, then 
\begin{equation}
    \eta_1(r_\mathrm{NS}) \approx \left(\dfrac{8}{15} - \dfrac{1}{5}\dfrac{R_\mathrm{NS}}{R_\mathrm{in}}\right)r_\mathrm{NS}^{-1}.
\end{equation}
It is interesting, that $R_\mathrm{in} = 0$ leads to the divergence of the $\eta_1$ and, therefore the divergence of the anomalous braking torque. However, this situation seems to be unphysical and would never be realized.

\subsection{Spin-down of a star with plasma-filled magnetosphere}
\label{sect:pulsars:plasma}

Much work has been done in the last two decades to obtain self-consistent solutions of pulsar magnetospheres 
in both axisymmetric  \cite{1999ApJ...511..351C, 2005PhRvL..94b1101G, 2006MNRAS.368L..30M, 
2006MNRAS.368.1055T, 2012MNRAS.423.1416P} and oblique \cite{2006ApJ...648L..51S, 2009A&A...496..495K, 
2013MNRAS.435L...1T, 2012MNRAS.424..605P,2014MNRAS.441.1879P, 2015ApJ...801L..19P} configurations. 
Significant progress has therefore been made in this area. Most models have assumed the force-free 
approximation, but radiative magnetospheres have also been considered (e.g. \cite{2020Univ....6...15P}). It 
has been shown that a magnetized spherical rotator with a plasma-filled magnetosphere is indeed affected by 
a torque proportional to $\mu^2/R^3_\mathrm{l}$.  However, the corresponding dimensionless coefficient is slightly different from that found in the vacuum approximation.

In his seminal paper Spitkovsky obtained a general numerical solution for the spin-down torque of an oblique rotator 
\cite{2006ApJ...648L..51S}: 
\begin{equation}
    K_\mathrm{sd} = \dfrac{\mu^2}{R^3_\mathrm{l}}(k_0 + k_1\sin^2\chi),
    \label{eq:pulsars:MHD}
\end{equation}
where $k_0\sim k_1\sim 1$. Later Philippov et al \cite{2014MNRAS.441.1879P} refined the parameters of the 
spin-down law:
\begin{equation}
    k_0 \approx 1 \mbox{ and }
    k_1 \approx 1.4 - 3.1r_\mathrm{NS}^2.
    \label{eq:pulsars:Philippov_Coeff}
\end{equation}
That is, in terms of 
(\ref{eq:pulsars:general_spindown}) we get $f_\perp = k_0 + k_1$ and $f_\parallel = k_0$. Since for most 
real pulsars $r_\mathrm{NS} \sim 10^{-4}-10^{-2}$ is small, a good approximation is to keep 
$k_0$ and $k_1$ constant. Further PIC modelling \cite{2015ApJ...801L..19P} also produced to a result that is consistent with the results of MHD simulations. Namely, it was found that $k_0 = 1.0\pm 0.1$ and $k_1 = 1.1\pm 0.1$. If the 
radiative magnetosphere is taken into account then $k_0 = 1.36-1.42$ and $k_1 = 1.73-1.76$ 
\cite{2020Univ....6...15P}.

The solution by Philippov et al. also contains the component responsible for the magnetic angle evolution. It is with the second term of (\ref{eq:pulsars:vacuum3Dtorque}) (or with scalar equation
(\ref{eq:pulsars:Euler_Pdot})) assuming $\eta_2 \equiv 1$. So, numerical consideration of pulsar magnetospheres predicts magnetic alignment of radio pulsars on spin-down timescales.

This result is the most complete and accurate theoretical description to date of the spin evolution of a single neutron star at the ejection stage. If one does not care about the value of the magnetic angle and its evolution, then spin-down losses can be estimated quantitatively simply by assuming an isotropic distribution of magnetic angles such that $\langle \sin^2\chi \rangle = 2/3$. Thus, assuming (\ref{eq:pulsars:Philippov_Coeff}), one finally obtains 
\begin{equation}
    \langle L_\mathrm{sd} \rangle \approx 
    \dfrac{\xi\mu^2\Omega^4}{c^3},
\end{equation}
where $\xi \approx 1.93$. For actual numerical analysis, $\xi = 2$ remains a good approximation.


\subsection{Observational verification}
\label{pulsars:verify}

In general, one has to accept that the theory of pulsar spin-down described above allows one to understand and predict the properties of the actual ejectors in much detail. On the other hand, the large number of neutron stars observed as active radio pulsars and the high precision of measurements of their timing parameters ($P$, $\dot P$, $\ddot P$, etc.) suggest that this theory can indeed be well verified in observations. But the reality is much more complicated. Despite many indirect arguments, there is no direct observational evidence to support the theory.

In this subsection, without pretending to provide a complete review, we briefly describe three relevant issues that remain unresolved and deserve to be discussed.

\subsubsection{Magnetic angle evolution}
The evolution of the magnetic angle $\chi$ during the spin-down of an isolated neutron star is an important part of its evolution. In a more general approach, the Euler equations (\ref{eq:pulsars:Euler_P}) and (\ref{eq:pulsars:Euler_Pdot}) can be rewritten in terms of the so-called symmetric and anti-symmetric components of the magnetospheric currents normalized to the Goldreich-Julian current: $i_\mathrm{s} = j_\mathrm{s}/j_\mathrm{GJ}$ and $i_\mathrm{a} = j_\mathrm{a}/j_\mathrm{GJ}$ respectively (see e.g. \cite{2018PhyU...61..353B}). In this approach one gets \cite{1993ppm..book.....B}:
\begin{equation}
    P\dot P \propto i_\mathrm{s} + (i_\mathrm{a} - i_\mathrm{s})\sin^2\chi
\end{equation}
and
\begin{equation}
    P^{-2}\dot\chi \propto (i_\mathrm{a} - i_\mathrm{s}) \sin\chi \cos\chi.
\end{equation}
It is clear from these equations, that the magnetic angle tends to evolve in such a way that the loss of rotational energy decreases with time. This means that if a perfectly aligned rotator ($\chi = 0$) will spin down less quickly relative to the perfectly orthogonal one ($\chi = 90^\circ$), then one should expect a magnetic alignment ($\dot\chi < 0$). On the other hand, if the physics of pulsar spin-down suggests that an orthogonal rotator loses rotational energy less efficiently, then magnetic orthogonalisation should occur ($\dot\chi > 0$).

As shown in Section~\ref{sect:pulsars:vacuum}, spin down is suppressed in vacuum magnetospheres when $\chi = 0$, and hence $\dot\chi < 0$ there. The same result has also been obtained in numerical simulations of plasma-filled magnetospheres. However, in the 1980s a theory was developed by Beskin, Gurevich and Istomin \cite{1984Ap&SS.102..301B, 1993ppm..book.....B, 2007Ap&SS.308..569B, 2013PhyU...56..164B} which assumes a non-vacuum magnetosphere of a neutron star, but predicts magnetic counter-alignment.
The MHD-based theory presented in the paper by Philippov \cite{2014MNRAS.441.1879P} and the analytical description by Beskin et al. (BGI theory) give clear predictions about the evolution of the magnetic angle from fundamental physics. To our knowledge, these are the only two such considerations published to date. And their predictions are not consistent with each other. On the other hand, there is still no observational evidence in favor of one of any of these approaches, although many analyses have been carried out \cite{1990ApJ...352..247R, 1998MNRAS.298..625T, 2006ApJ...643..332F, 2008MNRAS.387.1755W, 2010MNRAS.402.1317Y, 2013Sci...342..598L, 2015MNRAS.453.3540A, 2020MNRAS.494.3899N}. So this part of the rotational evolution of an isolated NS is not entirely clear.

\subsubsection{Pulsar timing irregularities and braking indices}
The spin-down equation (\ref{eq:pulsars:general_spindown}) assumes that an isolated ejector loses its rotational energy monotonically. However, the actual observed spin evolution of such objects is not like this. At least about 6 per cent of known radio pulsars exhibit sudden and rapid spin-up events, known as glitches \cite{2011MNRAS.414.1679E, 2015IJMPD..2430008H, 2022Univ....8..641Z}. However, the spin-down rate $\diff P/\diff t$ of almost every pulsar appears to be significantly contaminated by additional irregular, almost stochastic variations on short time scales of months to years, typically referred to as `timing noise' \cite{1972ApJ...175..217B, 1980ApJ...239..640C, 1995MNRAS.277.1033D, 2006MNRAS.370L..76U, 2010MNRAS.402.1027H, 2013ApJ...772...50N}.

The strength of this unmodelled component of the pulsar spin-down can be characterized numerically in various ways. The standard approach is to use either the second derivative of the pulsar spin period \cite{1994ApJ...422..671A} or the variance of the timing residuals (e.g. \cite{1980ApJ...239..640C}). These two measures are correlated. There is also a widely used dimensionless combination:
\begin{equation}
	n_\mathrm{br} = \dfrac{\ddot \Omega \Omega}{\dot \Omega^2} = 2 - \dfrac{\ddot P P}{\dot P^2}
	\label{eq:pulsars:bi_obs}
\end{equation}
-- the so-called `braking index'. It has a simple and clear physical meaning. In particular,
if the loss of rotational energy follows the expression $L_\mathrm{sd}=-K \Omega^{n+1}$ with $K=const$, then the combination (\ref{eq:pulsars:bi_obs}) is nothing other than $n$.

Since the spin-down law for a spherical star is $I\dot\Omega=-f(\chi)\mu^2\Omega^3/c^3$, by direct differentiation it is clear that
\begin{equation}
	n_\mathrm{br} = 3 -  2\left [2\dfrac{1}{\mu}\dfrac{\diff\mu}{\diff t} - \dfrac{1}{I}\dfrac{\diff I}{\diff t} + \dfrac{1}{f}\dfrac{\diff f}{\diff \chi} \dot\chi \right]\tau_\mathrm{ch},
	\label{eq:pulsars:bi}
\end{equation}
where the parameters of a neutron star are assumed to be variable and 
\begin{equation}
    \tau_\mathrm{ch}=-\dfrac{\Omega}{2\dot \Omega} = \dfrac{P}{2\dot P}
    \label{eq:pulsar:tauch}
\end{equation}
characteristic age of a pulsar. Classical vacuum losses ($f_\perp=2/3$, $f_\parallel=0$) with constant magnetic fields and moment of inertia give $n_\mathrm{br}=3+2\cot^2\chi \lesssim 10^3$ for real pulsars. Only the third term in the brackets is non-zero in this case due to $\dot\chi < 0$. The theory by Beskin et al. \cite{1984Ap&SS.102..301B} gives a similar result with $n_\mathrm{br}\approx1.93+1.5\tan^2\chi$. On the other hand, Spitkovsky's model \cite{2006ApJ...648L..51S} predicts a narrower interval for this quantity under the same assumptions (see also \cite{2016ApJ...823...34E}). Namely,
\begin{equation}
    n_\mathrm{br} = 3 + 2 \dfrac{\sin^2\chi \cos^2\chi}{(1 + \sin^2\chi)^2} \in 3-3.25.
\end{equation}
But a big problem is that the estimated values of $n_\mathrm{br}$ for hundreds of pulsars are surprisingly far from all these predictions. The values have been found to range from $\sim -10^6$ to $\sim 10^6$, and are negative for about half of the objects (e.g. \cite{2004MNRAS.353.1311H, 2012MNRAS.420..103B, 2012ApJ...761..102Z}). They are therefore unlikely to represent the secular spin evolution of radio pulsars. In addition, the braking indices of most pulsars do not seem to be stable from observation to observation\footnote{However, there are a few dozen `low noise' pulsars whose $n_\mathrm{br}$ are constant within spans of 10-30 years, but are still extremely anomalous \cite{2007AdSpR..40.1498B}.} There are only about a dozen sources that are accepted as having meaningful and stable $n_\mathrm{br} \in 0.03$-$3.15$ \cite{2016ApJ...819L..16A, 2016ApJ...827L..39M}.

Although the physics of pulsar timing irregularities remains generally unclear, the proposed solutions to the `anomalous braking index' problem can be qualitatively divided into two categories. On the one hand, there are theoretical models that assume the relatively slow variability of the NS parameters affecting equation (\ref{eq:pulsars:bi}). The authors of these ideas assume either the variability of the magnetic moment \cite{2012A&A...547A...9P, 2012ApJ...761..102Z, 2016MNRAS.457.3922O, 2020MNRAS.499.2826I} or the magnetic angle \cite{2000MNRAS.313..217M, 2013Sci...342..598L, 2015MNRAS.453.3540A}, or the effective moment of inertia \cite{2006A&A...457..611J, 2013ApJ...773L..17T, 2014MNRAS.437...21M, 2015PhRvD..91f3007H, 2016PhRvD..94f3012H} of the star. On the other hand, some researchers suggest the existence of an additional either quasi-periodic or purely stochastic external component in the spin-down torque $\dot \Omega_\mathrm{ext}$. The underlying physics has been proposed to be related either to magnetospheric perturbations \cite{1987ApJ...321..799C, 2006Sci...312..549K, 2007A&A...475..639C, 2010Sci...329..408L}, or the imprint of the anomalous vacuum torque discussed above \cite{2007AdSpR..40.1498B, 2010MNRAS.409.1077B, 2012MNRAS.420..103B}. 

Biryukov et al. \cite{2012MNRAS.420..103B} attempted to estimate $\dot\Omega_\mathrm{ext}$ from the statistics of the observed second derivatives of pulsar spin frequencies. Assuming that anomalous braking indices are due to long-term variations in pulsar spin-down (with typical timescales of hundreds and thousands of years), they found $0.5 < \varepsilon < 0.8$, where $\varepsilon = |\dot \Omega_\mathrm{ext}/\dot\Omega|$ -- is the relative strength of the effect. 

On the other hand, it has been shown in many papers that the observable parameters of Galactic pulsars can be well reproduced within a population synthesis adopting (\ref{eq:pulsars:general_spindown}) with various forms of $f_\perp$ and $f_\parallel$ \cite{2006ApJ...643..332F, 2010MNRAS.404.1081R, 2014MNRAS.443.1891G}  but neglecting the corrections due to $\varepsilon \neq 0$. Moreover, the model-independent analysis of pulsar kinetics in the $P-\dot P$ diagram even led to a reasonable value of their birth rate of $\sim 1$ per century, simply assuming $f_\perp=f_\parallel\equiv 1$ and $\varepsilon \equiv 0$ \citep{2008MNRAS.391.2009K, 2011MNRAS.410.2363V}. This ultimately means that the simple equation for the isolated ejector spin-down torque $K_\mathrm{sd}=\xi\mu^2\Omega^3/c^3$, where $\xi\sim 1$ is a `universal constant', is still meaningful within a numerical analysis.

\subsubsection{Isolated neutron stars birthrate}
Knowledge of the birth rate $\dot N_\mathrm{INS}$ of Galactic isolated neutron stars (INSs) is also important in the context of neutron star population analysis and, in particular, details of their spin down. In general, several types of objects contribute to the total value of $\dot N_\mathrm{INS}$: (i) ordinary rotation-powered pulsars (RPPs), (ii) rotating radio transients (RRATs), and (iii) young radio-quiet neutron stars (XDINS or the ``Magnificent Seven'')\footnote{Although few other objects have also been attributed to XDINS (e.g. \citep{2009A&A...498..233P,2019A&A...627A..69R}) in addition to seven classical sources \cite{2009ASSL..357..141T}.},
(iv) central compact objects in supernova remnants (CCOs) and (v) magnetars.

The birthrate of each of them, however, is not simply proportional to the number of observed sources but must be revealed assuming models of their spin-down, activity, magnetic and thermal evolution. For instance, despite the overwhelming number of ordinary pulsars, a significant fraction of neutron stars appear to be born as magnetars \cite{2000ASPC..202..699G, 2019MNRAS.487.1426B}.

However, the total INSs birthrate is not simply the sum of birthrates of their different types. Keane and Kramer have shown that being simply summarized they give the value $\gtrsim 5-10$ century$^{-1}$ \cite{2008MNRAS.391.2009K} which is in tension with a proposed rate of Galactic core-collapse supernova explosions
\begin{equation}
	\dot N_{\rm CCSN} = 1-3\mbox{ century}^{-1}
\end{equation}
as it has been constrained by different methods \cite{1994ApJS...92..487T, 2006Natur.439...45D, 2008MNRAS.391.2009K, 2014AN....335..935S}.

Within these calculations, the birthrate of RPPs, as the most numerous class of observed sources, can be estimated with formally small uncertainties. And understanding of their spin-down evolution is crucial, since $\dot N_\mathrm{RPP} \sim N_\mathrm{RPP}/t_\mathrm{ej}$.

So the analysis of pulsar kinetic current by Vrane{\v s}evi{\'c} et al. \cite{2004ApJ...617L.139V} initially gave $\dot N_{\mathrm{RPP}} \sim 1-2$ century$^{-1}$. Later Lorimer et al. \cite{2006MNRAS.372..777L} used the same approach and estimated $\dot N_{\mathrm{RPP}}$ to be $1.4\pm 0.2$ century$^{-1}$. And finally, Vrane{\v s}evi{\'c} \& Merlose \cite{2011MNRAS.410.2363V} perform an even more accurate consideration, taking into account the pulsar spin-down law in the more general form $P^{n-2}\dot P = $const. They found $\dot N_{\mathrm{RPP}} = 0.8\pm 0.3$ century$^{-1}$ for $n = 2.9\pm 0.1$. All these results are more or less consistent with each other.

Another way of measuring the birth rate of RPPs -- a population synthesis approach -- tends to give even higher values (probably due to selection effects being taken into account). For example, Faucher-Gigu{\`e}re and Kaspi \cite{2006ApJ...643..332F} reported $\dot N_{\mathrm{RPP}}=2.8\pm0.5$, Ridley and Lorimer found $\dot N_{\mathrm{RPP}}$ from $\sim 2$ to $5.3$ century$^{-1}$ depending on the details of the assumed spin-down model \cite{2010MNRAS.404.1081R}. Although a spin-down law of the form $P\dot P\propto\mu^2(k_0+k_1\sin^2\chi)$ was included in the calculations in the latter paper, the exponential decay of a pulsar obliquity was also assumed. which is inconsistent with the MHD model of pulsar rotational evolution. Finally, Gull{\'o}n et al. found $\dot N_{\mathrm{RPP}}=2.5\pm0.1$ century$^{-1}$ including pulsar magneto-thermal evolution in their simulations \cite{2014MNRAS.443.1891G}. Therefore, the pulsar population synthesis generally predicts values of $\dot N_{\mathrm{RPP}}$ within 2-3 century$^{-1}$ -- about one pulsar per century more than the kinetic current approach and which is close to the rate of core-collapse supernovae themselves.

The obvious solution to the paradox of the overestimated birthrates of Galactic neutron stars is to assume that their populations are connected. Their observable properties depend on their current spin period, age, magnetic field strength and surface temperature, while the transition between the populations is driven by the magneto-rotational evolution \cite{2010MNRAS.401.2675P, 2010PNAS..107.7147K, 2014MNRAS.444.1066I, 2015MNRAS.454..615G}. The latter is therefore crucial for understanding the properties and fate of INS. They need to be considered as a connected population of sources, unified by the common origin, rather than a set of separate classes\footnote{Kaspi proposed the term 'Grand unification' for this \cite{2010PNAS..107.7147K}.}

\section{Propeller stage}
\label{propeller}

The propeller stage is, probably, the least understood one. On one hand, from the observational point of view, it is not easy to study this phase of evolution. This is so, partly because an NS luminosity can be rather low at this stage. In addition, the stage can be relatively short in many cases. On the other hand, the theoretical analysis of propellers is complicated and results of modeling often are not conclusive. 
We start by presenting different theoretical approaches to model the propeller stage. And then, observational results are reviewed and discussed. 

\subsection{Theory}


The idea that a rapidly rotating magnetosphere can prevent penetration of matter down to the surface of an NS was initially proposed by Shvartzman \cite{1970R&QE...13.1428S}. However,  as the paper by Shvartsman has been published in a not-so-well-known journal, this idea became popular only later, mainly due to the paper by Illarionov and Sunyaev \cite{1975A&A....39..185I} who also proposed the name for the evolutionary stage: propeller. The existence of the centrifugal barrier for accretion onto NSs was also mentioned by \cite{1972A&A....21....1P}, and discussed in more detail by \cite{1973ApJ...179..585D}, who also provided an estimate for the spin-down torque: $K_\mathrm{sd}={\mu^2}/{R_\mathrm{m}^3}.$
Later on, a more detailed study of the propeller stage was presented by  \cite{1979MNRAS.186..779D}, \cite{1981MNRAS.196..209D}. 

Realistic 3D modeling of the interaction between the magnetosphere and surrounding matter is a very challenging task. Not many attempts have been made. That's why, early studies were analytical and typically limited by some simplified but physically motivated approaches.

At the propeller stage, a rapidly rotating magnetosphere transfers rotational energy and angular momentum to the medium. Thus, the matter is heated and moves away carrying some angular momentum, so that the NS spins down. 
Depending on the relative efficiency of heating and cooling, the properties of the shell around the magnetosphere can be different. External pressure at the magnetospheric boundary determines its radius, $R_\mathrm{m}$, which can significantly deviate from the Alfven radius, $R_\mathrm{A}$, see eq.~(\ref{ra}). Without a disc formation, typically $R_\mathrm{m}>R_\mathrm{A}$.

In the simplest approach, e.g. \cite{1975SvAL....1..223S}, it is assumed that the external matter is gravitationally captured
 and cools effectively. Thus, 
 
 $$\rho(R_\mathrm{m})=\rho_\infty \left(\frac{R_\mathrm{m}}{R_\mathrm{G}} \right)^{-3/2}.$$ 
 This provides a large density at the magnetospheric boundary. 
 The magnetospheric boundary is determined from 
 
 $$\frac{B(R_\mathrm{m})^2}{8\pi}=\frac12\rho(R_\mathrm{m})(\Omega R_\mathrm{m})^2.$$ 
  So, $R_\mathrm{m}\propto B^{4/13} P^{4/13}\rho_\infty^{-2/13} v^{6/13}$.  
 Then, it is assumed that the matter is expelled with the rotational velocity of the magnetosphere: $\Omega R_\mathrm{m}$. Thus, it carries away a huge amount of angular momentum. Altogether, this results in a very effective 
spin-down:

 \begin{equation}
     K_\mathrm{sd}=\dot M \Omega R_\mathrm{m}^2.
     \label{eq:shakura75}
 \end{equation}
So, the spin period increases rapidly. 

Both assumptions (effective cooling and large velocity of the expelled matter) in this approach were strongly criticized.  
At first, due to significant heating, the density at $R_\mathrm{m}$ might be much lower than 

$$\rho(R_\mathrm{m})=\rho_\infty \left(\frac{R_\mathrm{m}}{R_\mathrm{G}} \right)^{-3/2}.$$ 
Then, the velocity $\Omega R_\mathrm{m}$ is too high for the outflow. 

Davidson and Ostriker \cite{1973ApJ...179..585D} assumed that the angular momentum is carried away with the free-fall velocity:

\begin{equation}
    K_\mathrm{sd}= \dot M v_\mathrm{ff}(R_\mathrm{m}) R_\mathrm{m} .
\end{equation}

In the seminal paper \cite{1975A&A....39..185I} the authors define the magnetospheric boundary as the usual Alfven radius and it is assumed that rotational energy loss is $(1/2)\dot M v_\mathrm{ff}(R_\mathrm{m})^2$. 
Then the spin-down torque is:

\begin{equation}
    K_\mathrm{sd}= \dot M v_\mathrm{ff}(R_\mathrm{m}) R_\mathrm{m} \frac{v_\mathrm{ff}(R_\mathrm{m})}{\Omega R_\mathrm{m}}.
\end{equation}

Fabian \cite{1975MNRAS.173..161F} obtained the radius of the magnetospheric boundary by equating the magnetic pressure and stellar wind pressure. To calculate the spin-down torque he used energetic considerations. In this approach, the rotational energy losses $\frac{1}{2}I(\Omega_1^2-\Omega_2^2)$ are equal to the kinetic energy of escaping matter 
$$\frac12 \dot M \frac{2GM}{R_\mathrm{m}}.$$
In this model, the spin-down rate is identical to the one by Illarionov, Sunyayev \cite{1975A&A....39..185I}. 
It can be written in an alternative form:

\begin{equation}
    K_\mathrm{sd}=\dot M \frac{GM}{R_\mathrm{m}} \frac{1}{\Omega}.
\end{equation}
Note, that $R_\mathrm{m}$ can be defined in different ways. What is important, is that $K_\mathrm{sd}\propto \Omega^{-1}$. 
Then the characteristic time scale of spin-down $\tau_\mathrm{sd}\equiv P/\dot P \propto (P_1^{-2}-P_2^{-2})$. This means the time scale is defined by the initial period. 

In their very influential paper Davies and Pringle \cite{1981MNRAS.196..209D} discussed several regimes of the propeller. In their notation, the most standard case, typically discussed in the literature, is dubbed ``supersonic propeller''. In this regime, $R_\mathrm{m}>R_\mathrm{co}$ and also typically $R_\mathrm{m}\ll R_\mathrm{l}$. At the subsonic propeller stage $R_\mathrm{m}<R_\mathrm{co}$, accretion is still not possible as matter cannot cool efficiently. Finally, ``a very rapid rotator'' regime can be realized in the situation when $R_\mathrm{m}\lesssim R_\mathrm{l}$,  i.e. when the magnetospheric radius is just slightly below the light cylinder radius.

One of the main points, used and discussed by Davies and Pringle is related to the thermal balance in the envelope around the magnetosphere. The existence of a hot envelope results in a different pressure outside the magnetosphere with respect to the case of an efficiently cooled inflow assumed for calculation of $R_\mathrm{A}$. 

At the supersonic propeller stage, the magnetospheric radius is:

\begin{equation}
    R_\mathrm{m}=R_\mathrm{A}\left(\frac{ R_\mathrm{G} } {R_\mathrm{A}}\right)^{2/9}.
\end{equation}

It is assumed that at this stage the energy released at the magnetospheric boundary is carried away by convection or turbulent motions. The turbulent velocity at any radius is assumed to be equal to the sound velocity. These assumptions result in the polytropic index $n=1/2$ and $\rho \propto r^{-1/2}$. 

Thus, we obtain:

\begin{equation}
    K_\mathrm{sd}=\frac{1}{8\sqrt{2}}\left( \frac{R_\mathrm{co}}{R_\mathrm{m}} \right)^{3/2} \frac{\mu^2}{R_\mathrm{m}^3}.
\end{equation}
Matter is expelled from the magnetospheric boundary with the velocity significantly lower than the free-fall velocity.

At the phase of a very rapid rotator, the linear velocity of the rotating magnetosphere, $\Omega R_\mathrm{m}$, is very high. 
This heats up the surrounding matter significantly. Effectively, this means that pressure is constant in the envelope up to ``infinity'', i.e. up to $R_\mathrm{G}$ where properties are determined by the external medium (e.g., ISM in case of an isolated NS). This stage can be realized at low accretion rates or with high magnetic fields when the outer boundary of the heated envelope extends beyond $R_\mathrm{G}$. Otherwise, the pressure gradient becomes important and so, from the ejector stage the NS immediately come to the supersonic propeller stage.

At the very rapid rotator stage matter is expelled with the rotational velocity of the magnetosphere, but the density is low, so the spin-down rate is low, too:

\begin{equation}
    K_\mathrm{sd}\approx \frac{\mu^2}{R_\mathrm{m}^3}.
\end{equation}

In addition to the propeller stage for which $R_\mathrm{m}>R_\mathrm{co}$, Davies and Pringle \cite{1981MNRAS.196..209D} discussed also a ``subsonic propeller'' stage with $R_\mathrm{m}<R_\mathrm{co}$ (the stage was later re-considered e.g., in \cite{2001A&A...368L...5I}). Here, despite the centrifugal barrier does not exist, matter cannot effectively accrete due to its high temperature. For low accretion rates, the subsonic stage can be quite long, and accretion can be postponed. However, instabilities at the magnetospheric boundary might allow the matter to accrete. This situation was recently analyzed in the framework of the so-called settling accretion model \cite{2012MNRAS.420..216S}. 

Now, let us say a few words about the numerical modeling of the propeller stage. 
Wang and Robertson \cite{1985A&A...151..361W} used numerical simulations to model spin-down at the propeller stage. 
Specifically, they accounted for instabilities at the magnetospheric boundary. 
They obtained a very rapid spin-down:

\begin{equation}
    K_\mathrm{sd}=\dot M \frac{2 \zeta_\mathrm{WR}}{\pi \beta} \Omega R_\mathrm{m}^2.
\end{equation}
The pre-factor $({2 \zeta_\mathrm{WR}})/({\pi \beta})$ is expected to be $>1$.
Notice the similarity with the equation by Shakura \cite{1975SvAL....1..223S}, see eq. (\ref{eq:shakura75}). 

What is also important, in their model $R_\mathrm{m}$ is a function of the spin period, which is quite natural at the propeller stage, but this is typically not taken into account explicitly in many analytical models.

Numerical modeling of the propeller stage in the case of an isolated NS interacting with the interstellar medium was performed in \cite{2003ApJ...588..400R}, see also their earlier study in \cite{1999ApJ...517..906T}.
 Fitting of the numerical results provided the following scaling: $K_\mathrm{sd}\propto B^{0.8} \Omega^{1.3}$. 
This is a very rapid spin-down, similar to the one proposed in \cite{1975SvAL....1..223S}, where $K_\mathrm{sd}\propto  \Omega$.
Curiously, the dependence $\propto B^{4/5}$ can also be obtained analytically, as discussed in \cite{2003ApJ...588..400R}. 
At first, the authors define the radius of the magnetospheric boundary from the condition 

$$\frac{B^2}{8 \pi} \frac{1}{R_\mathrm{m}^6} = \rho (R_\mathrm{m}) (\Omega R_\mathrm{m})^2.$$ 
So, that 

$$R_\mathrm{m}=\left( \frac{f \mu^2}{2 \dot M \Omega} \right)^{1/5}.$$
And then, the spin-down torques is defined as 

$$I \frac{\diff \Omega}{\diff t}= - \frac{f \mu^2}{2 R_\mathrm{m}^3}.$$ 
Note, that here again $R_\mathrm{m}$ depends on the spin period. 

Francischelli and Wijers \cite{2002astro.ph..5212F} compared several variants of the outflow velocity at the propeller stage and application to fossil discs around young NSs and to the source SMC X-1. They concluded that the highest velocity -- $\Omega R_\mathrm{m}$, -- is not compatible with observations of SMC X-1.





\subsection{Observations}

Observational data about NSs at the propeller stage are not very abundant. Mainly, they are limited to compact objects in low-mass (LMXBs) or high-mass (HMXBs) X-ray binaries with a disc formation. 
 Unfortunately, measurements of the spin-down rate at the propeller stage are not done, yet. 
Still, it is possible to determine the critical luminosity at which the transition to this stage occurs. 

The first identification (not very secure at that time) of the propeller stage was probably made for the X-ray pulsar V0332+53 in
\cite{1986ApJ...308..669S}.  
Some other candidates have been also identified in this paper, including 4U 0115+63 which was later studied by \cite{2001ApJ...561..924C}.   
This source demonstrated a decrease of the X-ray flux by more than two orders of magnitude in nearly one-half of a day. 
Another example -- Aql X-1, -- was presented in \cite{1998ApJ...499L..65C}.  
This object is of special interest as it has a millisecond spin period. 

Several other sources demonstrate behavior which can be explained as on-set of the propeller stage: 
SAX J1808.4-3658 \cite{2008ApJ...684L..99C}, 
 ULX M82 X-2 \cite{2016MNRAS.457.1101T}, 
 GRO J1744-28, and GX 1+4 \cite{1997ApJ...482L.163C}. 
  
In \cite{2017A&A...608A..17T} 
 the authors present observations of GRO J1008-57. The initial goal was to detect the transition to the propeller stage. Surprisingly, at some value (larger than the expected critical number) the flux decrease stopped. Thus, the authors conclude that the source instead of reaching the propeller stage, changed the mode of accretion. 
In this case, the disc at $R_\mathrm{m}$ has temperature $T<6500$~K. Interestingly, this regime is valid only for pulsars with relatively long spin periods: $P>36.6 \, B_{12}^{0.49}$ s. For smaller periods a usual propeller regime exists. 

In \cite{2018A&A...610A..46C}  the authors consider various systems (LMXBs, HMXBs, cataclysmic variables, and young stars) for which transitions to/from the propeller stage have been proposed. 
Among the systems with NSs the authors discuss three LMXBs -- SAX J1808.4–3658, IGR J18245–2452, and GRO J1744–28, --
and three HMXBs -- 4U 0115+63, V 0332+53, and SMC X-2. In addition, three cataclysmic variables with accreting white dwarf and one young stellar object have been considered.

The authors tested the basic equation for the critical luminosity at which the transition to the propeller stage happens: $L_\mathrm{lim}\propto \mu^2 P^{-7/3} R^{-1}$. 
The analyzed systems provide the possibility to check this equation for a wide range of magnetic moments, spin periods, and radii of accretors. Surprisingly, the observed data perfectly confirm the dependences. 
In particular, the authors parametrized the equation as: 
$L_\mathrm{lim}\propto \mu^{\alpha} P^{-\beta}R^{-\gamma}$. And the fit gives: $\alpha=1.9^{+0.6}_{-0.3}$, $\beta=2.3^{+0.8}_{-0.4}$, $\gamma=0.9^{+0.3}_{-03}$. 


More recently, the source GRO J1750–27 was proposed as a system with accretion/propeller transition \cite{2019MNRAS.485..770L}. The authors observed a drastic drop in the X-ray flux accompanied by a decrease in the spin-up rate of the X-ray pulsar. Interpretation of this decrease in luminosity as transition to the propeller stage allows (for the known distance) determination of the magnetic field of the NS: $\sim 4\times10^{12}$~G.

An X-ray binary SWIFT J0850.8-4219 with a red supergiant donor is proposed to be at the propeller stage due to its low luminosity in comparison with expectations based on typical stellar wind mass losses from red supergiants \cite{2023arXiv230907833D}. 
 Still, this hypothesis deserves a more detailed analysis as the present-day data are not sufficient to make a firm conclusion.

Future instruments, like Athena \cite{2012arXiv1207.2745B}, with large collecting area might allow detecting pulsations at low fluxes corresponding to the propeller stage. This will give an opportunity to test directly theoretical models of spin-down.

\section{Accretion}
\label{accretion}

\subsection{General properties of accretion flows onto magnetized NSs}

NSs may accrete mass under various circumstances and from different sources, including ISM, gravitationally bound supernova ejecta (the case of fallback, see section~\ref{fallback}), or a donor star. 
Interaction with the accreting matter is determined by the set of characteristic radii introduced in section~\ref{introduction}: if both Shvartsman and magnetospheric radii are smaller than $R_{\rm G}$, mass is gravitationally captured.  
The flows inside the gravitational capture radius depend on the amount of linear and angular momentum in the matter captured by the star. 
If both are negligibly small, the star accretes spherically in the regime of Bondi accretion \citep{1952MNRAS.112..195B}. 
Such a regime would be realized if the star is immersed in a static uniform gas. 
The gravity of the star creates a radial transonic inflow of matter that in its inner parts is modified by the presence of the solid surface and magnetosphere.
Though unrealistic, Bondi accretion model allows to estimate the mass accretion rate in the case of subsonic motion through a uniform ambient medium according to formula (\ref{mdot}). 

Accretion from a \emph{supersonically} moving uniform media may be thought of as gravitational focussing of the flow: gravity changes the direction of a streamline with a shooting parameter $b$ by an angle $\sim R_{\rm G}/b$, resulting in a shock wave of a roughly conical shape that dissipates some of the kinetic energy of the flow and allows for mass accretion onto the star from the downstream direction. 
This is an idealized case of Bondi-Hoyle-Lyttleton accretion \citep{2004NewAR..48..843E}. 

An important factor that modifies the two scenarios mentioned above is the net angular momentum of the captured matter. 
For the net angular momentum of $j$ the size of the disc is determined by the circularization radius
\begin{equation}
    R_{\rm c}  = \frac{j^2}{GM}.
\end{equation}
If $R_{\rm c} \gg R_{\rm m}$, the dynamics of the magnetosphere are determined by its interaction with the disc. 
In the case of Roche lobe overflow, $j \sim \sqrt{GM_{\rm tot} a}$, where $a$ is the binary separation in the system, and $M_{\rm tot}$ is the total mass of the system, hence $ R_{\rm c}  \sim a$ (though usually several times smaller). 
If the NS captures the stellar wind of its companion, $j$ is much smaller, and the accretion flow is much better described by the Bondi-Hoyle-Lyttleton scenario (see however section~\ref{sec:acc:qsph}). 

All the accretion flows described above exist only outside the magnetosphere. 
Within the magnetosphere, magnetic stresses redirect the plasma along the field lines, creating \emph{funnel flows} that start at $R\sim R_{\rm m}$ in the regions where the field lines interact with the inflow. 
Making assumptions about the shape of the magnetic field lines allows us to predict the shape of the funnel flow and the regions of the NS surface where the accreting matter finally falls and releases its gravitational energy. 
For a dipolar magnetic field, these accretion sites are associated with the polar caps or rings or arcs around them. 
If a dipolar magnetosphere acquires mass in its equatorial regions at a distance of $R_{\rm m}$, the matter, following a field line, hits the NS surface at a polar angle $\sin \theta \sim \sqrt{R_{\rm NS}/R_{\rm m}} \sim 0.01$ (for the values typical for an accreting strongly magnetized NS, see section~\ref{sec:acc:zoo}).

Accretion onto magnetized stars is a subject of many numerical studies, that are often designed for young stellar objects but are also applicable for NS accretion with an identical set of dimensionless parameters. 
To describe the accretion flow geometry and torques acting on an NS with a dipolar magnetic field accreting from a thin disc, the required dimensionless parameters are the so-called fastness parameter $\omega = \Omega / \Omega_{\rm K}(R_{\rm m})$ (where $\Omega_{\rm K}(R) = GM/R^3$ is the local Keplerian frequency), $R_{\rm m} / R_{\rm NS}$, and magnetic angle $\chi$. 
The rotation of the NS is assumed aligned with the rotation in the disc. 
We will talk about such simulations in more detail later in Sec.~\ref{sec:acc:down}. 
Simulations of Bondi and Bondi-Hoyle-Lyttleton accretion onto magnetized stars are much less numerous (see however the following section \ref{sec:acc:ains}).

\subsection{Isolated neutron stars}\label{sec:acc:ains}

Accretion onto isolated NSs from the ISM was proposed more than half a century ago by Shvartsman
 \cite{1971SvA....14..662S, 1970Ap......6...56S} 
and Ostriker, Rees, and Silk  \cite{1970ApL.....6..179O} 
(the papers by Shvartsman were submitted slightly earlier).  
All of these authors correctly noticed that due to gradual spin-down or/and significant magnetic field decay an NS finally can start accreting the interstellar gas. In \cite{1970ApL.....6..179O} it is estimated that for a typical pulsar magnetic field accretion onto a low-velocity NS can start in a few hundred million years due to spin-down and corresponding decrease of the pulsar wind power which prevents accretion at earlier times.  

These authors applied Bondi accretion to estimate the luminosities of accreting isolated neutron stars (AINSs), see eq.~(\ref{mdot}). 
As they mainly considered low-velocity NSs (in 1969-1970 it was not known that compact objects could obtain large kicks), the estimated luminosity appeared to be $\sim10^{31}$~-~$10^{32}$~erg~s$^{-1}$. Shvartsman also accounted for a `negative feedback' related to the heating of the external medium by the NS emission. This somehow lowered the expected luminosity and shifted the spectrum from X-rays to UV. 

Interestingly, in the original papers \cite{1971SvA....14..662S,1970ApL.....6..179O} the authors did not discuss in detail the possibility of detecting AINSs as individual X-rays sources but mainly focused on the influence (heating and ionization) of the accretion luminosity on the surrounding medium. Later in the 1990s different authors intensively discussed observations of individual accreting isolated sources with space X-ray observatories.

Blaes et al. analyzed the observability of AINSs in several papers 
\cite{1991ApJ...381..210B, 1993ApJ...403..690B, 1994ApJ...423..748M, 1995ApJ...454..370B}.
In particular, in \cite{1993ApJ...403..690B} the authors calculated the number of AINSs potentially detectable by ROSAT, and obtained that about several thousand such sources can be found. 
This very optimistic prediction is mainly due to the fact that the initial velocities of NSs were underestimated at that time. Despite the authors discussed some evolutionary aspects, finally, for their estimates they assumed that all old NSs accrete at the Bondi rate. This is, of course, an oversimplification. 
Reasonable accounting for spin evolution for realistic velocities of NSs has been done only in 
 \cite{2000ApJ...530..896P} and the realistic magnetic field distribution was added later by \cite{2010MNRAS.407.1090B}.  
In \cite{1995ApJ...454..370B} the authors considered the influence of the AINS emission onto the surrounding medium and `negative feedback' on a deeper level than it has been done in early studies. Naturally, this process reduces the number of potentially observable AINSs. 

3D hydrodynamical simulations of accretion onto an INS were presented in \cite{2012ApJ...752...30B}.  
 Magnetic field was not explicitly included in calculations but the authors modeled accretors of different sizes comparable with magnetospheric radius. The accretion rate was found to be reduced by a factor $\sim 2$ in comparison with the Bondi-Hoyle-Littleton rate (eq.~\ref{mdot} with $\xi_\mathrm{acc}=\pi$).
 
 Accretion onto isolated magnetized NSs was studied numerically in a series of papers by Romanova and her colleagues \cite{1999ApJ...517..906T,2001ApJ...561..964T,2003ApJ...588..400R, 2012MNRAS.420..810T}. The main result presented in \cite{2012MNRAS.420..810T} is related to the dependence of the accretion rate on the magnetic field of the NS. 
 For magnetic field $\sim10^{12}$~G the authors obtain a reduction in the accretion rate by a factor $\sim 2$ with approximate dependence $\dot M \propto \mu^{-0.4}$.   
 
  An important question related to accretion onto INSs is about the spin properties of these objects. 
  Naively, one can expect that an AINS is just spinning down, and the braking torque is $\propto {\mu^2}/{R_\mathrm{co}^3}$ as the external medium does not bring, on average, angular momentum. 
 However, the picture is more complicated due to turbulence.
 Accreting matter brings fluctuating angular momentum which is equal to zero only after averaging over a relatively long time $\sim R_\mathrm{G}/v$. Up to our knowledge, for the first time, this situation was briefly considered in 
  \cite{1995ARep...39..632L}. 
  In this paper the authors proposed that an AINS does not spin down continuously, but can reach some quasi-equilibrium period $P_\mathrm{turb}\propto \mu^{2/3}$ around which the spin fluctuates. The situation was considered in more detail in a brief note \cite{2001astro.ph.10022P}.

 The evolutionary equation for the spin frequency can be written as:

 \begin{equation}
     \frac{\diff \mathbf{\Omega}}{\diff t } = \mathbf{F} + \mathbf{\Phi}, \, F=-\frac{k_\mathrm{t} \mu^2}{I\, R_\mathrm{co}^3}, \, \Phi\propto \frac{\dot M j}{I}.
 \end{equation}
Here $\mathbf{\Phi}$ is the turbulent torque and $\left< \mathbf{\Phi} \right>=0$. Specific angular momentum $j$ is determined as min$[v_t(R_\mathrm{G}) R_\mathrm{G}, v_\mathrm{K}(R_\mathrm{A}) R_\mathrm{A}]$, where $v_\mathrm{K}$ is Keplerian velocity and $v_t$ is turbulent velocity taken at the gravitational capture radius. The coefficient $k_\mathrm{t}$ is usually taken to be equal to 1/3. 

The evolution of the spin frequency can be represented as diffusion. Then for $\Omega_\mathrm{turb}=2 \pi / P_\mathrm{turb}$ we have:

\begin{equation}
    \Omega_\mathrm{turb}^2= \int_0^{\infty} \Omega^4 \exp{[-V(\Omega)/D]} \diff  \Omega / 
   \int_0^{\infty} \Omega^2 \exp{[-V(\Omega)/D]} \diff  \Omega.
\end{equation}
Here 
$$D=\frac 1 6 \left(\frac{\dot M j}{I}\right)^2 \frac{R_\mathrm{G}}{v}$$
and 
$$V(\Omega)=\frac{\mu^2}{3GMI}|\Omega |^3.$$ 
For $j=v_\mathrm{t(}R_\mathrm{G}) R_\mathrm{G}$ we obtain:

\begin{equation}
    P_\mathrm{turb} = 3.9\times 10^8 \, \mu_{30}^{2/3} n_1^{-2/3} I_{45}^{1/3} \left( \frac{M}{1.4\, M_\odot}\right)^{-26/9}\left(\frac{v}{100 \, \mathrm{km}\, {\mathrm s}^{-1}}\right)^{43/9} 
   \left( \frac{R_t}{2\times 10^{20} \mathrm{cm}}\right)^{2/9} \rm s. 
\end{equation}
Here $R_\mathrm{t}$ is the characteristic scale of interstellar turbulence for which $v_\mathrm{t}(R_\mathrm{t})=10$~km~s$^{-1}$ and the turbulent velocity depends on the spatial scale as $v_\mathrm{t}\propto r^{1/3}$. Normalized units are $I_{45}=I/10^{45}\,\mathrm{g}\, \mathrm{cm}^2$ and $\mu_{30}=\mu/10^{30} \mathrm{G}\, \mathrm{cm}^3$.

Numerically, the spin evolution of AINSs was studied in
\cite{2002A&A...381.1000P}. 
The authors studied the spin evolution at the stage of accretion of a population of isolated NSs with realistic velocity and magnetic field distributions, but for a constant ISM density. 
After the onset of accretion, an NS is spinning down: $\diff \Omega/\diff t \approx \mu^2/R_\mathrm{co}^3$. The influence of turbulence starts to be significant when the spin period reaches the value 

\begin{equation}
P_\mathrm{cr}=2\pi\frac{\mu}{\sqrt{GM\dot M j}}.
\end{equation}

Typically this happens in $\sim 10^5$~-~$10^7$~yrs. 
After that, the period on average continues to increase, finally reaching $P_\mathrm{turb}$ on a time scale $\gtrsim 10^9$~yrs, and fluctuating around this value. A typical time scale for $\dot P$ fluctuations is $R_\mathrm{G}/v.$ 
 The spin period distribution of AINSs appears to be very wide with typical values in the range $\sim 10^5$~-~$10^7$~s. 
 Note, that the authors did not discuss the influence of shocks around AINSs on the angular momentum transfer by turbulent accreted matter. 


As accretion onto INSs proceeds at low rates, the settling accretion regime might be applicable. 
 We describe this mode of accretion in Sec.~\ref{sec:acc:qsph}.
To the problem of AINSs this model was applied in \cite{2015MNRAS.447.2817P}.   
It was demonstrated that in the settling accretion regime, low-velocity NSs are expected to be transient sources with a typical duration of bursts of about a few hours and luminosity $\sim 10^{31}$~erg~s$^{-1}$. Persistent luminosities are expected to be three orders of magnitude lower, and so out of the reach of the present-day X-ray surveys. In addition, the authors proposed that inside convective envelopes AINSs can spin down much faster, and so they can reach quasi-equilibrium in a shorter time than it has been discussed in \cite{2002A&A...381.1000P}.



  In the early 1990s, high hopes to detect AINSs were related 
 to ROSAT observations \cite{1991A&A...241..107T}. 
  Step by step, it became clear that due to different effects, the number of observable AINSs cannot be very high (see e.g., \cite{1996MNRAS.278..577M} and references therein). 
  In the late 1990s, early enthusiasm was changed by pessimism related to the non-detection of any AINS by ROSAT (see a review in Treves et al. \cite{2000PASP..112..297T}). 
Finally, calculations presented in  \cite{2000ApJ...530..896P, 2000ApJ...544L..53P} demonstrated that it is impossible to expect even a few AINSs in the ROSAT data. 

One of the most detailed searches for INSs in the ROSAT data was presented in
\cite{2010ApJ...714.1424T}.  
 Up to now there are no known AINSs or good candidates.
 Still, the discovery of such sources will be a major breakthrough in understanding low-rate accretion and long-term evolution of isolated NSs. 

\subsection{Accretion in binaries}

\subsubsection{Variety of NSs in binary systems}\label{sec:acc:zoo}

Among the massive stars producing NSs, most are components of binary or multiple systems \citep{2012Sci...337..444S}. 
Existence of a binary companion may affect the evolution of the NS in multiple ways, from tidal torques adjusting the rotation of the progenitor to mass exchange. 
Observationally, NSs in binaries manifest themselves as X-ray binaries (XRBs) and as millisecond pulsars (MSP).
In both cases, the presence of a binary companion is crucial for the rotational properties of the NS. 
MSPs are radiopulsars thought to have accretion episodes in their past \citep{2017JApA...38...42M}. 
The connection between the two classes of systems is confirmed by the existence of accreting millisecond X-ray pulsars (AMXPs, \citet{2009Sci...324.1411A}). 
We will cover these classes of objects later in section \ref{sec:acc:msp}. 

Classification of XRBs may be done in different ways, using such criteria as mass transfer type (Roche-lobe overflow or wind capture) or magnetic field of the accretor, but the traditionally adopted distinction is between high-mass and low-mass XRBs (HMXBs and LMXBs, respectively), with the mass of the donor being the crucial parameter.
 Higher donor mass (several Solar masses or larger) implies a relatively young age of the NS, a large magnetic field, and often mass transfer through the wind. 
Low donor mass makes the formation of the binary an unlikely scenario, as progenitor systems should be disrupted by supernova explosions. 
Thus, many LMXBs, especially in dense stellar environments like globular clusters, should have been formed by exotic processes such as binary companion exchange \citep{2023hxga.book..120B}, or require a finely tuned natal kick. 
In terms of the number of active systems, the low formation probability is compensated by the abundance of low-mass stars and their longevity. 
The NSs in such systems are on average much older, which implies low magnetic field strengths. 
Also, accretion from a low-mass donor is usually associated with Roche-lobe-overflow mass transfer. 

Low magnetic fields of LMXBs make it difficult to study their rotational evolution. 
With the exception of AMXPs, their spin periods are inferred indirectly from the other variability patterns such as post-burst oscillations in X-ray bursters \citep{1996ApJ...469L...9S} or quasi-periodic oscillations in the sources lacking coherent periodic pulsations. 
The latter includes the so-called 'Z` and 'Atoll` sources introduced by \cite{1989A&A...225...79H}, containing most of the known LMXBs (see \cite{2023hxga.book..120B} for a review.) It is however unclear if the QPO frequencies are reliable indicators of the spins of these objects, see \citet{2007MNRAS.381..790M}). 

Because the matter accreted by a non-magnetized NS through an accretion disc has the net angular momentum equal to Keplerian at the inner disc edge $R_{\rm in} \simeq 10$km (determined either by the surface of the star or by its gravity field), angular momentum is gained at the time scales only several times smaller than the time scales of mass growth. 
For accretion at the Eddington limit ($\dot M \simeq 10^{18}\rm g \, s^{-1}$), spinning up to a several millisecond period would take millions of years. 

Strong magnetic field makes the spin-up process much more efficient, as a large magnetosphere intercepts accreting matter with a larger ($\propto \sqrt{R_{\rm m}}$, if accretion runs through a disc) net angular momentum. 
On the other hand, rapid rotation in combination with strong magnetic fields suggests pulsar torques and pulsar wind formation. 
Even more importantly, rapid rotation of the NS may lead to propeller regime and thus quenching of mass transfer (see section \ref{propeller}). 
Hence, the spin evolution in HMXBs is usually thought of as deviations from an equilibrium between spin-up and spin-down torques. 
The equilibrium point shifts with the changing mass accretion rate and some other parameters, likely related to the geometry of the accretion flows. 

Many HMXBs are observed as X-ray pulsars (XRPs, i. e., sources of coherent periodic variability in the X-ray range, clearly related to rotation). 
Studying X-ray pulsations allows to probe their rotational evolution in real time and on long archival data series.
One of the most important projects of this kind is the monitoring program held out with {\it Fermi/GBM} in the hard X-ray range \citep{2020ApJ...896...90M}. 
The series obtained by {\it Fermi/GBM} span more than 10 years of observations on 39 X-ray pulsars. 

The majority of X-ray pulsars are members of binaries with Be-class donors (Be/X-ray systems, \citet{2005A&AT...24..151R,2011Ap&SS.332....1R}).
The population of such systems is diverse, with the orbital periods in the range $20-200$d and NS spins $1-10^3$s. 
Orbital and spin periods are known to correlate with each other \citep{1984A&A...141...91C}, supporting the idea that that their rotation is close to equilibrium between different torques acting on the star (see section \ref{sec:acc:down}). 
The exact origin of this correlation is however unclear. 
Mass transfer rate in Be/X-ray systems is strongly modulated by orbital eccentricity and other factors such as the mass accumulation in the decretion disc of the star. 
This results in a complex phenomenology of periodic and aperiodic flares, during which the NS is spun up by accreting matter, and quiescent states where the rotation slows down. 
This picture is not universal and allows for objects that are, for example, spun up even during the quiescent state, suggesting that the equilibrium is reached beyond the time span of observations. 
For a given object, however, $\dot{P}$ is well correlated with the luminosity \citep{2020ApJ...896...90M}. 

Another important category of XRPs are objects accreting from the winds of early-type supergiants. 
The prototype and the best studied object of this class is Vela~X-1 \citep{2021A&A...652A..95K}. 
Accretion disc may form in such systems but not necessarily, as the circularization radius is not necessarily larger than the radius of the magnetosphere. 

In a homogeneous wind, the net angular momentum is (see for instance \citet{1975A&A....39..185I}) $j_{\rm 0} \sim \Omega_{\rm orb} R_{\rm G}^2$.
This angular momentum is aligned with the rotation of the binary. 
Strong inhomogeneity of the wind would create an additional randomly oriented net angular momentum component $j_{\rm w} \sim R_{\rm G}v_{\rm w} \gg j_0$, where $v_{\rm w} \gg R_{\rm G} \Omega_{\rm orb}$ is the velocity of the wind. 
Hence, for such systems, the angular momentum of the accreting matter is likely to be randomly aligned. 

\subsubsection{Spin-up by an aligned disc}

First, let us consider a planar thin conducting accretion disc interacting with the dipolar magnetic field of a star. 
Even in its simplified version, the problem of disc-magnetosphere interaction does not have a general analytical solution. 
A thorough review of the existing models and approaches to disc-magnetosphere interaction was made, for instance, by \citet{2014EPJWC..6401001L}. 

The general picture is dictated by the existence of a magnetosphere and the strong radial decrease in magnetic tensions.
One might expect magnetic field lines threading the disc within some depth $\Delta R$ that is determined by intermediate-scale processes such as turbulence and reconnections between the stellar magnetic fields and the fields frozen into the accreting matter. 
Large $\Delta R \gtrsim R$ models are usually referred to as `magnetically threaded disc', or MTD \citep{1979ApJ...234..296G, 1995ApJ...449L.153W} models.
Within the penetration region, the internal angular momentum flow in the disc, normally governed by effective viscosity forces \citep{1973A&A....24..337S}, is contributed by the torque from the distorted magnetospheric lines. 
Purely poloidal magnetic field of an aligned dipole does not produce any torque, but the relative motion of the magnetosphere with respect to the disc creates toroidal field components $B_\varphi^{\pm}$, having different signs on the different surfaces of the disc. 
Toroidal fields should also change sign at the corotation radius and are likely of the same order as poloidal fields. 
Magnetic field torque acting on the NS has two contributions that are not clearly distinguishable: the angular momentum coming with the accreting matter itself and the torque produced by Maxwell stresses $B_z B_\varphi^{\pm} / 4 \pi$ (where $B_p$ is the poloidal field).
As the flow inside the magnetosphere is bound to corotate with the NS, its angular momentum is also primarily transmitted by magnetic stresses. 
While the net angular momentum at the magnetospheric boundary is $\Omega R_{\rm m}^2$, its value near the surface of the NS is $\sim \Omega R_{\rm NS}^2$, that is normally several orders of magnitude smaller. 

The torque acting on the surface of the NS from the disc may be calculated as the star-disc interaction (SDI) torque, containing both spin-up and spin-down parts
\begin{equation}\label{E:binaries:Nsu}
    K_{\rm SDI} = - \int R B_z B_\varphi R \diff R \diff \varphi,
\end{equation}
where integration is performed over the surfaces of the disc threaded by magnetic field lines (usually, upper and lower surfaces). 
Straightforward integration of this expression requires knowledge about both field components. 
In general, the SDI torque may contain a spin-up part created by the field lines threading the disc inside the corotation radius, and a spin-down part from the lines extending outside (their tensions marked with green and red arrows in Fig.~\ref{fig:resmap}). 
Magnetic field opening and screening (see below) tend to decrease the contribution of the outer parts of the disc, making SDI torque essentially a spin-up torque.

The MTD approach assumes that the poloidal field interacting with the disc is equal or at least close to the original dipolar field in a broad range of radii $\Delta R \gtrsim R$, and the toroidal fields are generated by the difference in the rotation velocity between the star and the disc. 
The expression used for the toroidal fields has the form $B^\pm_\varphi(R) = \pm B_z(R) f(\Omega/\Omega_{\rm K})$, where the function $f(x)$ should meet a couple of requirements: $f(1)=0$, $f(x<1) <0$, $f(x>1)>0$, and $|f(x)| \lesssim 1$. 
The last condition is related to the stability of predominantly toroidal force-free magnetic field configurations. 
In these assumptions, the torque related to disc-magnetosphere interaction is convenient to normalize with the accretion torque spinning the NS up,
\begin{equation}\label{E:binaries:ksu}
    K_{\rm SDI} = \dot{M} \sqrt{GMR_{\rm m}} n(\omega).
\end{equation}
This kind of normalization may also be used in a more general case when many different aligned torques affect the rotation of the star at the same time.
The simplified picture where all the matter accreting through the disc gives its angular momentum at $R_{\rm m}$ to the star corresponds to $n(\omega) = 1$.
In general, the dimensionless factor $n(\omega)$ accounts for all the processes of angular momentum exchange between the disc and the star \citep{1995ApJ...449L.153W, 2015ApJ...813...91S}.

\begin{figure}
\includegraphics[width=0.8\columnwidth]{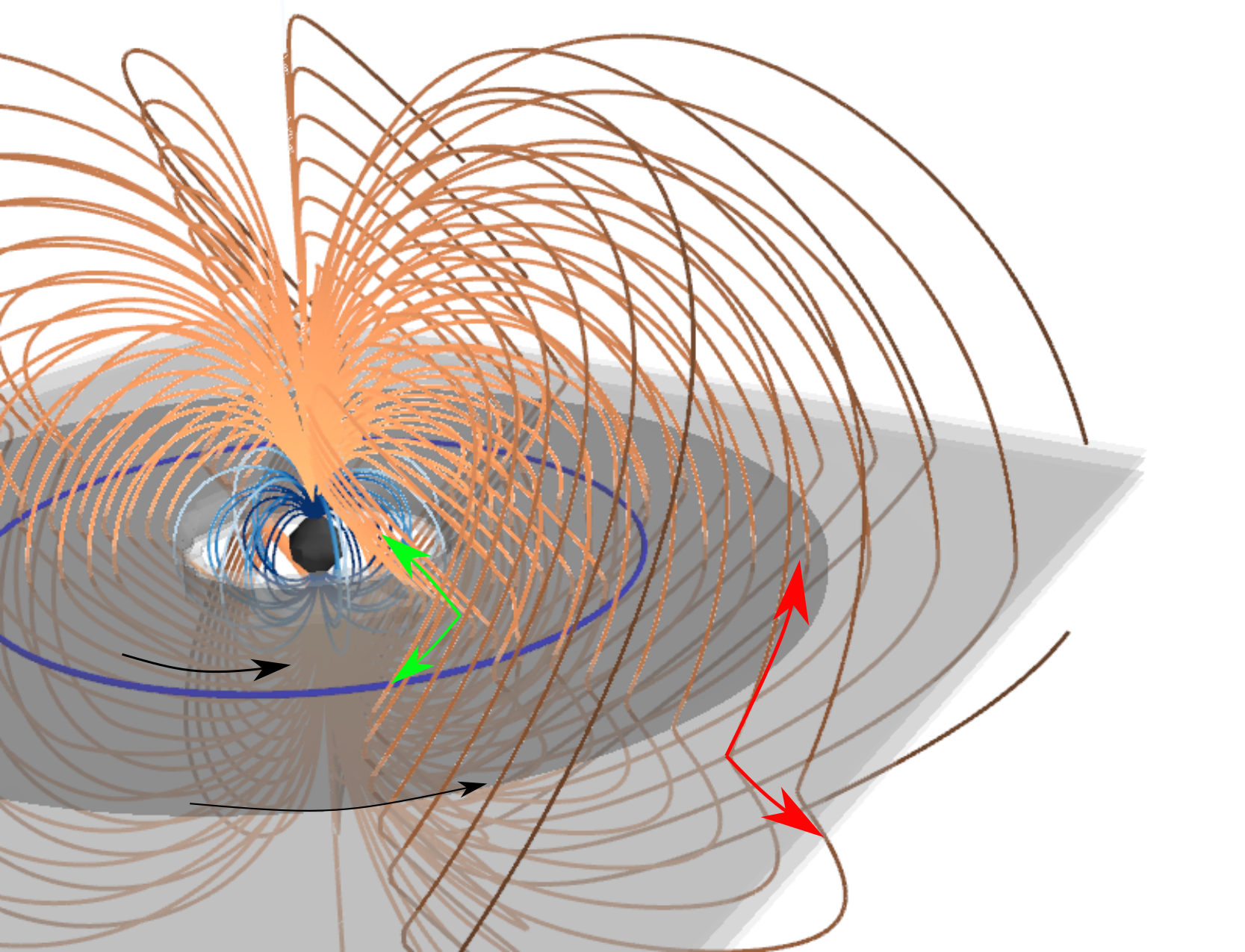}
 \caption{ Three-dimensional sketch illustrating the MTD model of field penetration into an aligned disc. The blue circle in the disc plane marks the corotation radius. The field lines have dipolar and toroidal components, the latter related to the motion of the disc and limited by the absolute value of the seed poloidal field. Arrows show the rotation of the disc and the tension along the field lines.}\label{fig:resmap}
\end{figure}

The classic MTD model assumes that the magnetic field retains its poloidal component well outside the radius of the magnetosphere, as one would expect if magnetic diffusivity in the disc were high. 
Rotation of the disc creates additional azimuthal components scaling with the poloidal dipolar field components $B_\varphi \sim B_{\rm p}$.
Integrating the torques over the surface of the disc, one yields for the torque (assuming $B_\varphi = \pm B_z$)
\begin{equation}\label{E:binaries:GL}
 K_{\rm SDI} = \frac{1}{3} \left( \frac{1}{R_{\rm m}^3} - \frac{2}{R_{\rm co}^3} \right) \mu^2.
\end{equation}
Such a configuration allows for a rotational equilibrium, where part of the disc (inside the corotation radius) spins the star up, while the outer regions spin the star down (see Figure~\ref{fig:resmap}). 
At a constant mass accretion rate, this kind of equilibrium keeps the corotation radius close to the magnetospheric radius. 

Decreasing the penetration depth restricts the integration range in (\ref{E:binaries:Nsu}) to a narrow strip where the rotation frequency of the disc is likely to be faster than $\Omega$. 
The spin-up torque should still be close to the accretion torque (\ref{E:binaries:ksu}), because of the angular momentum advected with the matter. 
However, as we show later in section~\ref{sec:acc:down}, this spin-up is opposed by other forces, not directly related to star-disc interaction. 

\subsubsection{Spin-up of an inclined dipole}

The very existence of an XRP phenomenon requires an NS with misaligned rotational and magnetic axes. 
Observational estimates of magnetic angles $\chi$ (see section \ref{sec:pulsars}) in X-ray binaries cover a broad range, from nearly aligned objects towards nearly orthogonal (see in particular the recent results from {\it IXPE} \citet{2022NatAs...6.1433D}). 
The observational estimates for magnetic angles in MSPs are strongly model-dependent \citep{2014ApJS..213....6J} but also favour a broad distribution.
For all these reasons, it is important to be able to relax the alignment condition for the dipolar magnetic field of the star. 
At the same time, its rotational axis is likely to be aligned with the spin-up torque, with a couple of important exceptions we consider later in section \ref{sec:acc:mis}. 
Relaxing the magnetic obliquity leads to several consequences. 
First, the condition for the magnetospheric boundary is now satisfied at different radii for different azimuthal angles. 
According to \citet{2018A&A...617A.126B}, this leads to a correction factor related to the magnetic field strength dependence on the polar angle. 
Even more importantly, magnetic inclination breaks the axial symmetry of the flow, and the funnel flow becomes concentrated to a couple of streams restricted in the azimuthal direction. 

An MTD approach allows calculating the spin-up torque, assuming that the magnetic field interaction with the plasma of the disc creates a non-dipolar component (generalization of $B_\varphi^\pm$ in the aligned case) and consequently angular momentum exchange between the disc and the star. 
Accurate calculation in the MTD approach was performed by \citep{1997ApJ...475L.135W} and involved three surfaces: upper, lower, and inner faces of the disc. 
The unperturbed magnetic field line of an inclined dipole was complemented by an additional field component directed along the relative velocity difference between the disc and the magnetosphere. 
Integration over all the surfaces leads to a torque that changes sign at some critical fastness parameter values close to unity.

\subsubsection{Spin-down torques and equilibrium}\label{sec:acc:down}

The MTD picture described above has a major caveat related to the opening of the field lines due to asynchronous rotation between the disc and the star \citep{1990A&A...227..473A,2014EPJWC..6401001L}. 
A more realistic picture involves field lines loaded with mass and extending to infinity, which creates possibilities for mass ejections.
 A lot of inspiration here comes from the studies initially aimed at magnetized young stellar objects that evolve under a similar combination of torques. 
In \citet{2005ApJ...632L.135M}, the existence of magnetized winds draining the star of angular momentum and carrying away some part of the accreting matter was proposed as an explanation for the spin period distribution in T~Tau stars. 
Many detailed 3D MHD simulations \citep{2013MNRAS.430..699R,2013A&A...550A..99Z,2022ApJ...929...65I}
designed for young star magnetospheres are scalable for the smaller and rapidly rotating magnetospheres of NSs with identical dimensionless parameters (fastness $\omega$, magnetic angle, and disc thickness), that allows to make implications about XRPs as long as relativistic effects are ignored. 
All the simulations reproduce a moderate penetration depth of the order disc thickness, field line opening and magnetized ejections. 
The estimated torque related to the ejections is of the order $K_{\rm sd} \sim \Omega R_{\rm m}^2$, similar to the braking torque given by the propeller theory $K_{\rm sd} \sim \mu^2 / R_{\rm co}^3$. 
The exact value contains correction factors depending on the fastness parameter \citep{2022ApJ...929...65I}.
Fully relativistic MHD simulations \citep{2023arXiv231104291P,2023arXiv231105301D} confirm the general picture, but are so far limited only to very compact magnetospheres.

In an accreting magnetosphere, some of the field lines remain open and free of the matter supplied from the disc.
 As these lines tend to corotate with the star and extend beyond the light cylinder, they create a braking torque similar to the pulsar torque considered in section~\ref{sec:pulsars}.
 Particle creation due to pair cascades is in this case probably suppressed, but the contribution from the vacuum losses remains as long as there are open magnetic field lines. 
On the other hand, interaction with the disc changes the field shape outside the radius of the magnetosphere, and the resulting electromagnetic torque is likely to be higher than in the case of an ejector. 
In the aligned case, \citet{2016ApJ...822...33P} estimate the value of the torque as
\begin{equation}
    K_{\rm Parfrey} \simeq \zeta^2 \left( \frac{R_{\rm l}}{R_{\rm m}}\right)^2 \frac{\mu^2}{R_\mathrm{l}^3}
\end{equation}
where $\zeta \sim 1$ is a dimensionless parameter, and $K_{\rm sd}$ corresponding to the pulsar braking torque given by equation (\ref{eq:pulsars:MHD}).
Three-dimensional MHD simulations confirm the picture drawn by \citet{2016ApJ...822...33P} and also show that the open field lines of the accreting magnetosphere behave like Poynting-dominated jets in accreting black hole systems powered by the rotation of the compact object \citep{2023arXiv231105301D,2023arXiv231104291P}.

It is unclear how applicable these results are to a more general geometry. 
In particular, is this spin-down torque aligned with the rotation of the NS or with the disc that opens up the field lines? 
How does this particular spin-down torque depend on the magnetic inclination of the rotator? 
Answering these questions would require full-scale simulations with non-zero magnetic and disc inclination angles. 

Combined, two or more torques of different signs, dependent on mass accretion rate and the spin on the NS, result in a rotational evolution equation 
\begin{equation}
    \frac{d}{dt}\left( I \Omega\right) = K_{\rm su} - K_{\rm sd} = n(\omega) K_{\rm su},
\end{equation}
which has a single equilibrium point at $n = 0$. 
Different models using different assumptions about the spin-down torques differ in the dimensionless factor $n(\omega)$. 
The solution $n(\omega)=0$ is usually close to the propeller limit $\omega=1$, or, in terms of spin period 
\begin{equation}\label{E:acc:peq}
    P_{\rm eq} \simeq 2\pi (GM)^{5/7} \mu^{6/7} \dot{M}^{-3/7} \simeq 5 \left( \frac{\mu}{10^{30}\rm G \, cm^3}\right)^{6/7}  \left( \frac{\dot M}{10^{16}\rm g \, s^{-1}}\right)^{-3/7}\rm s.
\end{equation}
Applying this expression to an XRP with a known spin period and a reliable estimate of the mass accretion rate (that is proportional to the bolometric accretion luminosity) allows to estimate the magnetic moment in the assumption of equilibrium. 
In a strong flare, when the spin-up torque dominates over the spin-down one, the magnetic moment may be probed by the observed spin-up rate. 
Several methods of magnetic field estimation using the spin-up and spin-down torques are described in  \citet{2012NewA...17..594C}. 

Depending on the variability of the mass supply rate and potentially other parameters, some objects might oscillate between accretor and propeller regimes, while others proceed to accrete in the low-accretion-rate state. 
There is observational evidence that some Be/X-ray systems enter the propeller stage (see section~\ref{propeller}), while others retain a small persistent mass accretion rate during quiescence \citep{2017A&A...608A..17T}. 
The reason for such a dichotomy is unclear and may be connected to additional parameters such as magnetic angle, or with the mass accretion rate variations in the disc. 
 
The time required to reach the equilibrium state may be estimated as 
\begin{equation}\label{E:acc:tsu}
    t_{\rm su} \simeq \frac{I\Omega_{\rm eq}}{K_{\rm su}} \sim \frac{I \Omega_{\rm eq}^{4/3}}{\dot{M} (GM)^{2/3}} \sim 10^3 \frac{I}{10^{45}\rm g\, cm^2 } \left(\frac{\dot{M}}{10^{18}\rm g \, s^{-1}}\right)^{-3/7} \rm yr,
\end{equation}
where the radius of the magnetosphere was assumed $R_{\rm m} \sim R_{\rm co}$ (that allows to exclude the value of $\Omega_{\rm eq} = 2\pi / P_{\rm eq}$ from the equation), that is a reasonable assumption near the equilibrium. 
Cases, when the spin period changes considerably during the observational period, are scarce and related to unique and probably catastrophic events. 
A bona fide example of such a case is the super-Eddington XRP NGC300~ULX1 \citep{2019MNRAS.488.5225V} which was spun up by about a factor of five during the period of observations.

Detailed timing studies of XRPs allow testing of the models of magnetospheric accretion flows. 
In particular, the flares of Be/X-ray systems are often thought of as the case when spin-up torque dominates over all the spin-down mechanisms, hence $\dot{\Omega} \propto K_{\rm su} \propto \dot{M} \sqrt{R_{\rm m}}$.
Classical scaling $R_{\rm m} \propto R_{\rm A} \propto \dot{M}^{-2/7}$, implying $\dot{\Omega} \propto \dot{M}^{6/7}$, is generally supported by observations \citep{1997ApJS..113..367B}. 
There are indications for a steeper dependence at higher luminosities \citep{1989ApJ...338..359P} that might be a signature for a more elaborate disc-magnetosphere interaction
\citep{2007ApJ...671.1990K,2017A&A...608A..17T}.

\subsubsection{Misaligned spin-up torques}\label{sec:acc:mis}

If the spin-up torque does not change its direction (in particular, if its direction is fixed by the orbital plane of the system), one would expect the rotation axis of the NS to align with the torque within the spin-up time (\ref{E:acc:tsu}).
But on some of the evolutionary stages and in certain accretion regimes in X-ray binaries the rotational axis of the star may deviate strongly from the external spin-up torque. 
First, before the mass exchange, the accreting NS is likely to have a rotational axis misaligned with the orbital plane due to the kick received during the supernova explosion \citep{2006ApJ...639.1007W} or a three-body process responsible for the creation of the system. 

Another important case is accretion from an inhomogeneous wind when the density and/or velocity gradients within the wind make the direction of the torque essentially decoupled from the orbital plane \citep{1981A&A...102...36W}. 
Numerical simulations of such an accretion regime were done, for instance, by \citet{2018MNRAS.475.3240E}.
 The spin of the NS in this case is lower than the equilibrium value, and its direction is involved in a random walk adjusting to the rapidly changing external torque. 
The observed spin derivative in this case should be weakly correlated with the flux, because of the relative orientation of the torque and the angular momentum. 
Such a behavior is observed for Vela~X-1 \citep{2021A&A...652A..95K} and other XRPs fed by supergiant winds. 

Misaligned torques are also important as a possible trigger for magnetic angle evolution. 
Adding the spin-up torque to Euler equations of rotational dynamics \citep{2021MNRAS.505.1775B} reveals an additional term proportional to $\sin \chi \cos\chi \sin\alpha \cos\alpha$, where $\alpha$ is the misalignment angle between the rotational axis of the NS and the external spin-up torque, that increases the magnetic angle during the alignment stage. 
As a result, the alignment of the NS rotational axis increases the magnetic angle, increasing the possibility of observing the NS as an XRP. 
Wind accretion with a randomly aligned spin-up torque results in a random walk evolution of $\chi$, increasing the number of strongly oblique pulsars. 

\subsubsection{Quasi-spherical accretion regime}\label{sec:acc:qsph}

If the circularization radius becomes smaller than $R_{\rm m}$, the Keplerian disc can not form, and the torques acting on the NS should be revised. 
The net angular momentum of the matter interacting with the magnetosphere is now much smaller than Keplerian at the magnetosphere radius, meaning that the spin-up torque is severely diminished. 
For a homogeneous wind, $j \sim \Omega_{\rm orb} R_{\rm G}^2$ \citep{1975A&A....39..185I}. 

The matter near the magnetospheric boundary is heated by shock waves to temperatures close to virial, which implies a quasi-spherical envelope around the magnetosphere. 
The rotation of the NS adjusts to the rotation of the envelope via magnetic stresses at the edge of the magnetosphere. 

The structure of the envelope depends on the presence of cooling and angular momentum transfer within it. 
In the case of weak cooling, an elaborate theoretical model was developed by \citet{2012MNRAS.420..216S}. 
As the main process responsible for angular momentum transfer, the authors considered an anisotropic turbulent velocity field. 
Anisotropy allows to change the direction of the angular momentum transfer \citep{1997PASJ...49..213K}.
In particular, predominantly radial motions lead to a flat distribution in angular momentum, meaning that the rotation frequency in the envelope $\Omega_{\rm env} \propto R^{-2}$. 
\citet{2012MNRAS.420..216S} adopt a more general power-law scaling for the rotation frequency. 

Equilibrium spin periods predicted by the quasi-spherical model are much longer than for disc accretors, roughly corresponding to the equal net angular momenta of the outer magnetosphere rim and the captured matter $\Omega R_{\rm m}^2 \simeq \Omega_{\rm orb}R_{\rm G}^2$, implying, up to a coefficient of the order unity, 
\begin{equation}
    P_{\rm eq, \, sph} \sim 10^3 \left( \frac{v}{10^8\rm cm\, s^{-1}}\right)^4 \left( \frac{\mu}{10^{30}\rm G \, cm^3}\right)^{8/7}  \left( \frac{\dot M}{10^{16}\rm g \, s^{-1}}\right)^{-4/7} \rm s. 
\end{equation}
For identical magnetic moments and mass accretion rates, this value is about two orders of magnitude longer than the equilibrium spin for disc accretion (compare equation \ref{E:acc:peq}). 

Such an accretion scenario may be applied to wind-fed HMXBs like Vela~X-1, to symbiotic wind-fed LMXBs like GX~1+4 \citep{2012A&A...537A..66G},
 and potentially to some of the Be/X-ray systems having very long spin periods that require suspiciously large magnetic fields if interpreted as disc accretors \citep{2012NewA...17..594C}. 

\subsubsection{Recycled and transitional pulsars}\label{sec:acc:msp}

About 10 percent of the RPP population belongs to millisecond pulsars (MSPs), a category of short-period, weakly magnetized NSs with the estimated ages of about Gyrs, well beyond the expected period of pulsar lifetime \citep{2017JApA...38...42M}.
This population is understood as recycled, i. e. spun up by accretion in a binary system to a period short enough to support a pulsar cascade. 
The majority of MSPs have binary companions, and some of them have masses in excess of $1.8$\Msun. 

Among the NSs with reliably measured rotation periods, the fastest rotator is PSR J1728-2446ad \citep{2006Sci...311.1901H} having a rotation frequency of more than 700Hz. 
Several other MSPs are rotating at similar rates $\sim 500-600$Hz. 
There are indications that some non-pulsating NSs in LMXBs have frequencies around 500-600Hz \citep{2017ApJ...850..106P}.
 These values are several times smaller than the Keplerian rotation rate ($\sim 2.5$kHz for a 2\Msun star with a radius of 10km), which is sometimes interpreted as a signature of an additional braking mechanism. 
If the excess $\sim 0.3-0.5\Msun$ of mass observed in MSPs was gained via disc accretion, the expected rotation frequencies should be pretty close to critical. 
For $I \simeq 10^{45} \rm g\, cm^2$ and total angular momentum gained via accretion $\Delta M \sqrt{GMR_{\rm NS}} \sim 10^{49}\rm g \, cm^2 s^{-1}$, this implies a spin period of $P_{\rm s}\lesssim 0.6$ms, or a frequency $\sim 1.5$kHz.  
One factor limiting neutron star rotation frequency is gravitational waves \citep{1998ApJ...501L..89B,1998PhRvD..58h4020O}.
The upper spin limit may be also related to the electromagnetic braking term (see \citet{2016ApJ...822...33P} and sec.~\ref{sec:acc:down}), or simply to the decrease in spin-up efficiency with the shrinkage of the magnetosphere \citep{2012ApJ...746....9P}.
Population synthesis simulations run by \citet{2012Sci...335..561T} shows that in certain assumptions the observed spin limit may be interpreted as a consequence of the braking torques slowing the NS down on the propeller stage. 

Our understanding of MSPs as a product of binary evolution is supported by the existence of LMXBs containing NSs with properties similar to MSPs (accretion-powered millisecond pulsars, AMXPs, see \citet{2021ASSL..461..143P}) and, most importantly, transitional MSPs (TMSPs) that combine the radio emission typical for RPPs with an evidence for an accretion disc \citep{2022ASSL..465..157P}. 
Observations show TMSPs switch between different activity states. 
The prototypical object PSR~J1023+0038, in particular, is observed either as a radiopulsar with an X-ray luminosity $L_{\rm X} < 10^{33}\rm erg \, s^{-1}$, or in an X-ray bright state during which it has pulsations in the X-ray range, but is not observed in radio \citep{2015ApJ...807...62A}.
X-ray pulsations are observed in both phases, but the pulse shapes are different. 
In the AMXP state, the pulse profiles are dominated by the first overtone. 
An obvious interpretation of this behavior is the transition between ejector and accretor states (see sec.~\ref{introduction}). 
Pulsating X-ray emission is interpreted as a signature of a funnel flow within the magnetosphere. 
An alternative explanation proposed by \citet{2019ApJ...884..144V} considers that the source is always rotation-powered, and the X-ray emission observed during the bright states is produced by the disc existing outside the light cylinder and reprocessing part of the pulsar wind power. 
Anisotropy of the pulsar wind leads to a modulation of the X-ray emission with a doubled spin frequency of the pulsar. 

\subsubsection{Magnetic field burial by accretion}

Ohmic magnetic field decay given by equation~(\ref{E:fallback:B}) predicts a late-time exponential decay on millions of years, but is definitely inapplicable to objects as old as hundreds of millions of years, such as MSPs and AMXPs. 
The situation is unclear, and most studies assume that Ohmic losses are negligibly small during late evolution (see discussion in \cite{2006ApJ...643..332F}).

The proposed long accretion histories and large accreted masses in MSPs imply that most of the crust material in these objects is renewed.
This should also affect the magnetic field strength and structure. 
The idea of the field being screened by a layer of accreted matter was proposed already in \citet{1974SvA....18..217B}. 
It is known to be an important effect in other circumstances, such as fallback accretion (see sec.~\ref{fallback}). 

Due to the presence of the magnetosphere, an XRP is primarily accreting onto a relatively small fraction of its surface, about the size of the two polar caps ($\sim 2\pi R_{\rm NS}^4 / R_{\rm m}^2$) or smaller. 
Magnetic field tensions are able to support only a small layer with a surface density of 
\begin{equation}
\Sigma \simeq \frac{B^2}{4\pi g} \sim 3\times 10^8 B_{12}^2\,{\rm g\, cm^{-2}},
\end{equation}
where $g$ is surface gravity. This led to the suggestion that the newly accreted matter spreads over the surface of the NS from the magnetic poles toward the equator, screening the field and distorting the shapes of the field lines \citep{2001PASA...18..421M}. 
In an idealized axisymmetric approach, the dipolar component of a buried magnetic field decays approximately quadratically with the accreted mass. 
For instance, \citet{2004MNRAS.351..569P} approximate their results with a power-law relation 

$$\mu \simeq \mu_{\rm i} \left( \frac{\Delta M}{4.6\times 10^{-5}\Msun}\right)^{-2.25\pm 0.22}.$$ 
Accretion of only about $10^{-4}\Msun$ would lead to a magnetic moment decrease by a factor of several. 
Local magnetic fields near the magnetic equator in this scenario are not buried and are even expected to grow due to flux conservation, forming a ``magnetic tutu''. 
The increase in field gradients, both vertical and latitudinal, increases Ohmic losses, which was also supposed to contribute to the long-term evolution of the field \citep{1999MNRAS.303..588K}. 

It should be noted that the spreading of the accreted matter over the surface of the NS is accompanied by a number of instabilities acting on different scales and caused by different processes \citep{2017JApA...38...42M}. 
Local interchange instabilities \citep{2001ApJ...553..788L,2020JPlPh..86f9002K} operate on very small time scales (several local dynamics) and effectively result in magnetic field diffusion in the favorable parts of the flow. 
Also, the strongly magnetized layer buried under a less-magnetized layer of accreted matter is susceptible to Parker instability \citep{2007MNRAS.376..609P}. 
The resulting magnetic fields are suppressed, decayed, and have a much more complex topology than the initial dipole.

\section{Exotic and hypothetical stages}
\label{exotic}

In this section, we briefly discuss several stages of magneto-rotational evolution that remain hypothetical, i.e. no direct evidence in favor of their existence is known.

\subsection{Georotator}
 The name of this stage is related to the fact that the shape of the magnetosphere and some processes are similar to the case of the Earth magnetosphere in the Solar wind. 
The stage with $R_\mathrm{m}<R_\mathrm{co} $ but $R_\mathrm{m}>R_\mathrm{G}$ was mentioned already in \cite{1975A&A....39..185I}. The authors just note that in this case matter flows around the magnetosphere as the gravitational influence of the compact object is too small. Such a situation can be realized if the velocity of matter relative to the NS is too high (e.g., due to very fast stellar wind in a wide binary, or due to a large spatial velocity of the compact object in the case of an INS), if the magnetic field is too large, or if matter density is too low (this can happen e.g., for an INS far above the Galactic plane). 

It is worth to note, that the condition $R_\mathrm{m}<R_\mathrm{co} $ is important. Correspondingly, the georotator stage is an alternative to the stage of accretion, not to the propeller stage. Thus, one can consider evolution as Ejector$\rightarrow$Propeller$\rightarrow$Georotator, or switching from accretor to georotator. The latter scenario can be realized e.g., while an INS moves in the Galactic potential rising high above the Galactic plane and then coming back.  

Also, we note that the magnetospheric radius is defined by different equations in the case of accretion, at the propellers stage, and in the case when $R_\mathrm{m}>R_\mathrm{G}$. That's why the critical condition for transition to/from the georotator stage might be carefully calculated accounting for necessary effects. 
If we consider a typical INS with $\mu \approx 10^{30}$~G~cm$^3$ and the ISM density $n\approx 1$~cm$^{-3}$, then the critical velocity for transition to the georotator stage is $\sim 400-500$~km~s$^{-1}$. Note, that for such parameters duration of the ejector plus propeller stages is longer than the Galactic age. Still, e.g. INSs with large magnetic fields can become georotators. 

To our knowledge, spin-down at the georotator stage was never analyzed in detail. From general considerations, it can have some similarities with the subsonic propeller stage \cite{1981MNRAS.196..209D} or/and with the settling accretion \cite{2012MNRAS.420..216S}. When a long spin period is reached, the influence of turbulence on the rotational frequency might become important \cite{2002A&A...381.1000P, 2001astro.ph.10022P}, similar to the case of accreting INSs.

The case $R_\mathrm{m}>R_\mathrm{G}$ was studied numerically for non-rotating INSs in \cite{2001ApJ...561..964T}. The authors dub this stage a `magnetic plow'. 
In this study, just the case of aligned magnetic axis and spatial velocity vector was analyzed. 
It is demonstrated that a tiny amount of matter still can accrete onto the NS surface, but the energy release is small and hardly can be detected. Instead, the authors focus on the energy dissipation in the bow shock and in the magnetic tail. 

In the bow shock, the energy release can be estimated as:

\begin{equation}
    \dot E_\mathrm{shock}\approx \frac{\pi}{2}R_\mathrm{m}^2 \rho v^3\sim 10^{21} B_{12}^{2/3}n^{2/3}\frac{v}{200\, \mathrm{km}\, \mathrm{s}^{-1}} \, \mathrm{erg}\, \mathrm{s}^{-1}.
\end{equation}
It is expected that the energy is emitted in X-ray and, maybe, in softer bands. 

The case of energy dissipation in the magnetic tail is more interesting as different regimes can be realized. 
If the tail is not expanding, i.e. even far from the NS its cross-section has the size $\sim R_\mathrm{m}$, 
then the total energy is:

\begin{equation}
    E_\mathrm{tot}\sim 10^{27} B_{12} n^{1/2} \frac{v}{200\, \mathrm{km}\, \mathrm{s}^{-1}} \frac{l}{100\, R_\mathrm{m}}\, \mathrm{erg}.
\end{equation}
Here $l$ is the length of the tail. 
This energy can be dissipated due to reconnection on the dynamical time scale 

$$t_\mathrm{dyn}=l/v\sim 10^6 B_{12}^{1/3} n^{-1/6} \frac{v}{200\, \mathrm{km}\, \mathrm{s}^{-1}} \frac{l}{100\, R_\mathrm{m}} \, \mathrm{s}.$$
Then

\begin{equation}
    \dot E_\mathrm{rec} = E_\mathrm{tot}/t_\mathrm{dyn}\sim 10^{21} B_{12}^{2/3} n^{2/3}\left(\frac{v}{200\, \mathrm{km}\, \mathrm{s}^{-1}}\right)^{7/3} \, \mathrm{erg}\, \mathrm{s}^{-1}.
\end{equation}

Energy can be also dissipated in bursts in a volume $\sim \pi R_\mathrm{m}^3$, so the emitted amount is:
\begin{equation}
    E_\mathrm{rec}\approx \frac{B(R_\mathrm{m})^2}{8\pi} \times \pi R_\mathrm{m}^3.
\end{equation}
 Then the power is about $E_\mathrm{rec}/t_\mathrm{A}$, where $t_\mathrm{A}\approx R_\mathrm{m}/v_\mathrm{A}$ and the Alfven velocity $v_\mathrm{A}=B(R_\mathrm{m})/\sqrt{4\pi\rho}$.
 In \cite{2001ApJ...561..964T} Toropina et al. consider two cases: large density ($n\sim 1$~cm$^{-3}$) and low density ($v_\mathrm{A}\approx c$). 
In the first case
\begin{equation}
    \dot E_\mathrm{rec}\sim 10^{21}  B_{12}^{2/3} \left(\frac{v}{200\, \mathrm{km}\, \mathrm{s}^{-1}}\right)^{7/3} \, \mathrm{erg}\, \mathrm{s}^{-1}.
\end{equation}
In the second, the luminosity is larger due to larger characteristic velocity:

\begin{equation} 
    \dot E_\mathrm{rec}\sim 1.6 \times 10^{24}  B_{12}^{2/3} n^{2/3} \left(\frac{v}{200\, \mathrm{km}\, \mathrm{s}^{-1}}\right)^{4/3} \, \mathrm{erg}\, \mathrm{s}^{-1}.
 \end{equation}
But notice the dependence on $n$ which is low. 

In the case of reconnection in the tail, the energy might go into accelerating electrons and then it can be emitted in the radio band. 

Potentially, the tail can even reach the light cylinder. So, a region of opened field lines can be formed. 
It is tempting to call such an object a geoejector. However, a particle-producing cascade in the magnetosphere might be switched off as the NS left the ejector stage long before it reached the georotator stage. Still, elongated structures due to large spatial velocities can accompany radio pulsar activity, see a brief discussion in \cite{2001ApJ...561..964T}, sec. 6.5.

\subsection{Pulsating magnetospheres and other exotic regimes}

Two exotic transient regimes with a pulsating magnetosphere can be realized at the propeller and the ejector stages. 

It was hypothesized (see \cite{1992ans..book.....L} and references therein) that for effective cooling, a dense envelope can grow around the magnetosphere at the propeller stage.  
Equilibrium at the magnetospheric boundary can be described as:
\begin{equation}
    \frac{\mu^2}{8\pi R^6} = \frac{G M_\mathrm{sh} M}{4 \pi R^4}. 
\end{equation}
Here $M_\mathrm{sh}$ is the mass of the accumulating shell. As it grows, the magnetosphere shrinks until it reaches $R_\mathrm{co}$. Then the whole envelope collapses on the free-fall time scale. 

A similar regime of transient accretion can be realized also in the framework of the settling accretion model \cite{2015MNRAS.447.2817P}.
In this case, episodes of accretion are due to the development of the Rayleigh-Taylor instability at the magnetospheric boundary.

Another transient regime proposed in \cite{1984Ap&SS..98..221L}, can be related to the stage of ejector. It can be realized e.g., in a dense stellar wind of the companion in a binary system. 
Relativistic wind produced by a pulsar can be `captured' by the stellar wind. In this case, a cavern is formed around the NS. With time the pressure inside the cavern grows and so it is expanding. Pressure in the surrounding medium is not uniform and is decreasing outwards. Thus, at some point, the cavern opens. This might produce a modest radio flare. 
This scenario was proposed to explain properties of the Galactic center radio transient GCRT 1745-3009 \cite{2008arXiv0812.4587P}.  
Such pulsating caverns can be also formed around isolated NSs.

Finally, we want to mention again a regime of enhanced rotational energy losses at the radio pulsar stage when the configuration of magnetic field lines is modified by an accretion flow.
 This model was proposed by Parfrey et al. \cite{2016ApJ...822...33P} and we described it in some detail above (Sec. 5).
 This is again a hybrid stage as a relativistic wind from the compact object and an accretion disc inside a light cylinder around coexist. It might be possible to identify this stage (if it is realized in nature) in binary systems.


\section{Discussion and conclusions}

 Spin properties of NSs -- period, period derivatives, their correlations with other parameters e.g., flux, etc. -- are often well-measured and provide a model-independent source of information about these compact objects. Thus, it is important to know how these properties are related to other parameters (e.g., magnetic field, accretion rate, etc.) and how they evolve in time. 

 Unfortunately now, even in a clearer case of radio pulsars, we do not have a complete understanding of how spin parameters are related to other properties. 
Thus, above we provided a review of different models of spin behavior of NSs at various stages. 

Lack of understanding is related partly to complicated physical processes at work, and partly to the absence of detailed observational data.  Theoretical questions are related to the unknown equation of state (EOS) of NS interiors (see e.g., a review in \cite{2022arXiv220700033S} and references therein), to poorly known topology and evolution of the magnetic field (see reviews \cite{2021Univ....7..351I, 2022Symm...14..130G}), and to many uncertainties in plasma-magnetic field interaction (see \cite{2016AdAst2016E...3W, 2022arXiv220107262A}). From the observational side,  for example, there is no final description of the radio pulsar stage. On the other hand, there are no detected isolated accreting NSs or NSs at the stage of georotator. 
This situation results in many different approaches to describe the spin evolution of NSs. 

It is tempting to say that in the near future, thanks to advances in 3D MHD modeling of processes around NSs and with the appearance of new observational facilities many puzzles of magneto-rotational behavior and evolution of NSs will be solved. However, more than half a century of studies in this field of research suggests that we have to be more realistic. It seems that many approaches suggested already in the 1970s despite all the uncertainties and simplifications will be used in the forthcoming years to derive properties of NSs from spin measurements. 

We conclude by stating that despite our understanding of the spin behavior of NSs is far from being complete, there are many useful approaches suggested by many authors, and step by step we approach a better description of the magneto-rotational evolution of NSs. 
 
\vspace{6pt} 



\authorcontributions{PA made the main contribution to Sec. 5, except 5.2, AB -- to Sec. 3, and SP -- to Secs. 1, 2, 4, 5.2, 6, and 7. The authors contributed equally to the final editing and discussion of the manuscript.}

\funding{SP acknowledges support from Simons Foundation. PA was supported by Simons Foundation grant 00001470. }

\dataavailability{All data owned by the authors can be obtained on a reasonable request.} 

\acknowledgments{We thank anonymous referees for useful comments. }

\conflictsofinterest{The authors declare no conflict of interest. The funders had no role in the design of the study; in the collection, analyses, or interpretation of data; in the writing of the manuscript; or in the decision to publish the results.} 

\abbreviations{Abbreviations}{
The following abbreviations are used in this manuscript:\\

\noindent 
\begin{tabular}{@{}ll}
AINS & Accreting isolated neutron star\\
AMXP & Accreting millisecond X-ray pulsar \\
CCO & Central compact object\\
EOS & Equation of state\\
HMXB & High mass X-ray binary\\
INS & Isolated neutron star\\
ISM & Interstellar medium\\
LXMB & Low mass X-ray binary\\
MHD & Magnetohydrodynamic\\
MSP & Millisecond pulsar \\
MTD & Magnetically threaded disc \\
NS & Neutron star\\
PIC & Particle-in-cell\\
RPP & Rotation powered pulsar \\
SDI & Spin-disc interaction \\
SN & Supernova\\
SNR & Supernova remnant\\
TMSP & Transitional millisecond pulsar \\
UV & Ultra-violet \\
XRP & X-ray pulsar
\end{tabular}
}

\appendixtitles{no} 



\begin{adjustwidth}{-\extralength}{0cm}

\reftitle{References}

\externalbibliography{yes}
\bibliography{bibl}

\begin{thebibliography}{999}

\bibitem[{Lipunov}(1992)]{1992ans..book.....L}
{Lipunov}, V.M.
\newblock {\em {Astrophysics of Neutron Stars}}; Astronomy and Astrophysics
  Library, Springer-Verlag: Berlin Heidelberg,  1992.

\bibitem[{Lyutikov}(2023)]{2023MNRAS.520.4315L}
{Lyutikov}, M.
\newblock {Centrifugal barriers in magnetospheric accretion}.
\newblock {\em \mnras} {\bf 2023}, {\em 520},~4315--4323,
  \href{http://xxx.lanl.gov/abs/2210.00300}{{\normalfont
  [arXiv:astro-ph.HE/2210.00300]}}.
\newblock {\url{https://doi.org/10.1093/mnras/stad284}}.

\bibitem[{Abolmasov} and {Biryukov}(2020)]{2020MNRAS.496...13A}
{Abolmasov}, P.; {Biryukov}, A.
\newblock {Inertial oscillation modes of an inclined dipolar magnetosphere as a
  source of band-limited noise in X-ray pulsars}.
\newblock {\em \mnras} {\bf 2020}, {\em 496},~13--18,
  \href{http://xxx.lanl.gov/abs/2005.13508}{{\normalfont
  [arXiv:astro-ph.HE/2005.13508]}}.
\newblock {\url{https://doi.org/10.1093/mnras/staa1544}}.

\bibitem[{Igoshev} \em{et~al.}(2021){Igoshev}, {Popov}, and
  {Hollerbach}]{2021Univ....7..351I}
{Igoshev}, A.P.; {Popov}, S.B.; {Hollerbach}, R.
\newblock {Evolution of Neutron Star Magnetic Fields}.
\newblock {\em Universe} {\bf 2021}, {\em 7},~351,
  \href{http://xxx.lanl.gov/abs/2109.05584}{{\normalfont
  [arXiv:astro-ph.HE/2109.05584]}}.
\newblock {\url{https://doi.org/10.3390/universe7090351}}.

\bibitem[{Colgate}(1971)]{1971ApJ...163..221C}
{Colgate}, S.A.
\newblock {Neutron-Star Formation, Thermonuclear Supernovae, and Heavy-Element
  Reimplosion}.
\newblock {\em \apj} {\bf 1971}, {\em 163},~221.
\newblock {\url{https://doi.org/10.1086/150760}}.

\bibitem[{Zel'dovich} \em{et~al.}(1972){Zel'dovich}, {Ivanova}, and
  {Nadezhin}]{1972SvA....16..209Z}
{Zel'dovich}, Y.B.; {Ivanova}, L.N.; {Nadezhin}, D.K.
\newblock {Nonstationary Hydrodynamical Accretion onto a Neutron Star.}
\newblock {\em \sovast} {\bf 1972}, {\em 16},~209.

\bibitem[{Chevalier}(1989)]{1989ApJ...346..847C}
{Chevalier}, R.A.
\newblock {Neutron Star Accretion in a Supernova}.
\newblock {\em \apj} {\bf 1989}, {\em 346},~847.
\newblock {\url{https://doi.org/10.1086/168066}}.

\bibitem[{Metzger} \em{et~al.}(2018){Metzger}, {Beniamini}, and
  {Giannios}]{2018ApJ...857...95M}
{Metzger}, B.D.; {Beniamini}, P.; {Giannios}, D.
\newblock {Effects of Fallback Accretion on Protomagnetar Outflows in Gamma-Ray
  Bursts and Superluminous Supernovae}.
\newblock {\em \apj} {\bf 2018}, {\em 857},~95,
  \href{http://xxx.lanl.gov/abs/1802.07750}{{\normalfont
  [arXiv:astro-ph.HE/1802.07750]}}.
\newblock {\url{https://doi.org/10.3847/1538-4357/aab70c}}.

\bibitem[{Muslimov} and {Page}(1995)]{1995ApJ...440L..77M}
{Muslimov}, A.; {Page}, D.
\newblock {Delayed Switch-on of Pulsars}.
\newblock {\em \apjl} {\bf 1995}, {\em 440},~L77.
\newblock {\url{https://doi.org/10.1086/187765}}.

\bibitem[{Geppert} \em{et~al.}(1999){Geppert}, {Page}, and
  {Zannias}]{1999A&A...345..847G}
{Geppert}, U.; {Page}, D.; {Zannias}, T.
\newblock {Submergence and re-diffusion of the neutron star magnetic field
  after the supernova}.
\newblock {\em \aap} {\bf 1999}, {\em 345},~847--854.

\bibitem[{Ho}(2011)]{2011MNRAS.414.2567H}
{Ho}, W.C.G.
\newblock {Evolution of a buried magnetic field in the central compact object
  neutron stars}.
\newblock {\em \mnras} {\bf 2011}, {\em 414},~2567--2575,
  \href{http://xxx.lanl.gov/abs/1102.4870}{{\normalfont
  [arXiv:astro-ph.HE/1102.4870]}}.
\newblock {\url{https://doi.org/10.1111/j.1365-2966.2011.18576.x}}.

\bibitem[{Vigan{\`o}} and {Pons}(2012)]{2012MNRAS.425.2487V}
{Vigan{\`o}}, D.; {Pons}, J.A.
\newblock {Central compact objects and the hidden magnetic field scenario}.
\newblock {\em \mnras} {\bf 2012}, {\em 425},~2487--2492,
  \href{http://xxx.lanl.gov/abs/1206.2014}{{\normalfont
  [arXiv:astro-ph.SR/1206.2014]}}.
\newblock {\url{https://doi.org/10.1111/j.1365-2966.2012.21679.x}}.

\bibitem[{Pons} \em{et~al.}(2012){Pons}, {Vigan{\`o}}, and
  {Geppert}]{2012A&A...547A...9P}
{Pons}, J.A.; {Vigan{\`o}}, D.; {Geppert}, U.
\newblock {Pulsar timing irregularities and the imprint of magnetic field
  evolution}.
\newblock {\em \aap} {\bf 2012}, {\em 547},~A9,
  \href{http://xxx.lanl.gov/abs/1209.2273}{{\normalfont
  [arXiv:astro-ph.SR/1209.2273]}}.
\newblock {\url{https://doi.org/10.1051/0004-6361/201220091}}.

\bibitem[{Igoshev} \em{et~al.}(2016){Igoshev}, {Elfritz}, and
  {Popov}]{2016MNRAS.462.3689I}
{Igoshev}, A.P.; {Elfritz}, J.G.; {Popov}, S.B.
\newblock {Post-fall-back evolution of multipolar magnetic fields and radio
  pulsar activation}.
\newblock {\em \mnras} {\bf 2016}, {\em 462},~3689--3702,
  \href{http://xxx.lanl.gov/abs/1608.08806}{{\normalfont
  [arXiv:astro-ph.HE/1608.08806]}}.
\newblock {\url{https://doi.org/10.1093/mnras/stw1902}}.

\bibitem[{Bernal} \em{et~al.}(2010){Bernal}, {Lee}, and
  {Page}]{2010RMxAA..46..309B}
{Bernal}, C.G.; {Lee}, W.H.; {Page}, D.
\newblock {Hypercritical accretion onto a magnetized neutron star surface: a
  numerical approach}.
\newblock {\em \rmxaa} {\bf 2010}, {\em 46},~309--322.

\bibitem[{Bernal} \em{et~al.}(2013){Bernal}, {Page}, and
  {Lee}]{2013ApJ...770..106B}
{Bernal}, C.G.; {Page}, D.; {Lee}, W.H.
\newblock {Hypercritical Accretion onto a Newborn Neutron Star and Magnetic
  Field Submergence}.
\newblock {\em \apj} {\bf 2013}, {\em 770},~106,
  \href{http://xxx.lanl.gov/abs/1212.0464}{{\normalfont
  [arXiv:astro-ph.HE/1212.0464]}}.
\newblock {\url{https://doi.org/10.1088/0004-637X/770/2/106}}.

\bibitem[{Fraija} \em{et~al.}(2018){Fraija}, {Bernal}, {Morales}, and
  {Negreiros}]{2018PhRvD..98h3012F}
{Fraija}, N.; {Bernal}, C.G.; {Morales}, G.; {Negreiros}, R.
\newblock {Hypercritical accretion scenario in central compact objects
  accompanied with an expected neutrino burst}.
\newblock {\em \prd} {\bf 2018}, {\em 98},~083012,
  \href{http://xxx.lanl.gov/abs/1809.07057}{{\normalfont
  [arXiv:astro-ph.HE/1809.07057]}}.
\newblock {\url{https://doi.org/10.1103/PhysRevD.98.083012}}.

\bibitem[{Shigeyama} and {Kashiyama}(2018)]{2018PASJ...70..107S}
{Shigeyama}, T.; {Kashiyama}, K.
\newblock {Repulsion of fallback matter due to central energy source in
  supernova}.
\newblock {\em \pasj} {\bf 2018}, {\em 70},~107,
  \href{http://xxx.lanl.gov/abs/1809.00487}{{\normalfont
  [arXiv:astro-ph.SR/1809.00487]}}.
\newblock {\url{https://doi.org/10.1093/pasj/psy108}}.

\bibitem[{Zhong} \em{et~al.}(2021){Zhong}, {Kashiyama}, {Shigeyama}, and
  {Takasao}]{2021ApJ...917...71Z}
{Zhong}, Y.; {Kashiyama}, K.; {Shigeyama}, T.; {Takasao}, S.
\newblock {A Necessary Condition for Supernova Fallback Invading Newborn
  Neutron-star Magnetosphere}.
\newblock {\em \apj} {\bf 2021}, {\em 917},~71,
  \href{http://xxx.lanl.gov/abs/2103.09461}{{\normalfont
  [arXiv:astro-ph.HE/2103.09461]}}.
\newblock {\url{https://doi.org/10.3847/1538-4357/ac0a74}}.

\bibitem[{Parfrey} \em{et~al.}(2016){Parfrey}, {Spitkovsky}, and
  {Beloborodov}]{2016ApJ...822...33P}
{Parfrey}, K.; {Spitkovsky}, A.; {Beloborodov}, A.M.
\newblock {Torque Enhancement, Spin Equilibrium, and Jet Power from
  Disk-Induced Opening of Pulsar Magnetic Fields}.
\newblock {\em \apj} {\bf 2016}, {\em 822},~33,
  \href{http://xxx.lanl.gov/abs/1507.08627}{{\normalfont
  [arXiv:astro-ph.HE/1507.08627]}}.
\newblock {\url{https://doi.org/10.3847/0004-637X/822/1/33}}.

\bibitem[{Wang} \em{et~al.}(2006){Wang}, {Chakrabarty}, and
  {Kaplan}]{2006Natur.440..772W}
{Wang}, Z.; {Chakrabarty}, D.; {Kaplan}, D.L.
\newblock {A debris disk around an isolated young neutron star}.
\newblock {\em \nat} {\bf 2006}, {\em 440},~772--775,
  \href{http://xxx.lanl.gov/abs/astro-ph/0604076}{{\normalfont
  [arXiv:astro-ph/astro-ph/0604076]}}.
\newblock {\url{https://doi.org/10.1038/nature04669}}.

\bibitem[{Chatterjee} \em{et~al.}(2000){Chatterjee}, {Hernquist}, and
  {Narayan}]{2000ApJ...534..373C}
{Chatterjee}, P.; {Hernquist}, L.; {Narayan}, R.
\newblock {An Accretion Model for Anomalous X-Ray Pulsars}.
\newblock {\em \apj} {\bf 2000}, {\em 534},~373--379,
  \href{http://xxx.lanl.gov/abs/astro-ph/9912137}{{\normalfont
  [arXiv:astro-ph/astro-ph/9912137]}}.
\newblock {\url{https://doi.org/10.1086/308748}}.

\bibitem[{Jones}(2007)]{2007MNRAS.382..871J}
{Jones}, P.B.
\newblock {Interaction and ablation of fall-back discs in isolated neutron
  stars}.
\newblock {\em \mnras} {\bf 2007}, {\em 382},~871--878,
  \href{http://xxx.lanl.gov/abs/0709.3730}{{\normalfont
  [arXiv:astro-ph/0709.3730]}}.
\newblock {\url{https://doi.org/10.1111/j.1365-2966.2007.12428.x}}.

\bibitem[{Benli} and {Ertan}(2016)]{2016MNRAS.457.4114B}
{Benli}, O.; {Ertan}, {\"U}.
\newblock {Long-term evolution of anomalous X-ray pulsars and soft gamma
  repeaters}.
\newblock {\em \mnras} {\bf 2016}, {\em 457},~4114--4122,
  \href{http://xxx.lanl.gov/abs/1601.07846}{{\normalfont
  [arXiv:astro-ph.HE/1601.07846]}}.
\newblock {\url{https://doi.org/10.1093/mnras/stw235}}.

\bibitem[{Janka} \em{et~al.}(2022){Janka}, {Wongwathanarat}, and
  {Kramer}]{2022ApJ...926....9J}
{Janka}, H.T.; {Wongwathanarat}, A.; {Kramer}, M.
\newblock {Supernova Fallback as Origin of Neutron Star Spins and Spin-kick
  Alignment}.
\newblock {\em \apj} {\bf 2022}, {\em 926},~9,
  \href{http://xxx.lanl.gov/abs/2104.07493}{{\normalfont
  [arXiv:astro-ph.HE/2104.07493]}}.
\newblock {\url{https://doi.org/10.3847/1538-4357/ac403c}}.

\bibitem[{Popov} and {Turolla}(2012)]{2012Ap&SS.341..457P}
{Popov}, S.B.; {Turolla}, R.
\newblock {Initial spin periods of neutron stars in supernova remnants}.
\newblock {\em \apss} {\bf 2012}, {\em 341},~457--464,
  \href{http://xxx.lanl.gov/abs/1204.0632}{{\normalfont
  [arXiv:astro-ph.HE/1204.0632]}}.
\newblock {\url{https://doi.org/10.1007/s10509-012-1100-z}}.

\bibitem[{Fu} and {Li}(2013)]{2013ApJ...775..124F}
{Fu}, L.; {Li}, X.D.
\newblock {Population Synthesis of Young Isolated Neutron Stars: The Effect of
  Fallback Disk Accretion and Magnetic Field Evolution}.
\newblock {\em \apj} {\bf 2013}, {\em 775},~124,
  \href{http://xxx.lanl.gov/abs/1308.3011}{{\normalfont
  [arXiv:astro-ph.HE/1308.3011]}}.
\newblock {\url{https://doi.org/10.1088/0004-637X/775/2/124}}.

\bibitem[{Caleb} \em{et~al.}(2022){Caleb}, {Heywood}, {Rajwade}, {Malenta},
  {Stappers}, {Barr}, {Chen}, {Morello}, {Sanidas}, {van den Eijnden},
  {Kramer}, {Buckley}, {Brink}, {Motta}, {Woudt}, {Weltevrede}, {Jankowski},
  {Surnis}, {Buchner}, {Bezuidenhout}, {Driessen}, and
  {Fender}]{2022NatAs...6..828C}
{Caleb}, M.; {Heywood}, I.; {Rajwade}, K.; {Malenta}, M.; {Stappers}, B.W.;
  {Barr}, E.; {Chen}, W.; {Morello}, V.; {Sanidas}, S.; {van den Eijnden}, J.;
  et~al.
\newblock {Discovery of a radio-emitting neutron star with an ultra-long spin
  period of 76 s}.
\newblock {\em Nature Astronomy} {\bf 2022}, {\em 6},~828--836,
  \href{http://xxx.lanl.gov/abs/2206.01346}{{\normalfont
  [arXiv:astro-ph.HE/2206.01346]}}.
\newblock {\url{https://doi.org/10.1038/s41550-022-01688-x}}.

\bibitem[{Hurley-Walker} \em{et~al.}(2022){Hurley-Walker}, {Zhang},
  {Bahramian}, {McSweeney}, {O'Doherty}, {Hancock}, {Morgan}, {Anderson},
  {Heald}, and {Galvin}]{2022Natur.601..526H}
{Hurley-Walker}, N.; {Zhang}, X.; {Bahramian}, A.; {McSweeney}, S.J.;
  {O'Doherty}, T.N.; {Hancock}, P.J.; {Morgan}, J.S.; {Anderson}, G.E.;
  {Heald}, G.H.; {Galvin}, T.J.
\newblock {A radio transient with unusually slow periodic emission}.
\newblock {\em \nat} {\bf 2022}, {\em 601},~526--530.
\newblock {\url{https://doi.org/10.1038/s41586-021-04272-x}}.

\bibitem[{Hurley-Walker} \em{et~al.}(2023){Hurley-Walker}, {Rea}, {McSweeney},
  {Meyers}, {Lenc}, {Heywood}, {Hyman}, {Men}, {Clarke}, {Coti Zelati},
  {Price}, {Horv{\'a}th}, {Galvin}, {Anderson}, {Bahramian}, {Barr}, {Bhat},
  {Caleb}, {Dall'Ora}, {de Martino}, {Giacintucci}, {Morgan}, {Rajwade},
  {Stappers}, and {Williams}]{2023Natur.619..487H}
{Hurley-Walker}, N.; {Rea}, N.; {McSweeney}, S.J.; {Meyers}, B.W.; {Lenc}, E.;
  {Heywood}, I.; {Hyman}, S.D.; {Men}, Y.P.; {Clarke}, T.E.; {Coti Zelati}, F.;
   et~al.
\newblock {A long-period radio transient active for three decades}.
\newblock {\em \nat} {\bf 2023}, {\em 619},~487--490.
\newblock {\url{https://doi.org/10.1038/s41586-023-06202-5}}.

\bibitem[{Ronchi} \em{et~al.}(2022){Ronchi}, {Rea}, {Graber}, and
  {Hurley-Walker}]{2022ApJ...934..184R}
{Ronchi}, M.; {Rea}, N.; {Graber}, V.; {Hurley-Walker}, N.
\newblock {Long-period Pulsars as Possible Outcomes of Supernova Fallback
  Accretion}.
\newblock {\em \apj} {\bf 2022}, {\em 934},~184,
  \href{http://xxx.lanl.gov/abs/2201.11704}{{\normalfont
  [arXiv:astro-ph.HE/2201.11704]}}.
\newblock {\url{https://doi.org/10.3847/1538-4357/ac7cec}}.

\bibitem[{Aguilera} \em{et~al.}(2008){Aguilera}, {Pons}, and
  {Miralles}]{2008A&A...486..255A}
{Aguilera}, D.N.; {Pons}, J.A.; {Miralles}, J.A.
\newblock {2D Cooling of magnetized neutron stars}.
\newblock {\em \aap} {\bf 2008}, {\em 486},~255--271,
  \href{http://xxx.lanl.gov/abs/0710.0854}{{\normalfont
  [arXiv:astro-ph/0710.0854]}}.
\newblock {\url{https://doi.org/10.1051/0004-6361:20078786}}.

\bibitem[{Manchester} \em{et~al.}(2005){Manchester}, {Hobbs}, {Teoh}, and
  {Hobbs}]{2005AJ....129.1993M}
{Manchester}, R.N.; {Hobbs}, G.B.; {Teoh}, A.; {Hobbs}, M.
\newblock {The Australia Telescope National Facility Pulsar Catalogue}.
\newblock {\em \aj} {\bf 2005}, {\em 129},~1993--2006,
  \href{http://xxx.lanl.gov/abs/astro-ph/0412641}{{\normalfont
  [arXiv:astro-ph/astro-ph/0412641]}}.
\newblock {\url{https://doi.org/10.1086/428488}}.

\bibitem[{Beskin} \em{et~al.}(2015){Beskin}, {Chernov}, {Gwinn}, and
  {Tchekhovskoy}]{2015SSRv..191..207B}
{Beskin}, V.S.; {Chernov}, S.V.; {Gwinn}, C.R.; {Tchekhovskoy}, A.A.
\newblock {Radio Pulsars}.
\newblock {\em \ssr} {\bf 2015}, {\em 191},~207--237,
  \href{http://xxx.lanl.gov/abs/1506.07881}{{\normalfont
  [arXiv:astro-ph.HE/1506.07881]}}.
\newblock {\url{https://doi.org/10.1007/s11214-015-0173-8}}.

\bibitem[{Beskin}(2018)]{2018PhyU...61..353B}
{Beskin}, V.S.
\newblock {Radio pulsars: already fifty years!}
\newblock {\em Physics Uspekhi} {\bf 2018}, {\em 61},~353--380,
  \href{http://xxx.lanl.gov/abs/1807.08528}{{\normalfont
  [arXiv:astro-ph.HE/1807.08528]}}.
\newblock {\url{https://doi.org/10.3367/UFNe.2017.10.038216}}.

\bibitem[{Lorimer} and {Kramer}(2012)]{2012hpa..book.....L}
{Lorimer}, D.R.; {Kramer}, M.
\newblock {\em {Handbook of Pulsar Astronomy}}; Cambridge University Press,
  2012.

\bibitem[{Shklovskii}(1970)]{1970SvA....13..562S}
{Shklovskii}, I.S.
\newblock {Possible Causes of the Secular Increase in Pulsar Periods.}
\newblock {\em \sovast} {\bf 1970}, {\em 13},~562.

\bibitem[{Beskin} and {Istomin}(2022)]{2022MNRAS.516.5084B}
{Beskin}, V.S.; {Istomin}, A.Y.
\newblock {Pulsar death line revisited - II. 'The death valley'}.
\newblock {\em \mnras} {\bf 2022}, {\em 516},~5084--5091,
  \href{http://xxx.lanl.gov/abs/2207.04723}{{\normalfont
  [arXiv:astro-ph.HE/2207.04723]}}.
\newblock {\url{https://doi.org/10.1093/mnras/stac2423}}.

\bibitem[{Krolik}(1991)]{1991ApJ...373L..69K}
{Krolik}, J.H.
\newblock {Multipolar Magnetic Fields in Neutron Stars}.
\newblock {\em \apjl} {\bf 1991}, {\em 373},~L69.
\newblock {\url{https://doi.org/10.1086/186053}}.

\bibitem[{Kantor} and {Tsygan}(2004)]{2004ARep...48.1029K}
{Kantor}, E.M.; {Tsygan}, A.I.
\newblock {The Death Lines of Radio Pulsars for Dipolar and Asymmetric Magnetic
  Fields}.
\newblock {\em Astronomy Reports} {\bf 2004}, {\em 48},~1029--1036.
\newblock {\url{https://doi.org/10.1134/1.1836026}}.

\bibitem[{P{\'e}tri}(2015)]{2015MNRAS.450..714P}
{P{\'e}tri}, J.
\newblock {Multipolar electromagnetic fields around neutron stars: exact vacuum
  solutions and related properties}.
\newblock {\em \mnras} {\bf 2015}, {\em 450},~714--742,
  \href{http://xxx.lanl.gov/abs/1503.05307}{{\normalfont
  [arXiv:astro-ph.HE/1503.05307]}}.
\newblock {\url{https://doi.org/10.1093/mnras/stv598}}.

\bibitem[{P{\'e}tri}(2017)]{2017MNRAS.472.3304P}
{P{\'e}tri}, J.
\newblock {Multipolar electromagnetic fields around neutron stars:
  general-relativistic vacuum solutions}.
\newblock {\em \mnras} {\bf 2017}, {\em 472},~3304--3336,
  \href{http://xxx.lanl.gov/abs/1702.03172}{{\normalfont
  [arXiv:astro-ph.HE/1702.03172]}}.
\newblock {\url{https://doi.org/10.1093/mnras/stx2147}}.

\bibitem[{Sturrock}(1971)]{1971ApJ...164..529S}
{Sturrock}, P.A.
\newblock {A Model of Pulsars}.
\newblock {\em \apj} {\bf 1971}, {\em 164},~529.
\newblock {\url{https://doi.org/10.1086/150865}}.

\bibitem[{Ruderman} and {Sutherland}(1975)]{1975ApJ...196...51R}
{Ruderman}, M.A.; {Sutherland}, P.G.
\newblock {Theory of pulsars: polar gaps, sparks, and coherent microwave
  radiation.}
\newblock {\em \apj} {\bf 1975}, {\em 196},~51--72.
\newblock {\url{https://doi.org/10.1086/153393}}.

\bibitem[{Spitkovsky}(2004)]{2004IAUS..218..357S}
{Spitkovsky}, A.
\newblock {Electrodynamics of Pulsar Magnetospheres}.
\newblock In Proceedings of the Young Neutron Stars and Their Environments;
  {Camilo}, F.; {Gaensler}, B.M., Eds.,  2004, Vol. 218, p. 357,
  \href{http://xxx.lanl.gov/abs/astro-ph/0310731}{{\normalfont
  [arXiv:astro-ph/astro-ph/0310731]}}.
\newblock {\url{https://doi.org/10.48550/arXiv.astro-ph/0310731}}.

\bibitem[{Goldreich} and {Julian}(1969)]{1969ApJ...157..869G}
{Goldreich}, P.; {Julian}, W.H.
\newblock {Pulsar Electrodynamics}.
\newblock {\em \apj} {\bf 1969}, {\em 157},~869.
\newblock {\url{https://doi.org/10.1086/150119}}.

\bibitem[{Kou} and {Tong}(2015)]{2015MNRAS.450.1990K}
{Kou}, F.F.; {Tong}, H.
\newblock {Rotational evolution of the Crab pulsar in the wind braking model}.
\newblock {\em \mnras} {\bf 2015}, {\em 450},~1990--1998,
  \href{http://xxx.lanl.gov/abs/1501.01534}{{\normalfont
  [arXiv:astro-ph.HE/1501.01534]}}.
\newblock {\url{https://doi.org/10.1093/mnras/stv734}}.

\bibitem[{Harding} \em{et~al.}(1999){Harding}, {Contopoulos}, and
  {Kazanas}]{1999ApJ...525L.125H}
{Harding}, A.K.; {Contopoulos}, I.; {Kazanas}, D.
\newblock {Magnetar Spin-Down}.
\newblock {\em \apjl} {\bf 1999}, {\em 525},~L125--L128,
  \href{http://xxx.lanl.gov/abs/astro-ph/9908279}{{\normalfont
  [arXiv:astro-ph/astro-ph/9908279]}}.
\newblock {\url{https://doi.org/10.1086/312339}}.

\bibitem[{Xu} and {Qiao}(2001)]{2001ApJ...561L..85X}
{Xu}, R.X.; {Qiao}, G.J.
\newblock {Pulsar Braking Index: A Test of Emission Models?}
\newblock {\em \apjl} {\bf 2001}, {\em 561},~L85--L88,
  \href{http://xxx.lanl.gov/abs/astro-ph/0108235}{{\normalfont
  [arXiv:astro-ph/astro-ph/0108235]}}.
\newblock {\url{https://doi.org/10.1086/324381}}.

\bibitem[{Contopoulos} and {Spitkovsky}(2006)]{2006ApJ...643.1139C}
{Contopoulos}, I.; {Spitkovsky}, A.
\newblock {Revised Pulsar Spin-down}.
\newblock {\em \apj} {\bf 2006}, {\em 643},~1139--1145,
  \href{http://xxx.lanl.gov/abs/astro-ph/0512002}{{\normalfont
  [arXiv:astro-ph/astro-ph/0512002]}}.
\newblock {\url{https://doi.org/10.1086/501161}}.

\bibitem[{Chen} and {Ruderman}(1993)]{1993ApJ...402..264C}
{Chen}, K.; {Ruderman}, M.
\newblock {Pulsar Death Lines and Death Valley}.
\newblock {\em \apj} {\bf 1993}, {\em 402},~264.
\newblock {\url{https://doi.org/10.1086/172129}}.

\bibitem[{Muslimov} and {Tsygan}(1992)]{1992MNRAS.255...61M}
{Muslimov}, A.G.; {Tsygan}, A.I.
\newblock {General relativistic electric potential drops above pulsar polar
  caps}.
\newblock {\em \mnras} {\bf 1992}, {\em 255},~61--70.
\newblock {\url{https://doi.org/10.1093/mnras/255.1.61}}.

\bibitem[{Zhang} \em{et~al.}(2000){Zhang}, {Harding}, and
  {Muslimov}]{2000ApJ...531L.135Z}
{Zhang}, B.; {Harding}, A.K.; {Muslimov}, A.G.
\newblock {Radio Pulsar Death Line Revisited: Is PSR J2144-3933 Anomalous?}
\newblock {\em \apjl} {\bf 2000}, {\em 531},~L135--L138,
  \href{http://xxx.lanl.gov/abs/astro-ph/0001341}{{\normalfont
  [arXiv:astro-ph/astro-ph/0001341]}}.
\newblock {\url{https://doi.org/10.1086/312542}}.

\bibitem[{Faucher-Gigu{\`e}re} and {Kaspi}(2006)]{2006ApJ...643..332F}
{Faucher-Gigu{\`e}re}, C.A.; {Kaspi}, V.M.
\newblock {Birth and Evolution of Isolated Radio Pulsars}.
\newblock {\em \apj} {\bf 2006}, {\em 643},~332--355,
  \href{http://xxx.lanl.gov/abs/astro-ph/0512585}{{\normalfont
  [arXiv:astro-ph/astro-ph/0512585]}}.
\newblock {\url{https://doi.org/10.1086/501516}}.

\bibitem[{Rawley} \em{et~al.}(1986){Rawley}, {Taylor}, and
  {Davis}]{1986Natur.319..383R}
{Rawley}, L.A.; {Taylor}, J.H.; {Davis}, M.M.
\newblock {Period derivative and orbital eccentricity of binary pulsar 1953 +
  29}.
\newblock {\em Nature} {\bf 1986}, {\em 319},~383.
\newblock {\url{https://doi.org/10.1038/319383a0}}.

\bibitem[{Gull{\'o}n} \em{et~al.}(2014){Gull{\'o}n}, {Miralles}, {Vigan{\`o}},
  and {Pons}]{2014MNRAS.443.1891G}
{Gull{\'o}n}, M.; {Miralles}, J.A.; {Vigan{\`o}}, D.; {Pons}, J.A.
\newblock {Population synthesis of isolated neutron stars with
  magneto-rotational evolution}.
\newblock {\em \mnras} {\bf 2014}, {\em 443},~1891--1899,
  \href{http://xxx.lanl.gov/abs/1406.6794}{{\normalfont
  [arXiv:astro-ph.HE/1406.6794]}}.
\newblock {\url{https://doi.org/10.1093/mnras/stu1253}}.

\bibitem[{Graber} \em{et~al.}(2023){Graber}, {Ronchi}, {Pardo-Araujo}, and
  {Rea}]{2023arXiv231214848G}
{Graber}, V.; {Ronchi}, M.; {Pardo-Araujo}, C.; {Rea}, N.
\newblock {Isolated pulsar population synthesis with simulation-based
  inference}.
\newblock {\em arXiv e-prints} {\bf 2023}, p. arXiv:2312.14848,
  \href{http://xxx.lanl.gov/abs/2312.14848}{{\normalfont
  [arXiv:astro-ph.HE/2312.14848]}}.
\newblock {\url{https://doi.org/10.48550/arXiv.2312.14848}}.

\bibitem[{Beskin} and {Litvinov}(2022)]{2022MNRAS.510.2572B}
{Beskin}, V.S.; {Litvinov}, P.E.
\newblock {Pulsar death line revisited - I. Almost vacuum gap}.
\newblock {\em \mnras} {\bf 2022}, {\em 510},~2572--2582,
  \href{http://xxx.lanl.gov/abs/2201.02875}{{\normalfont
  [arXiv:astro-ph.HE/2201.02875]}}.
\newblock {\url{https://doi.org/10.1093/mnras/stab3575}}.

\bibitem[{Spitkovsky}(2006)]{2006ApJ...648L..51S}
{Spitkovsky}, A.
\newblock {Time-dependent Force-free Pulsar Magnetospheres: Axisymmetric and
  Oblique Rotators}.
\newblock {\em \apjl} {\bf 2006}, {\em 648},~L51--L54,
  \href{http://xxx.lanl.gov/abs/astro-ph/0603147}{{\normalfont
  [arXiv:astro-ph/astro-ph/0603147]}}.
\newblock {\url{https://doi.org/10.1086/507518}}.

\bibitem[{Beskin} \em{et~al.}(1993){Beskin}, {Gurevich}, and
  {Istomin}]{1993ppm..book.....B}
{Beskin}, V.S.; {Gurevich}, A.V.; {Istomin}, Y.N.
\newblock {\em {Physics of the pulsar magnetosphere}}; Cambridge University
  Press,  1993.

\bibitem[{Novoselov} \em{et~al.}(2020){Novoselov}, {Beskin}, {Galishnikova},
  {Rashkovetskyi}, and {Biryukov}]{2020MNRAS.494.3899N}
{Novoselov}, E.M.; {Beskin}, V.S.; {Galishnikova}, A.K.; {Rashkovetskyi}, M.M.;
  {Biryukov}, A.V.
\newblock {Orthogonal pulsars as a key test for pulsar evolution}.
\newblock {\em \mnras} {\bf 2020}, {\em 494},~3899--3911,
  \href{http://xxx.lanl.gov/abs/2004.03211}{{\normalfont
  [arXiv:astro-ph.HE/2004.03211]}}.
\newblock {\url{https://doi.org/10.1093/mnras/staa904}}.

\bibitem[{Hewish} \em{et~al.}(1968){Hewish}, {Bell}, {Pilkington}, {Scott}, and
  {Collins}]{1968Natur.217..709H}
{Hewish}, A.; {Bell}, S.J.; {Pilkington}, J.D.H.; {Scott}, P.F.; {Collins},
  R.A.
\newblock {Observation of a Rapidly Pulsating Radio Source}.
\newblock {\em \nat} {\bf 1968}, {\em 217},~709--713.
\newblock {\url{https://doi.org/10.1038/217709a0}}.

\bibitem[{Pacini}(1967)]{1967Natur.216..567P}
{Pacini}, F.
\newblock {Energy Emission from a Neutron Star}.
\newblock {\em \nat} {\bf 1967}, {\em 216},~567--568.
\newblock {\url{https://doi.org/10.1038/216567a0}}.

\bibitem[{Deutsch}(1955)]{1955AnAp...18....1D}
{Deutsch}, A.J.
\newblock {The electromagnetic field of an idealized star in rigid rotation in
  vacuo}.
\newblock {\em Annales d'Astrophysique} {\bf 1955}, {\em 18},~1.

\bibitem[{Satherley} and {Gordon}(2022)]{2022PASA...39...38S}
{Satherley}, J.C.; {Gordon}, C.
\newblock {A pedagogical review of the vacuum retarded dipole model of pulsar
  spin down}.
\newblock {\em \pasa} {\bf 2022}, {\em 39},~e038,
  \href{http://xxx.lanl.gov/abs/2103.10617}{{\normalfont
  [arXiv:astro-ph.HE/2103.10617]}}.
\newblock {\url{https://doi.org/10.1017/pasa.2022.35}}.

\bibitem[{Ostriker} and {Gunn}(1969)]{1969ApJ...157.1395O}
{Ostriker}, J.P.; {Gunn}, J.E.
\newblock {On the Nature of Pulsars. I. Theory}.
\newblock {\em \apj} {\bf 1969}, {\em 157},~1395.
\newblock {\url{https://doi.org/10.1086/150160}}.

\bibitem[{Davis} and {Goldstein}(1970)]{1970ApJ...159L..81D}
{Davis}, L.; {Goldstein}, M.
\newblock {Magnetic-Dipole Alignment in Pulsars}.
\newblock {\em \apjl} {\bf 1970}, {\em 159},~L81--L86.
\newblock {\url{https://doi.org/10.1086/180482}}.

\bibitem[{Good} and {Ng}(1985)]{1985ApJ...299..706G}
{Good}, M.L.; {Ng}, K.K.
\newblock {Electromagnetic torques, secular alignment, and spin-down of neutron
  stars}.
\newblock {\em \apj} {\bf 1985}, {\em 299},~706--722.
\newblock {\url{https://doi.org/10.1086/163736}}.

\bibitem[{Melatos}(2000)]{2000MNRAS.313..217M}
{Melatos}, A.
\newblock {Radiative precession of an isolated neutron star}.
\newblock {\em \mnras} {\bf 2000}, {\em 313},~217--228,
  \href{http://xxx.lanl.gov/abs/astro-ph/0004035}{{\normalfont
  [arXiv:astro-ph/astro-ph/0004035]}}.
\newblock {\url{https://doi.org/10.1046/j.1365-8711.2000.03031.x}}.

\bibitem[{Biryukov} \em{et~al.}(2012){Biryukov}, {Beskin}, and
  {Karpov}]{2012MNRAS.420..103B}
{Biryukov}, A.; {Beskin}, G.; {Karpov}, S.
\newblock {Monotonic and cyclic components of radio pulsar spin-down}.
\newblock {\em \mnras} {\bf 2012}, {\em 420},~103--117,
  \href{http://xxx.lanl.gov/abs/1105.5019}{{\normalfont
  [arXiv:astro-ph.HE/1105.5019]}}.
\newblock {\url{https://doi.org/10.1111/j.1365-2966.2011.20005.x}}.

\bibitem[{Goldreich}(1970)]{1970ApJ...160L..11G}
{Goldreich}, P.
\newblock {Neutron Star Crusts and Alignment of Magnetic Axes in Pulsars}.
\newblock {\em \apjl} {\bf 1970}, {\em 160},~L11.
\newblock {\url{https://doi.org/10.1086/180513}}.

\bibitem[{Beskin} and {Zheltoukhov}(2014)]{2014PhyU...57..799B}
{Beskin}, V.S.; {Zheltoukhov}, A.A.
\newblock {Anomalous torque applied to a rotating magnetized sphere in a
  vacuum}.
\newblock {\em Physics Uspekhi} {\bf 2014}, {\em 57},~799--806,
  \href{http://xxx.lanl.gov/abs/1411.2107}{{\normalfont
  [arXiv:astro-ph.HE/1411.2107]}}.
\newblock {\url{https://doi.org/10.3367/UFNe.0184.201408e.0865}}.

\bibitem[{Contopoulos} \em{et~al.}(1999){Contopoulos}, {Kazanas}, and
  {Fendt}]{1999ApJ...511..351C}
{Contopoulos}, I.; {Kazanas}, D.; {Fendt}, C.
\newblock {The Axisymmetric Pulsar Magnetosphere}.
\newblock {\em \apj} {\bf 1999}, {\em 511},~351--358,
  \href{http://xxx.lanl.gov/abs/astro-ph/9903049}{{\normalfont
  [arXiv:astro-ph/astro-ph/9903049]}}.
\newblock {\url{https://doi.org/10.1086/306652}}.

\bibitem[{Gruzinov}(2005)]{2005PhRvL..94b1101G}
{Gruzinov}, A.
\newblock {Power of an Axisymmetric Pulsar}.
\newblock {\em \prl} {\bf 2005}, {\em 94},~021101,
  \href{http://xxx.lanl.gov/abs/astro-ph/0407279}{{\normalfont
  [arXiv:astro-ph/astro-ph/0407279]}}.
\newblock {\url{https://doi.org/10.1103/PhysRevLett.94.021101}}.

\bibitem[{McKinney}(2006)]{2006MNRAS.368L..30M}
{McKinney}, J.C.
\newblock {Relativistic force-free electrodynamic simulations of neutron star
  magnetospheres}.
\newblock {\em \mnras} {\bf 2006}, {\em 368},~L30--L34,
  \href{http://xxx.lanl.gov/abs/astro-ph/0601411}{{\normalfont
  [arXiv:astro-ph/astro-ph/0601411]}}.
\newblock {\url{https://doi.org/10.1111/j.1745-3933.2006.00150.x}}.

\bibitem[{Timokhin}(2006)]{2006MNRAS.368.1055T}
{Timokhin}, A.N.
\newblock {On the force-free magnetosphere of an aligned rotator}.
\newblock {\em \mnras} {\bf 2006}, {\em 368},~1055--1072,
  \href{http://xxx.lanl.gov/abs/astro-ph/0511817}{{\normalfont
  [arXiv:astro-ph/astro-ph/0511817]}}.
\newblock {\url{https://doi.org/10.1111/j.1365-2966.2006.10192.x}}.

\bibitem[{Parfrey} \em{et~al.}(2012){Parfrey}, {Beloborodov}, and
  {Hui}]{2012MNRAS.423.1416P}
{Parfrey}, K.; {Beloborodov}, A.M.; {Hui}, L.
\newblock {Introducing PHAEDRA: a new spectral code for simulations of
  relativistic magnetospheres}.
\newblock {\em \mnras} {\bf 2012}, {\em 423},~1416--1436,
  \href{http://xxx.lanl.gov/abs/1110.6669}{{\normalfont
  [arXiv:astro-ph.HE/1110.6669]}}.
\newblock {\url{https://doi.org/10.1111/j.1365-2966.2012.20969.x}}.

\bibitem[{Kalapotharakos} and {Contopoulos}(2009)]{2009A&A...496..495K}
{Kalapotharakos}, C.; {Contopoulos}, I.
\newblock {Three-dimensional numerical simulations of the pulsar magnetosphere:
  preliminary results}.
\newblock {\em \aap} {\bf 2009}, {\em 496},~495--502,
  \href{http://xxx.lanl.gov/abs/0811.2863}{{\normalfont
  [arXiv:astro-ph/0811.2863]}}.
\newblock {\url{https://doi.org/10.1051/0004-6361:200810281}}.

\bibitem[{Tchekhovskoy} \em{et~al.}(2013){Tchekhovskoy}, {Spitkovsky}, and
  {Li}]{2013MNRAS.435L...1T}
{Tchekhovskoy}, A.; {Spitkovsky}, A.; {Li}, J.G.
\newblock {Time-dependent 3D magnetohydrodynamic pulsar magnetospheres: oblique
  rotators.}
\newblock {\em \mnras} {\bf 2013}, {\em 435},~L1--L5,
  \href{http://xxx.lanl.gov/abs/1211.2803}{{\normalfont
  [arXiv:astro-ph.HE/1211.2803]}}.
\newblock {\url{https://doi.org/10.1093/mnrasl/slt076}}.

\bibitem[{P{\'e}tri}(2012)]{2012MNRAS.424..605P}
{P{\'e}tri}, J.
\newblock {The pulsar force-free magnetosphere linked to its striped wind:
  time-dependent pseudo-spectral simulations}.
\newblock {\em \mnras} {\bf 2012}, {\em 424},~605--619,
  \href{http://xxx.lanl.gov/abs/1205.0889}{{\normalfont
  [arXiv:astro-ph.HE/1205.0889]}}.
\newblock {\url{https://doi.org/10.1111/j.1365-2966.2012.21238.x}}.

\bibitem[{Philippov} \em{et~al.}(2014){Philippov}, {Tchekhovskoy}, and
  {Li}]{2014MNRAS.441.1879P}
{Philippov}, A.; {Tchekhovskoy}, A.; {Li}, J.G.
\newblock {Time evolution of pulsar obliquity angle from 3D simulations of
  magnetospheres}.
\newblock {\em \mnras} {\bf 2014}, {\em 441},~1879--1887,
  \href{http://xxx.lanl.gov/abs/1311.1513}{{\normalfont
  [arXiv:astro-ph.HE/1311.1513]}}.
\newblock {\url{https://doi.org/10.1093/mnras/stu591}}.

\bibitem[{Philippov} \em{et~al.}(2015){Philippov}, {Spitkovsky}, and
  {Cerutti}]{2015ApJ...801L..19P}
{Philippov}, A.A.; {Spitkovsky}, A.; {Cerutti}, B.
\newblock {Ab Initio Pulsar Magnetosphere: Three-dimensional Particle-in-cell
  Simulations of Oblique Pulsars}.
\newblock {\em \apjl} {\bf 2015}, {\em 801},~L19,
  \href{http://xxx.lanl.gov/abs/1412.0673}{{\normalfont
  [arXiv:astro-ph.HE/1412.0673]}}.
\newblock {\url{https://doi.org/10.1088/2041-8205/801/1/L19}}.

\bibitem[{P{\'e}tri}(2020)]{2020Univ....6...15P}
{P{\'e}tri}, J.
\newblock {Electrodynamics and Radiation from Rotating Neutron Star
  Magnetospheres}.
\newblock {\em Universe} {\bf 2020}, {\em 6},~15,
  \href{http://xxx.lanl.gov/abs/2001.03422}{{\normalfont
  [arXiv:astro-ph.HE/2001.03422]}}.
\newblock {\url{https://doi.org/10.3390/universe6010015}}.

\bibitem[{Beskin} \em{et~al.}(1984){Beskin}, {Gurevich}, and
  {Istomin}]{1984Ap&SS.102..301B}
{Beskin}, V.S.; {Gurevich}, A.V.; {Istomin}, I.N.
\newblock {Spin-Down of Pulsars by the Current - Comparison of Theory with
  Observations}.
\newblock {\em \apss} {\bf 1984}, {\em 102},~301--326.
\newblock {\url{https://doi.org/10.1007/BF00650179}}.

\bibitem[{Beskin} and {Nokhrina}(2007)]{2007Ap&SS.308..569B}
{Beskin}, V.S.; {Nokhrina}, E.E.
\newblock {On the role of the current loss in radio pulsar evolution}.
\newblock {\em \apss} {\bf 2007}, {\em 308},~569--573,
  \href{http://xxx.lanl.gov/abs/astro-ph/0608689}{{\normalfont
  [arXiv:astro-ph/astro-ph/0608689]}}.
\newblock {\url{https://doi.org/10.1007/s10509-007-9307-0}}.

\bibitem[{Beskin} \em{et~al.}(2013){Beskin}, {Istomin}, and
  {Philippov}]{2013PhyU...56..164B}
{Beskin}, V.S.; {Istomin}, Y.N.; {Philippov}, A.A.
\newblock {Radio pulsars: the search for truth}.
\newblock {\em Physics Uspekhi} {\bf 2013}, {\em 56},~164--179,
  \href{http://xxx.lanl.gov/abs/1305.1740}{{\normalfont
  [arXiv:astro-ph.HE/1305.1740]}}.
\newblock {\url{https://doi.org/10.3367/UFNe.0183.201302e.0179}}.

\bibitem[{Rankin}(1990)]{1990ApJ...352..247R}
{Rankin}, J.M.
\newblock {Toward an Empirical Theory of Pulsar Emission. IV. Geometry of the
  Core Emission Region}.
\newblock {\em \apj} {\bf 1990}, {\em 352},~247.
\newblock {\url{https://doi.org/10.1086/168530}}.

\bibitem[{Tauris} and {Manchester}(1998)]{1998MNRAS.298..625T}
{Tauris}, T.M.; {Manchester}, R.N.
\newblock {On the Evolution of Pulsar Beams}.
\newblock {\em \mnras} {\bf 1998}, {\em 298},~625--636.
\newblock {\url{https://doi.org/10.1046/j.1365-8711.1998.01369.x}}.

\bibitem[{Weltevrede} and {Johnston}(2008)]{2008MNRAS.387.1755W}
{Weltevrede}, P.; {Johnston}, S.
\newblock {The population of pulsars with interpulses and the implications for
  beam evolution}.
\newblock {\em \mnras} {\bf 2008}, {\em 387},~1755--1760,
  \href{http://xxx.lanl.gov/abs/0804.4318}{{\normalfont
  [arXiv:astro-ph/0804.4318]}}.
\newblock {\url{https://doi.org/10.1111/j.1365-2966.2008.13382.x}}.

\bibitem[{Young} \em{et~al.}(2010){Young}, {Chan}, {Burman}, and
  {Blair}]{2010MNRAS.402.1317Y}
{Young}, M.D.T.; {Chan}, L.S.; {Burman}, R.R.; {Blair}, D.G.
\newblock {Pulsar magnetic alignment and the pulsewidth-age relation}.
\newblock {\em \mnras} {\bf 2010}, {\em 402},~1317--1329,
  \href{http://xxx.lanl.gov/abs/0911.0502}{{\normalfont
  [arXiv:astro-ph.HE/0911.0502]}}.
\newblock {\url{https://doi.org/10.1111/j.1365-2966.2009.15972.x}}.

\bibitem[{Lyne} \em{et~al.}(2013){Lyne}, {Graham-Smith}, {Weltevrede},
  {Jordan}, {Stappers}, {Bassa}, and {Kramer}]{2013Sci...342..598L}
{Lyne}, A.; {Graham-Smith}, F.; {Weltevrede}, P.; {Jordan}, C.; {Stappers}, B.;
  {Bassa}, C.; {Kramer}, M.
\newblock {Evolution of the Magnetic Field Structure of the Crab Pulsar}.
\newblock {\em Science} {\bf 2013}, {\em 342},~598--601,
  \href{http://xxx.lanl.gov/abs/1311.0408}{{\normalfont
  [arXiv:astro-ph.HE/1311.0408]}}.
\newblock {\url{https://doi.org/10.1126/science.1243254}}.

\bibitem[{Arzamasskiy} \em{et~al.}(2015){Arzamasskiy}, {Philippov}, and
  {Tchekhovskoy}]{2015MNRAS.453.3540A}
{Arzamasskiy}, L.; {Philippov}, A.; {Tchekhovskoy}, A.
\newblock {Evolution of non-spherical pulsars with plasma-filled
  magnetospheres}.
\newblock {\em \mnras} {\bf 2015}, {\em 453},~3540--3553,
  \href{http://xxx.lanl.gov/abs/1504.06626}{{\normalfont
  [arXiv:astro-ph.HE/1504.06626]}}.
\newblock {\url{https://doi.org/10.1093/mnras/stv1818}}.

\bibitem[{Espinoza} \em{et~al.}(2011){Espinoza}, {Lyne}, {Stappers}, and
  {Kramer}]{2011MNRAS.414.1679E}
{Espinoza}, C.M.; {Lyne}, A.G.; {Stappers}, B.W.; {Kramer}, M.
\newblock {A study of 315 glitches in the rotation of 102 pulsars}.
\newblock {\em \mnras} {\bf 2011}, {\em 414},~1679--1704,
  \href{http://xxx.lanl.gov/abs/1102.1743}{{\normalfont
  [arXiv:astro-ph.HE/1102.1743]}}.
\newblock {\url{https://doi.org/10.1111/j.1365-2966.2011.18503.x}}.

\bibitem[{Haskell} and {Melatos}(2015)]{2015IJMPD..2430008H}
{Haskell}, B.; {Melatos}, A.
\newblock {Models of pulsar glitches}.
\newblock {\em International Journal of Modern Physics D} {\bf 2015}, {\em
  24},~1530008,  \href{http://xxx.lanl.gov/abs/1502.07062}{{\normalfont
  [arXiv:astro-ph.SR/1502.07062]}}.
\newblock {\url{https://doi.org/10.1142/S0218271815300086}}.

\bibitem[{Zhou} \em{et~al.}(2022){Zhou}, {G{\"u}gercino{\u{g}}lu}, {Yuan},
  {Ge}, and {Yu}]{2022Univ....8..641Z}
{Zhou}, S.; {G{\"u}gercino{\u{g}}lu}, E.; {Yuan}, J.; {Ge}, M.; {Yu}, C.
\newblock {Pulsar Glitches: A Review}.
\newblock {\em Universe} {\bf 2022}, {\em 8},~641,
  \href{http://xxx.lanl.gov/abs/2211.13885}{{\normalfont
  [arXiv:astro-ph.HE/2211.13885]}}.
\newblock {\url{https://doi.org/10.3390/universe8120641}}.

\bibitem[{Boynton} \em{et~al.}(1972){Boynton}, {Groth}, {Hutchinson}, {Nanos},
  {Partridge}, and {Wilkinson}]{1972ApJ...175..217B}
{Boynton}, P.E.; {Groth}, E.J.; {Hutchinson}, D.P.; {Nanos}, G.~P., J.;
  {Partridge}, R.B.; {Wilkinson}, D.T.
\newblock {Optical Timing of the Crab Pulsar, NP 0532}.
\newblock {\em \apj} {\bf 1972}, {\em 175},~217.
\newblock {\url{https://doi.org/10.1086/151550}}.

\bibitem[{Cordes} and {Helfand}(1980)]{1980ApJ...239..640C}
{Cordes}, J.M.; {Helfand}, D.J.
\newblock {Pulsar timing. III - Timing noise of 50 pulsars}.
\newblock {\em \apj} {\bf 1980}, {\em 239},~640--650.
\newblock {\url{https://doi.org/10.1086/158150}}.

\bibitem[{D'Alessandro} \em{et~al.}(1995){D'Alessandro}, {McCulloch},
  {Hamilton}, and {Deshpande}]{1995MNRAS.277.1033D}
{D'Alessandro}, F.; {McCulloch}, P.M.; {Hamilton}, P.A.; {Deshpande}, A.A.
\newblock {The timing noise of 45 southern pulsars}.
\newblock {\em \mnras} {\bf 1995}, {\em 277},~1033--1046.
\newblock {\url{https://doi.org/10.1093/mnras/277.3.1033}}.

\bibitem[{Urama} \em{et~al.}(2006){Urama}, {Link}, and
  {Weisberg}]{2006MNRAS.370L..76U}
{Urama}, J.O.; {Link}, B.; {Weisberg}, J.M.
\newblock {A strong correlation in radio pulsars with implications for torque
  variations}.
\newblock {\em \mnras} {\bf 2006}, {\em 370},~L76--L79,
  \href{http://xxx.lanl.gov/abs/astro-ph/0605380}{{\normalfont
  [astro-ph/0605380]}}.
\newblock {\url{https://doi.org/10.1111/j.1745-3933.2006.00192.x}}.

\bibitem[{Hobbs} \em{et~al.}(2010){Hobbs}, {Lyne}, and
  {Kramer}]{2010MNRAS.402.1027H}
{Hobbs}, G.; {Lyne}, A.G.; {Kramer}, M.
\newblock {An analysis of the timing irregularities for 366 pulsars}.
\newblock {\em \mnras} {\bf 2010}, {\em 402},~1027--1048,
  \href{http://xxx.lanl.gov/abs/0912.4537}{{\normalfont [0912.4537]}}.
\newblock {\url{https://doi.org/10.1111/j.1365-2966.2009.15938.x}}.

\bibitem[{Nice} \em{et~al.}(2013){Nice}, {Altiere}, {Bogdanov}, {Cordes},
  {Farrington}, {Hessels}, {Kaspi}, {Lyne}, {Popa}, {Ransom}, {Sanpa-arsa},
  {Stappers}, {Wang}, {Allen}, {Bhat}, {Brazier}, {Camilo}, {Champion},
  {Chatterjee}, {Crawford}, {Deneva}, {Desvignes}, {Freire}, {Jenet},
  {Knispel}, {Lazarus}, {Lee}, {van Leeuwen}, {Lorimer}, {Lynch}, {McLaughlin},
  {Scholz}, {Siemens}, {Stairs}, {Stovall}, {Venkataraman}, and
  {Zhu}]{2013ApJ...772...50N}
{Nice}, D.J.; {Altiere}, E.; {Bogdanov}, S.; {Cordes}, J.M.; {Farrington}, D.;
  {Hessels}, J.W.T.; {Kaspi}, V.M.; {Lyne}, A.G.; {Popa}, L.; {Ransom}, S.M.;
  et~al.
\newblock {Timing and Interstellar Scattering of 35 Distant Pulsars Discovered
  in the PALFA Survey}.
\newblock {\em \apj} {\bf 2013}, {\em 772},~50,
  \href{http://xxx.lanl.gov/abs/1304.7370}{{\normalfont
  [arXiv:astro-ph.SR/1304.7370]}}.
\newblock {\url{https://doi.org/10.1088/0004-637X/772/1/50}}.

\bibitem[{Arzoumanian} \em{et~al.}(1994){Arzoumanian}, {Nice}, {Taylor}, and
  {Thorsett}]{1994ApJ...422..671A}
{Arzoumanian}, Z.; {Nice}, D.J.; {Taylor}, J.H.; {Thorsett}, S.E.
\newblock {Timing Behavior of 96 Radio Pulsars}.
\newblock {\em \apj} {\bf 1994}, {\em 422},~671.
\newblock {\url{https://doi.org/10.1086/173760}}.

\bibitem[{Ek{\c s}i} \em{et~al.}(2016){Ek{\c s}i}, {Anda{\c c}}, {{\c
  C}{\i}k{\i}nto{\u g}lu}, {G{\"u}gercino{\u g}lu}, {Vahdat Motlagh}, and
  {K{\i}z{\i}ltan}]{2016ApJ...823...34E}
{Ek{\c s}i}, K.Y.; {Anda{\c c}}, I.C.; {{\c C}{\i}k{\i}nto{\u g}lu}, S.;
  {G{\"u}gercino{\u g}lu}, E.; {Vahdat Motlagh}, A.; {K{\i}z{\i}ltan}, B.
\newblock {The Inclination Angle and Evolution of the Braking Index of Pulsars
  with Plasma-filled Magnetosphere: Application to the High Braking Index of
  PSR J1640-4631}.
\newblock {\em \apj} {\bf 2016}, {\em 823},~34,
  \href{http://xxx.lanl.gov/abs/1603.01487}{{\normalfont
  [arXiv:astro-ph.HE/1603.01487]}}.
\newblock {\url{https://doi.org/10.3847/0004-637X/823/1/34}}.

\bibitem[{Hobbs} \em{et~al.}(2004){Hobbs}, {Lyne}, {Kramer}, {Martin}, and
  {Jordan}]{2004MNRAS.353.1311H}
{Hobbs}, G.; {Lyne}, A.G.; {Kramer}, M.; {Martin}, C.E.; {Jordan}, C.
\newblock {Long-term timing observations of 374 pulsars}.
\newblock {\em \mnras} {\bf 2004}, {\em 353},~1311--1344.
\newblock {\url{https://doi.org/10.1111/j.1365-2966.2004.08157.x}}.

\bibitem[{Zhang} and {Xie}(2012)]{2012ApJ...761..102Z}
{Zhang}, S.N.; {Xie}, Y.
\newblock {Why Do the Braking Indices of Pulsars Span a Range of More Than 100
  Millions?}
\newblock {\em \apj} {\bf 2012}, {\em 761},~102,
  \href{http://xxx.lanl.gov/abs/1209.2478}{{\normalfont
  [arXiv:astro-ph.HE/1209.2478]}}.
\newblock {\url{https://doi.org/10.1088/0004-637X/761/2/102}}.

\bibitem[{Biryukov} \em{et~al.}(2007){Biryukov}, {Beskin}, {Karpov}, and
  {Chmyreva}]{2007AdSpR..40.1498B}
{Biryukov}, A.; {Beskin}, G.; {Karpov}, S.; {Chmyreva}, L.
\newblock {Evidence of long-term cyclic evolution of radio pulsar periods}.
\newblock {\em Advances in Space Research} {\bf 2007}, {\em 40},~1498--1504,
  \href{http://xxx.lanl.gov/abs/0709.2549}{{\normalfont [0709.2549]}}.
\newblock {\url{https://doi.org/10.1016/j.asr.2007.06.051}}.

\bibitem[{Archibald} \em{et~al.}(2016){Archibald}, {Gotthelf}, {Ferdman},
  {Kaspi}, {Guillot}, {Harrison}, {Keane}, {Pivovaroff}, {Stern}, {Tendulkar},
  and {Tomsick}]{2016ApJ...819L..16A}
{Archibald}, R.F.; {Gotthelf}, E.V.; {Ferdman}, R.D.; {Kaspi}, V.M.; {Guillot},
  S.; {Harrison}, F.A.; {Keane}, E.F.; {Pivovaroff}, M.J.; {Stern}, D.;
  {Tendulkar}, S.P.;  et~al.
\newblock {A High Braking Index for a Pulsar}.
\newblock {\em \apjl} {\bf 2016}, {\em 819},~L16,
  \href{http://xxx.lanl.gov/abs/1603.00305}{{\normalfont
  [arXiv:astro-ph.HE/1603.00305]}}.
\newblock {\url{https://doi.org/10.3847/2041-8205/819/1/L16}}.

\bibitem[{Marshall} \em{et~al.}(2016){Marshall}, {Guillemot}, {Harding},
  {Martin}, and {Smith}]{2016ApJ...827L..39M}
{Marshall}, F.E.; {Guillemot}, L.; {Harding}, A.K.; {Martin}, P.; {Smith}, D.A.
\newblock {A New, Low Braking Index for the LMC Pulsar B0540-69}.
\newblock {\em \apjl} {\bf 2016}, {\em 827},~L39,
  \href{http://xxx.lanl.gov/abs/1608.01901}{{\normalfont
  [arXiv:astro-ph.HE/1608.01901]}}.
\newblock {\url{https://doi.org/10.3847/2041-8205/827/2/L39}}.

\bibitem[{Ou} \em{et~al.}(2016){Ou}, {Tong}, {Kou}, and
  {Ding}]{2016MNRAS.457.3922O}
{Ou}, Z.W.; {Tong}, H.; {Kou}, F.F.; {Ding}, G.Q.
\newblock {Fluctuating neutron star magnetosphere: braking indices of eight
  pulsars, frequency second derivatives of 222 pulsars and 15 magnetars}.
\newblock {\em \mnras} {\bf 2016}, {\em 457},~3922--3933,
  \href{http://xxx.lanl.gov/abs/1512.01679}{{\normalfont
  [arXiv:astro-ph.HE/1512.01679]}}.
\newblock {\url{https://doi.org/10.1093/mnras/stw227}}.

\bibitem[{Igoshev} and {Popov}(2020)]{2020MNRAS.499.2826I}
{Igoshev}, A.P.; {Popov}, S.B.
\newblock {Braking indices of young radio pulsars: theoretical perspective}.
\newblock {\em \mnras} {\bf 2020}, {\em 499},~2826--2835,
  \href{http://xxx.lanl.gov/abs/2008.11737}{{\normalfont
  [arXiv:astro-ph.HE/2008.11737]}}.
\newblock {\url{https://doi.org/10.1093/mnras/staa3070}}.

\bibitem[{Janssen} and {Stappers}(2006)]{2006A&A...457..611J}
{Janssen}, G.H.; {Stappers}, B.W.
\newblock {30 glitches in slow pulsars}.
\newblock {\em \aap} {\bf 2006}, {\em 457},~611--618,
  \href{http://xxx.lanl.gov/abs/astro-ph/0607260}{{\normalfont
  [arXiv:astro-ph/astro-ph/0607260]}}.
\newblock {\url{https://doi.org/10.1051/0004-6361:20065267}}.

\bibitem[{Tsang} and {Gourgouliatos}(2013)]{2013ApJ...773L..17T}
{Tsang}, D.; {Gourgouliatos}, K.N.
\newblock {Timing Noise in Pulsars and Magnetars and the Magnetospheric Moment
  of Inertia}.
\newblock {\em \apjl} {\bf 2013}, {\em 773},~L17,
  \href{http://xxx.lanl.gov/abs/1302.4448}{{\normalfont
  [arXiv:astro-ph.HE/1302.4448]}}.
\newblock {\url{https://doi.org/10.1088/2041-8205/773/1/L17}}.

\bibitem[{Melatos} and {Link}(2014)]{2014MNRAS.437...21M}
{Melatos}, A.; {Link}, B.
\newblock {Pulsar timing noise from superfluid turbulence}.
\newblock {\em \mnras} {\bf 2014}, {\em 437},~21--31,
  \href{http://xxx.lanl.gov/abs/1310.3108}{{\normalfont
  [arXiv:astro-ph.HE/1310.3108]}}.
\newblock {\url{https://doi.org/10.1093/mnras/stt1828}}.

\bibitem[{Hamil} \em{et~al.}(2015){Hamil}, {Stone}, {Urbanec}, and
  {Urbancov{\'a}}]{2015PhRvD..91f3007H}
{Hamil}, O.; {Stone}, J.R.; {Urbanec}, M.; {Urbancov{\'a}}, G.
\newblock {Braking index of isolated pulsars}.
\newblock {\em \prd} {\bf 2015}, {\em 91},~063007,
  \href{http://xxx.lanl.gov/abs/1608.01383}{{\normalfont
  [arXiv:astro-ph.HE/1608.01383]}}.
\newblock {\url{https://doi.org/10.1103/PhysRevD.91.063007}}.

\bibitem[{Hamil} \em{et~al.}(2016){Hamil}, {Stone}, and
  {Stone}]{2016PhRvD..94f3012H}
{Hamil}, O.; {Stone}, N.J.; {Stone}, J.R.
\newblock {Braking index of isolated pulsars. II. A novel two-dipole model of
  pulsar magnetism}.
\newblock {\em \prd} {\bf 2016}, {\em 94},~063012.
\newblock {\url{https://doi.org/10.1103/PhysRevD.94.063012}}.

\bibitem[{Cheng}(1987)]{1987ApJ...321..799C}
{Cheng}, K.S.
\newblock {Outer magnetospheric fluctuations and pulsar timing noise}.
\newblock {\em \apj} {\bf 1987}, {\em 321},~799--804.
\newblock {\url{https://doi.org/10.1086/165672}}.

\bibitem[{Kramer} \em{et~al.}(2006){Kramer}, {Lyne}, {O'Brien}, {Jordan}, and
  {Lorimer}]{2006Sci...312..549K}
{Kramer}, M.; {Lyne}, A.G.; {O'Brien}, J.T.; {Jordan}, C.A.; {Lorimer}, D.R.
\newblock {A Periodically Active Pulsar Giving Insight into Magnetospheric
  Physics}.
\newblock {\em Science} {\bf 2006}, {\em 312},~549--551,
  \href{http://xxx.lanl.gov/abs/astro-ph/0604605}{{\normalfont
  [astro-ph/0604605]}}.
\newblock {\url{https://doi.org/10.1126/science.1124060}}.

\bibitem[{Contopoulos}(2007)]{2007A&A...475..639C}
{Contopoulos}, I.
\newblock {A note on the cyclic evolution of the pulsar magnetosphere}.
\newblock {\em \aap} {\bf 2007}, {\em 475},~639--642,
  \href{http://xxx.lanl.gov/abs/0709.3957}{{\normalfont
  [arXiv:astro-ph/0709.3957]}}.
\newblock {\url{https://doi.org/10.1051/0004-6361:20078108}}.

\bibitem[{Lyne} \em{et~al.}(2010){Lyne}, {Hobbs}, {Kramer}, {Stairs}, and
  {Stappers}]{2010Sci...329..408L}
{Lyne}, A.; {Hobbs}, G.; {Kramer}, M.; {Stairs}, I.; {Stappers}, B.
\newblock {Switched Magnetospheric Regulation of Pulsar Spin-Down}.
\newblock {\em Science} {\bf 2010}, {\em 329},~408,
  \href{http://xxx.lanl.gov/abs/1006.5184}{{\normalfont
  [arXiv:astro-ph.GA/1006.5184]}}.
\newblock {\url{https://doi.org/10.1126/science.1186683}}.

\bibitem[{Barsukov} and {Tsygan}(2010)]{2010MNRAS.409.1077B}
{Barsukov}, D.P.; {Tsygan}, A.I.
\newblock {The influence of nondipolar magnetic field and neutron star
  precession on braking indices of radiopulsars}.
\newblock {\em \mnras} {\bf 2010}, {\em 409},~1077--1087,
  \href{http://xxx.lanl.gov/abs/1003.0808}{{\normalfont
  [arXiv:astro-ph.SR/1003.0808]}}.
\newblock {\url{https://doi.org/10.1111/j.1365-2966.2010.17365.x}}.

\bibitem[{Ridley} and {Lorimer}(2010)]{2010MNRAS.404.1081R}
{Ridley}, J.P.; {Lorimer}, D.R.
\newblock {Isolated pulsar spin evolution on the diagram}.
\newblock {\em \mnras} {\bf 2010}, {\em 404},~1081--1088,
  \href{http://xxx.lanl.gov/abs/1001.2483}{{\normalfont [1001.2483]}}.
\newblock {\url{https://doi.org/10.1111/j.1365-2966.2010.16342.x}}.

\bibitem[{Keane} and {Kramer}(2008)]{2008MNRAS.391.2009K}
{Keane}, E.F.; {Kramer}, M.
\newblock {On the birthrates of Galactic neutron stars}.
\newblock {\em \mnras} {\bf 2008}, {\em 391},~2009--2016,
  \href{http://xxx.lanl.gov/abs/0810.1512}{{\normalfont [0810.1512]}}.
\newblock {\url{https://doi.org/10.1111/j.1365-2966.2008.14045.x}}.

\bibitem[{Vrane{\v s}evi{\'c}} and {Melrose}(2011)]{2011MNRAS.410.2363V}
{Vrane{\v s}evi{\'c}}, N.; {Melrose}, D.B.
\newblock {Pulsar current revisited}.
\newblock {\em \mnras} {\bf 2011}, {\em 410},~2363--2369,
  \href{http://xxx.lanl.gov/abs/1009.0311}{{\normalfont
  [arXiv:astro-ph.SR/1009.0311]}}.
\newblock {\url{https://doi.org/10.1111/j.1365-2966.2010.17612.x}}.

\bibitem[{Pires} \em{et~al.}(2009){Pires}, {Motch}, {Turolla}, {Treves}, and
  {Popov}]{2009A&A...498..233P}
{Pires}, A.M.; {Motch}, C.; {Turolla}, R.; {Treves}, A.; {Popov}, S.B.
\newblock {The isolated neutron star candidate 2XMM J104608.7-594306}.
\newblock {\em \aap} {\bf 2009}, {\em 498},~233--240,
  \href{http://xxx.lanl.gov/abs/0812.4151}{{\normalfont
  [arXiv:astro-ph/0812.4151]}}.
\newblock {\url{https://doi.org/10.1051/0004-6361/200810966}}.

\bibitem[{Rigoselli} \em{et~al.}(2019){Rigoselli}, {Mereghetti}, {Suleimanov},
  {Potekhin}, {Turolla}, {Taverna}, and {Pintore}]{2019A&A...627A..69R}
{Rigoselli}, M.; {Mereghetti}, S.; {Suleimanov}, V.; {Potekhin}, A.Y.;
  {Turolla}, R.; {Taverna}, R.; {Pintore}, F.
\newblock {XMM-Newton observations of PSR J0726-2612, a radio-loud XDINS}.
\newblock {\em \aap} {\bf 2019}, {\em 627},~A69,
  \href{http://xxx.lanl.gov/abs/1906.01372}{{\normalfont
  [arXiv:astro-ph.HE/1906.01372]}}.
\newblock {\url{https://doi.org/10.1051/0004-6361/201935485}}.

\bibitem[{Turolla}(2009)]{2009ASSL..357..141T}
{Turolla}, R.
\newblock {Isolated Neutron Stars: The Challenge of Simplicity}.
\newblock In Proceedings of the Astrophysics and Space Science Library;
  {Becker}, W., Ed.,  2009, Vol. 357, {\em Astrophysics and Space Science
  Library}, p. 141.
\newblock {\url{https://doi.org/10.1007/978-3-540-76965-1_7}}.

\bibitem[{Gotthelf} and {Vasisht}(2000)]{2000ASPC..202..699G}
{Gotthelf}, E.V.; {Vasisht}, G.
\newblock {A New View on Young Pulsars in Supernova Remnants: Slow Radio-quiet
  \& X-ray Bright}.
\newblock In Proceedings of the IAU Colloq. 177: Pulsar Astronomy - 2000 and
  Beyond; {Kramer}, M.; {Wex}, N.; {Wielebinski}, R., Eds.,  2000, Vol. 202,
  {\em Astronomical Society of the Pacific Conference Series}, p. 699,
  \href{http://xxx.lanl.gov/abs/astro-ph/9911344}{{\normalfont
  [arXiv:astro-ph/astro-ph/9911344]}}.
\newblock {\url{https://doi.org/10.48550/arXiv.astro-ph/9911344}}.

\bibitem[{Beniamini} \em{et~al.}(2019){Beniamini}, {Hotokezaka}, {van der
  Horst}, and {Kouveliotou}]{2019MNRAS.487.1426B}
{Beniamini}, P.; {Hotokezaka}, K.; {van der Horst}, A.; {Kouveliotou}, C.
\newblock {Formation rates and evolution histories of magnetars}.
\newblock {\em \mnras} {\bf 2019}, {\em 487},~1426--1438,
  \href{http://xxx.lanl.gov/abs/1903.06718}{{\normalfont
  [arXiv:astro-ph.HE/1903.06718]}}.
\newblock {\url{https://doi.org/10.1093/mnras/stz1391}}.

\bibitem[{Tammann} \em{et~al.}(1994){Tammann}, {Loeffler}, and
  {Schroeder}]{1994ApJS...92..487T}
{Tammann}, G.A.; {Loeffler}, W.; {Schroeder}, A.
\newblock {The Galactic supernova rate}.
\newblock {\em \apjs} {\bf 1994}, {\em 92},~487--493.
\newblock {\url{https://doi.org/10.1086/192002}}.

\bibitem[{Diehl} \em{et~al.}(2006){Diehl}, {Halloin}, {Kretschmer}, {Lichti},
  {Sch{\"o}nfelder}, {Strong}, {von Kienlin}, {Wang}, {Jean}, {Kn{\"o}dlseder},
  {Roques}, {Weidenspointner}, {Schanne}, {Hartmann}, {Winkler}, and
  {Wunderer}]{2006Natur.439...45D}
{Diehl}, R.; {Halloin}, H.; {Kretschmer}, K.; {Lichti}, G.G.;
  {Sch{\"o}nfelder}, V.; {Strong}, A.W.; {von Kienlin}, A.; {Wang}, W.; {Jean},
  P.; {Kn{\"o}dlseder}, J.;  et~al.
\newblock {Radioactive $^{26}$Al from massive stars in the Galaxy}.
\newblock {\em \nat} {\bf 2006}, {\em 439},~45--47,
  \href{http://xxx.lanl.gov/abs/astro-ph/0601015}{{\normalfont
  [astro-ph/0601015]}}.
\newblock {\url{https://doi.org/10.1038/nature04364}}.

\bibitem[{Schmidt} \em{et~al.}(2014){Schmidt}, {Hohle}, and
  {Neuh{\"a}user}]{2014AN....335..935S}
{Schmidt}, J.G.; {Hohle}, M.M.; {Neuh{\"a}user}, R.
\newblock {Determination of a temporally and spatially resolved supernova rate
  from OB stars within 5 kpc}.
\newblock {\em Astronomische Nachrichten} {\bf 2014}, {\em 335},~935--948,
  \href{http://xxx.lanl.gov/abs/1409.3357}{{\normalfont
  [arXiv:astro-ph.SR/1409.3357]}}.
\newblock {\url{https://doi.org/10.1002/asna.201312070}}.

\bibitem[{Vranesevic} \em{et~al.}(2004){Vranesevic}, {Manchester}, {Lorimer},
  {Hobbs}, {Lyne}, {Kramer}, {Camilo}, {Stairs}, {Kaspi}, {D'Amico},
  {Possenti}, {Crawford}, {Faulkner}, and {McLaughlin}]{2004ApJ...617L.139V}
{Vranesevic}, N.; {Manchester}, R.N.; {Lorimer}, D.R.; {Hobbs}, G.B.; {Lyne},
  A.G.; {Kramer}, M.; {Camilo}, F.; {Stairs}, I.H.; {Kaspi}, V.M.; {D'Amico},
  N.;  et~al.
\newblock {Pulsar Birthrates from the Parkes Multibeam Survey}.
\newblock {\em \apjl} {\bf 2004}, {\em 617},~L139--L142,
  \href{http://xxx.lanl.gov/abs/astro-ph/0310201}{{\normalfont
  [arXiv:astro-ph/astro-ph/0310201]}}.
\newblock {\url{https://doi.org/10.1086/427208}}.

\bibitem[{Lorimer} \em{et~al.}(2006){Lorimer}, {Faulkner}, {Lyne},
  {Manchester}, {Kramer}, {McLaughlin}, {Hobbs}, {Possenti}, {Stairs},
  {Camilo}, {Burgay}, {D'Amico}, {Corongiu}, and
  {Crawford}]{2006MNRAS.372..777L}
{Lorimer}, D.R.; {Faulkner}, A.J.; {Lyne}, A.G.; {Manchester}, R.N.; {Kramer},
  M.; {McLaughlin}, M.A.; {Hobbs}, G.; {Possenti}, A.; {Stairs}, I.H.;
  {Camilo}, F.;  et~al.
\newblock {The Parkes Multibeam Pulsar Survey - VI. Discovery and timing of 142
  pulsars and a Galactic population analysis}.
\newblock {\em \mnras} {\bf 2006}, {\em 372},~777--800,
  \href{http://xxx.lanl.gov/abs/astro-ph/0607640}{{\normalfont
  [arXiv:astro-ph/astro-ph/0607640]}}.
\newblock {\url{https://doi.org/10.1111/j.1365-2966.2006.10887.x}}.

\bibitem[{Popov} \em{et~al.}(2010){Popov}, {Pons}, {Miralles}, {Boldin}, and
  {Posselt}]{2010MNRAS.401.2675P}
{Popov}, S.B.; {Pons}, J.A.; {Miralles}, J.A.; {Boldin}, P.A.; {Posselt}, B.
\newblock {Population synthesis studies of isolated neutron stars with magnetic
  field decay}.
\newblock {\em MNRAS} {\bf 2010}, {\em 401},~2675--2686,
  \href{http://xxx.lanl.gov/abs/0910.2190}{{\normalfont
  [arXiv:astro-ph.HE/0910.2190]}}.
\newblock {\url{https://doi.org/10.1111/j.1365-2966.2009.15850.x}}.

\bibitem[{Kaspi}(2010)]{2010PNAS..107.7147K}
{Kaspi}, V.M.
\newblock {Grand unification of neutron stars}.
\newblock {\em Proceedings of the National Academy of Science} {\bf 2010}, {\em
  107},~7147--7152,  \href{http://xxx.lanl.gov/abs/1005.0876}{{\normalfont
  [arXiv:astro-ph.HE/1005.0876]}}.
\newblock {\url{https://doi.org/10.1073/pnas.1000812107}}.

\bibitem[{Igoshev} and {Popov}(2014)]{2014MNRAS.444.1066I}
{Igoshev}, A.P.; {Popov}, S.B.
\newblock {Modified pulsar current analysis: probing magnetic field evolution}.
\newblock {\em MNRAS} {\bf 2014}, {\em 444},~1066--1076,
  \href{http://xxx.lanl.gov/abs/1407.6269}{{\normalfont
  [arXiv:astro-ph.HE/1407.6269]}}.
\newblock {\url{https://doi.org/10.1093/mnras/stu1496}}.

\bibitem[{Gull{\'o}n} \em{et~al.}(2015){Gull{\'o}n}, {Pons}, {Miralles},
  {Vigan{\`o}}, {Rea}, and {Perna}]{2015MNRAS.454..615G}
{Gull{\'o}n}, M.; {Pons}, J.A.; {Miralles}, J.A.; {Vigan{\`o}}, D.; {Rea}, N.;
  {Perna}, R.
\newblock {Population synthesis of isolated neutron stars with
  magneto-rotational evolution - II. From radio-pulsars to magnetars}.
\newblock {\em \mnras} {\bf 2015}, {\em 454},~615--625,
  \href{http://xxx.lanl.gov/abs/1507.05452}{{\normalfont
  [arXiv:astro-ph.HE/1507.05452]}}.
\newblock {\url{https://doi.org/10.1093/mnras/stv1644}}.

\bibitem[{Shvartsman}(1970)]{1970R&QE...13.1428S}
{Shvartsman}, V.F.
\newblock {Two generations of pulsars}.
\newblock {\em Radiophysics and Quantum Electronics} {\bf 1970}, {\em
  13},~1428--1440.
\newblock {\url{https://doi.org/10.1007/BF01032996}}.

\bibitem[{Illarionov} and {Sunyaev}(1975)]{1975A&A....39..185I}
{Illarionov}, A.F.; {Sunyaev}, R.A.
\newblock {Why the Number of Galactic X-ray Stars Is so Small?}
\newblock {\em \aap} {\bf 1975}, {\em 39},~185.

\bibitem[{Pringle} and {Rees}(1972)]{1972A&A....21....1P}
{Pringle}, J.E.; {Rees}, M.J.
\newblock {Accretion Disc Models for Compact X-Ray Sources}.
\newblock {\em \aap} {\bf 1972}, {\em 21},~1.

\bibitem[{Davidson} and {Ostriker}(1973)]{1973ApJ...179..585D}
{Davidson}, K.; {Ostriker}, J.P.
\newblock {Neutron-Star Accretion in a Stellar Wind: Model for a Pulsed X-Ray
  Source}.
\newblock {\em \apj} {\bf 1973}, {\em 179},~585--598.
\newblock {\url{https://doi.org/10.1086/151897}}.

\bibitem[{Davies} \em{et~al.}(1979){Davies}, {Fabian}, and
  {Pringle}]{1979MNRAS.186..779D}
{Davies}, R.E.; {Fabian}, A.C.; {Pringle}, J.E.
\newblock {Spindown of neutron stars in close binary systems.}
\newblock {\em \mnras} {\bf 1979}, {\em 186},~779--782.
\newblock {\url{https://doi.org/10.1093/mnras/186.4.779}}.

\bibitem[{Davies} and {Pringle}(1981)]{1981MNRAS.196..209D}
{Davies}, R.E.; {Pringle}, J.E.
\newblock {Spindown of neutron stars in close binary systems - II.}
\newblock {\em \mnras} {\bf 1981}, {\em 196},~209--224.
\newblock {\url{https://doi.org/10.1093/mnras/196.2.209}}.

\bibitem[{Shakura}(1975)]{1975SvAL....1..223S}
{Shakura}, N.I.
\newblock {The long-period X-ray pulsar 3U 0900-40 as a neutron star with an
  abnormally strong magnetic field.}
\newblock {\em Soviet Astronomy Letters} {\bf 1975}, {\em 1},~223--225.

\bibitem[{Fabian}(1975)]{1975MNRAS.173..161F}
{Fabian}, A.C.
\newblock {Slowly rotating neutron stars and transient X-ray sources.}
\newblock {\em \mnras} {\bf 1975}, {\em 173},~161--165.
\newblock {\url{https://doi.org/10.1093/mnras/173.1.161}}.

\bibitem[{Ikhsanov}(2001)]{2001A&A...368L...5I}
{Ikhsanov}, N.R.
\newblock {On the duration of the subsonic propeller state of neutron stars in
  wind-fed mass-exchange close binary systems}.
\newblock {\em \aap} {\bf 2001}, {\em 368},~L5--L7,
  \href{http://xxx.lanl.gov/abs/astro-ph/0111505}{{\normalfont
  [arXiv:astro-ph/astro-ph/0111505]}}.
\newblock {\url{https://doi.org/10.1051/0004-6361:20010140}}.

\bibitem[{Shakura} \em{et~al.}(2012){Shakura}, {Postnov}, {Kochetkova}, and
  {Hjalmarsdotter}]{2012MNRAS.420..216S}
{Shakura}, N.; {Postnov}, K.; {Kochetkova}, A.; {Hjalmarsdotter}, L.
\newblock {Theory of quasi-spherical accretion in X-ray pulsars}.
\newblock {\em \mnras} {\bf 2012}, {\em 420},~216--236,
  \href{http://xxx.lanl.gov/abs/1110.3701}{{\normalfont
  [arXiv:astro-ph.HE/1110.3701]}}.
\newblock {\url{https://doi.org/10.1111/j.1365-2966.2011.20026.x}}.

\bibitem[{Wang} and {Robertson}(1985)]{1985A&A...151..361W}
{Wang}, Y.M.; {Robertson}, J.A.
\newblock {'Propeller' action by rotating neutron stars}.
\newblock {\em \aap} {\bf 1985}, {\em 151},~361--371.

\bibitem[{Romanova} \em{et~al.}(2003){Romanova}, {Toropina}, {Toropin}, and
  {Lovelace}]{2003ApJ...588..400R}
{Romanova}, M.M.; {Toropina}, O.D.; {Toropin}, Y.M.; {Lovelace}, R.V.E.
\newblock {Magnetohydrodynamic Simulations of Accretion onto a Star in the
  ``Propeller'' Regime}.
\newblock {\em \apj} {\bf 2003}, {\em 588},~400--407,
  \href{http://xxx.lanl.gov/abs/astro-ph/0209548}{{\normalfont
  [arXiv:astro-ph/astro-ph/0209548]}}.
\newblock {\url{https://doi.org/10.1086/373990}}.

\bibitem[{Toropin} \em{et~al.}(1999){Toropin}, {Toropina}, {Savelyev},
  {Romanova}, {Chechetkin}, and {Lovelace}]{1999ApJ...517..906T}
{Toropin}, Y.M.; {Toropina}, O.D.; {Savelyev}, V.V.; {Romanova}, M.M.;
  {Chechetkin}, V.M.; {Lovelace}, R.V.E.
\newblock {Spherical Bondi Accretion onto a Magnetic Dipole}.
\newblock {\em \apj} {\bf 1999}, {\em 517},~906--918,
  \href{http://xxx.lanl.gov/abs/astro-ph/9811272}{{\normalfont
  [arXiv:astro-ph/astro-ph/9811272]}}.
\newblock {\url{https://doi.org/10.1086/307229}}.

\bibitem[{Francischelli} and {Wijers}(2002)]{2002astro.ph..5212F}
{Francischelli}, G.J.; {Wijers}, R.A.M.J.
\newblock {On Fossil Disk Models of Anomalous X-Ray Pulsars}.
\newblock {\em arXiv e-prints} {\bf 2002}, pp. astro--ph/0205212,
  \href{http://xxx.lanl.gov/abs/astro-ph/0205212}{{\normalfont
  [arXiv:astro-ph/astro-ph/0205212]}}.
\newblock {\url{https://doi.org/10.48550/arXiv.astro-ph/0205212}}.

\bibitem[{Stella} \em{et~al.}(1986){Stella}, {White}, and
  {Rosner}]{1986ApJ...308..669S}
{Stella}, L.; {White}, N.E.; {Rosner}, R.
\newblock {Intermittent Stellar Wind Acceleration and the Long-Term Activity of
  Population I Binary Systems Containing an X-Ray Pulsar}.
\newblock {\em \apj} {\bf 1986}, {\em 308},~669.
\newblock {\url{https://doi.org/10.1086/164538}}.

\bibitem[{Campana} \em{et~al.}(2001){Campana}, {Gastaldello}, {Stella},
  {Israel}, {Colpi}, {Pizzolato}, {Orlandini}, and {Dal
  Fiume}]{2001ApJ...561..924C}
{Campana}, S.; {Gastaldello}, F.; {Stella}, L.; {Israel}, G.L.; {Colpi}, M.;
  {Pizzolato}, F.; {Orlandini}, M.; {Dal Fiume}, D.
\newblock {The Transient X-Ray Pulsar 4U 0115+63 from Quiescence to Outburst
  through the Centrifugal Transition}.
\newblock {\em \apj} {\bf 2001}, {\em 561},~924--929,
  \href{http://xxx.lanl.gov/abs/astro-ph/0107236}{{\normalfont
  [arXiv:astro-ph/astro-ph/0107236]}}.
\newblock {\url{https://doi.org/10.1086/323317}}.

\bibitem[{Campana} \em{et~al.}(1998){Campana}, {Stella}, {Mereghetti}, {Colpi},
  {Tavani}, {Ricci}, {Dal Fiume}, and {Belloni}]{1998ApJ...499L..65C}
{Campana}, S.; {Stella}, L.; {Mereghetti}, S.; {Colpi}, M.; {Tavani}, M.;
  {Ricci}, D.; {Dal Fiume}, D.; {Belloni}, T.
\newblock {Aquila X-1 from Outburst to Quiescence: The Onset of the Propeller
  Effect and Signs of a Turned-on Rotation-powered Pulsar}.
\newblock {\em \apjl} {\bf 1998}, {\em 499},~L65--L68,
  \href{http://xxx.lanl.gov/abs/astro-ph/9803303}{{\normalfont
  [arXiv:astro-ph/astro-ph/9803303]}}.
\newblock {\url{https://doi.org/10.1086/311357}}.

\bibitem[{Campana} \em{et~al.}(2008){Campana}, {Stella}, and
  {Kennea}]{2008ApJ...684L..99C}
{Campana}, S.; {Stella}, L.; {Kennea}, J.A.
\newblock {Swift Observations of SAX J1808.4-3658: Monitoring the Return to
  Quiescence}.
\newblock {\em \apjl} {\bf 2008}, {\em 684},~L99,
  \href{http://xxx.lanl.gov/abs/0807.4444}{{\normalfont
  [arXiv:astro-ph/0807.4444]}}.
\newblock {\url{https://doi.org/10.1086/592002}}.

\bibitem[{Tsygankov} \em{et~al.}(2016){Tsygankov}, {Mushtukov}, {Suleimanov},
  and {Poutanen}]{2016MNRAS.457.1101T}
{Tsygankov}, S.S.; {Mushtukov}, A.A.; {Suleimanov}, V.F.; {Poutanen}, J.
\newblock {Propeller effect in action in the ultraluminous accreting magnetar
  M82 X-2}.
\newblock {\em \mnras} {\bf 2016}, {\em 457},~1101--1106,
  \href{http://xxx.lanl.gov/abs/1507.08288}{{\normalfont
  [arXiv:astro-ph.HE/1507.08288]}}.
\newblock {\url{https://doi.org/10.1093/mnras/stw046}}.

\bibitem[{Cui}(1997)]{1997ApJ...482L.163C}
{Cui}, W.
\newblock {Evidence for ``Propeller'' Effects in X-Ray Pulsars GX 1+4 and GRO
  J1744-28}.
\newblock {\em \apjl} {\bf 1997}, {\em 482},~L163--L166,
  \href{http://xxx.lanl.gov/abs/astro-ph/9704084}{{\normalfont
  [arXiv:astro-ph/astro-ph/9704084]}}.
\newblock {\url{https://doi.org/10.1086/310712}}.

\bibitem[{Tsygankov} \em{et~al.}(2017){Tsygankov}, {Mushtukov}, {Suleimanov},
  {Doroshenko}, {Abolmasov}, {Lutovinov}, and {Poutanen}]{2017A&A...608A..17T}
{Tsygankov}, S.S.; {Mushtukov}, A.A.; {Suleimanov}, V.F.; {Doroshenko}, V.;
  {Abolmasov}, P.K.; {Lutovinov}, A.A.; {Poutanen}, J.
\newblock {Stable accretion from a cold disc in highly magnetized neutron
  stars}.
\newblock {\em \aap} {\bf 2017}, {\em 608},~A17,
  \href{http://xxx.lanl.gov/abs/1703.04528}{{\normalfont
  [arXiv:astro-ph.HE/1703.04528]}}.
\newblock {\url{https://doi.org/10.1051/0004-6361/201630248}}.

\bibitem[{Campana} \em{et~al.}(2018){Campana}, {Stella}, {Mereghetti}, and {de
  Martino}]{2018A&A...610A..46C}
{Campana}, S.; {Stella}, L.; {Mereghetti}, S.; {de Martino}, D.
\newblock {A universal relation for the propeller mechanisms in magnetic
  rotating stars at different scales}.
\newblock {\em \aap} {\bf 2018}, {\em 610},~A46,
  \href{http://xxx.lanl.gov/abs/1711.08233}{{\normalfont
  [arXiv:astro-ph.HE/1711.08233]}}.
\newblock {\url{https://doi.org/10.1051/0004-6361/201730769}}.

\bibitem[{Lutovinov} \em{et~al.}(2019){Lutovinov}, {Tsygankov}, {Karasev},
  {Molkov}, and {Doroshenko}]{2019MNRAS.485..770L}
{Lutovinov}, A.A.; {Tsygankov}, S.S.; {Karasev}, D.I.; {Molkov}, S.V.;
  {Doroshenko}, V.
\newblock {GRO J1750-27: A neutron star far behind the Galactic Center
  switching into the propeller regime}.
\newblock {\em \mnras} {\bf 2019}, {\em 485},~770--776,
  \href{http://xxx.lanl.gov/abs/1902.05153}{{\normalfont
  [arXiv:astro-ph.HE/1902.05153]}}.
\newblock {\url{https://doi.org/10.1093/mnras/stz437}}.

\bibitem[{De} \em{et~al.}(2023){De}, {Daly}, and {Soria}]{2023arXiv230907833D}
{De}, K.; {Daly}, F.A.; {Soria}, R.
\newblock {Infrared spectroscopy of SWIFT J0850.8-4219: Identification of the
  second red supergiant X-ray binary in the Milky Way}.
\newblock {\em arXiv e-prints} {\bf 2023}, p. arXiv:2309.07833,
  \href{http://xxx.lanl.gov/abs/2309.07833}{{\normalfont
  [arXiv:astro-ph.SR/2309.07833]}}.

\bibitem[{Barcons} \em{et~al.}(2012){Barcons}, {Barret}, {Decourchelle}, {den
  Herder}, {Dotani}, {Fabian}, {Fraga-Encinas}, {Kunieda}, {Lumb}, {Matt},
  {Nandra}, {Piro}, {Rando}, {Sciortino}, {Smith}, {Str{\"u}der}, {Watson},
  {White}, and {Willingale}]{2012arXiv1207.2745B}
{Barcons}, X.; {Barret}, D.; {Decourchelle}, A.; {den Herder}, J.W.; {Dotani},
  T.; {Fabian}, A.C.; {Fraga-Encinas}, R.; {Kunieda}, H.; {Lumb}, D.; {Matt},
  G.;  et~al.
\newblock {Athena (Advanced Telescope for High ENergy Astrophysics) Assessment
  Study Report for ESA Cosmic Vision 2015-2025}.
\newblock {\em arXiv e-prints} {\bf 2012}, p. arXiv:1207.2745,
  \href{http://xxx.lanl.gov/abs/1207.2745}{{\normalfont
  [arXiv:astro-ph.HE/1207.2745]}}.
\newblock {\url{https://doi.org/10.48550/arXiv.1207.2745}}.

\bibitem[{Bondi}(1952)]{1952MNRAS.112..195B}
{Bondi}, H.
\newblock {On spherically symmetrical accretion}.
\newblock {\em \mnras} {\bf 1952}, {\em 112},~195.
\newblock {\url{https://doi.org/10.1093/mnras/112.2.195}}.

\bibitem[{Edgar}(2004)]{2004NewAR..48..843E}
{Edgar}, R.
\newblock {A review of Bondi-Hoyle-Lyttleton accretion}.
\newblock {\em \nar} {\bf 2004}, {\em 48},~843--859,
  \href{http://xxx.lanl.gov/abs/astro-ph/0406166}{{\normalfont
  [arXiv:astro-ph/astro-ph/0406166]}}.
\newblock {\url{https://doi.org/10.1016/j.newar.2004.06.001}}.

\bibitem[{Shvartsman}(1971)]{1971SvA....14..662S}
{Shvartsman}, V.G.
\newblock {Ionization Zones around Neutron Stars: H{\ensuremath{\alpha}}
  Emission, Heating of the Interstellar Medium, and the Influence on
  Accretion.}
\newblock {\em \sovast} {\bf 1971}, {\em 14},~662.

\bibitem[{Shvartsman}(1970)]{1970Ap......6...56S}
{Shvartsman}, V.F.
\newblock {Gamma and radio emission of neutrons stars in the state of
  accretion}.
\newblock {\em Astrophysics} {\bf 1970}, {\em 6},~56--62.
\newblock {\url{https://doi.org/10.1007/BF01002575}}.

\bibitem[{Ostriker} \em{et~al.}(1970){Ostriker}, {Rees}, and
  {Silk}]{1970ApL.....6..179O}
{Ostriker}, J.P.; {Rees}, M.J.; {Silk}, J.
\newblock {Some Observable Consequences of Accretion by Defunct Pulsars}.
\newblock {\em \aplett} {\bf 1970}, {\em 6},~179.

\bibitem[{Blaes} and {Rajagopal}(1991)]{1991ApJ...381..210B}
{Blaes}, O.; {Rajagopal}, M.
\newblock {The Statistics of Slow Interstellar Accretion onto Neutron Stars}.
\newblock {\em \apj} {\bf 1991}, {\em 381},~210.
\newblock {\url{https://doi.org/10.1086/170642}}.

\bibitem[{Blaes} and {Madau}(1993)]{1993ApJ...403..690B}
{Blaes}, O.; {Madau}, P.
\newblock {Can We Observe Accreting, Isolated Neutron Stars?}
\newblock {\em \apj} {\bf 1993}, {\em 403},~690.
\newblock {\url{https://doi.org/10.1086/172240}}.

\bibitem[{Madau} and {Blaes}(1994)]{1994ApJ...423..748M}
{Madau}, P.; {Blaes}, O.
\newblock {Constraints on Accreting, Isolated Neutron Stars from the ROSAT and
  EUVE Surveys}.
\newblock {\em \apj} {\bf 1994}, {\em 423},~748.
\newblock {\url{https://doi.org/10.1086/173854}}.

\bibitem[{Blaes} \em{et~al.}(1995){Blaes}, {Warren}, and
  {Madau}]{1995ApJ...454..370B}
{Blaes}, O.; {Warren}, O.; {Madau}, P.
\newblock {Accreting, Isolated Neutron Stars. III. Preheating of Infalling Gas
  and Cometary H II Regions}.
\newblock {\em \apj} {\bf 1995}, {\em 454},~370.
\newblock {\url{https://doi.org/10.1086/176488}}.

\bibitem[{Popov} \em{et~al.}(2000){Popov}, {Colpi}, {Treves}, {Turolla},
  {Lipunov}, and {Prokhorov}]{2000ApJ...530..896P}
{Popov}, S.B.; {Colpi}, M.; {Treves}, A.; {Turolla}, R.; {Lipunov}, V.M.;
  {Prokhorov}, M.E.
\newblock {The Neutron Star Census}.
\newblock {\em \apj} {\bf 2000}, {\em 530},~896--903,
  \href{http://xxx.lanl.gov/abs/astro-ph/9910114}{{\normalfont
  [arXiv:astro-ph/astro-ph/9910114]}}.
\newblock {\url{https://doi.org/10.1086/308408}}.

\bibitem[{Boldin} and {Popov}(2010)]{2010MNRAS.407.1090B}
{Boldin}, P.A.; {Popov}, S.B.
\newblock {The evolution of isolated neutron stars until accretion: the role of
  the initial magnetic field}.
\newblock {\em \mnras} {\bf 2010}, {\em 407},~1090--1097,
  \href{http://xxx.lanl.gov/abs/1004.4805}{{\normalfont
  [arXiv:astro-ph.HE/1004.4805]}}.
\newblock {\url{https://doi.org/10.1111/j.1365-2966.2010.16910.x}}.

\bibitem[{Blondin} and {Raymer}(2012)]{2012ApJ...752...30B}
{Blondin}, J.M.; {Raymer}, E.
\newblock {Hoyle-Lyttleton Accretion in Three Dimensions}.
\newblock {\em \apj} {\bf 2012}, {\em 752},~30,
  \href{http://xxx.lanl.gov/abs/1204.0717}{{\normalfont
  [arXiv:astro-ph.SR/1204.0717]}}.
\newblock {\url{https://doi.org/10.1088/0004-637X/752/1/30}}.

\bibitem[{Toropina} \em{et~al.}(2001){Toropina}, {Romanova}, {Toropin}, and
  {Lovelace}]{2001ApJ...561..964T}
{Toropina}, O.D.; {Romanova}, M.M.; {Toropin}, Y.M.; {Lovelace}, R.V.E.
\newblock {Propagation of Magnetized Neutron Stars through the Interstellar
  Medium}.
\newblock {\em \apj} {\bf 2001}, {\em 561},~964--979,
  \href{http://xxx.lanl.gov/abs/astro-ph/0105422}{{\normalfont
  [arXiv:astro-ph/astro-ph/0105422]}}.
\newblock {\url{https://doi.org/10.1086/323233}}.

\bibitem[{Toropina} \em{et~al.}(2012){Toropina}, {Romanova}, and
  {Lovelace}]{2012MNRAS.420..810T}
{Toropina}, O.D.; {Romanova}, M.M.; {Lovelace}, R.V.E.
\newblock {Bondi-Hoyle accretion on to a magnetized neutron star}.
\newblock {\em \mnras} {\bf 2012}, {\em 420},~810--816,
  \href{http://xxx.lanl.gov/abs/1111.2460}{{\normalfont
  [arXiv:astro-ph.HE/1111.2460]}}.
\newblock {\url{https://doi.org/10.1111/j.1365-2966.2011.20093.x}}.

\bibitem[{Lipunov} and {Popov}(1995)]{1995ARep...39..632L}
{Lipunov}, V.M.; {Popov}, S.B.
\newblock {Evolution of the periods of isolated neutron stars: A spindown
  theorem}.
\newblock {\em Astronomy Reports} {\bf 1995}, {\em 39},~632--637.

\bibitem[{Popov} \em{et~al.}(2001){Popov}, {Prokhorov}, {Khoperskov}, and
  {Lipunov}]{2001astro.ph.10022P}
{Popov}, S.B.; {Prokhorov}, M.E.; {Khoperskov}, A.V.; {Lipunov}, V.M.
\newblock {Stochastic spin evolution of neutron stars}.
\newblock {\em arXiv e-prints} {\bf 2001}, pp. astro--ph/0110022,
  \href{http://xxx.lanl.gov/abs/astro-ph/0110022}{{\normalfont
  [arXiv:astro-ph/astro-ph/0110022]}}.
\newblock {\url{https://doi.org/10.48550/arXiv.astro-ph/0110022}}.

\bibitem[{Prokhorov} \em{et~al.}(2002){Prokhorov}, {Popov}, and
  {Khoperskov}]{2002A&A...381.1000P}
{Prokhorov}, M.E.; {Popov}, S.B.; {Khoperskov}, A.V.
\newblock {The period distribution of old accreting isolated neutron stars}.
\newblock {\em \aap} {\bf 2002}, {\em 381},~1000--1006,
  \href{http://xxx.lanl.gov/abs/astro-ph/0108503}{{\normalfont
  [arXiv:astro-ph/astro-ph/0108503]}}.
\newblock {\url{https://doi.org/10.1051/0004-6361:20011529}}.

\bibitem[{Popov} \em{et~al.}(2015){Popov}, {Postnov}, and
  {Shakura}]{2015MNRAS.447.2817P}
{Popov}, S.B.; {Postnov}, K.A.; {Shakura}, N.I.
\newblock {Settling accretion on to isolated neutron stars from interstellar
  medium}.
\newblock {\em \mnras} {\bf 2015}, {\em 447},~2817--2820,
  \href{http://xxx.lanl.gov/abs/1412.4066}{{\normalfont
  [arXiv:astro-ph.HE/1412.4066]}}.
\newblock {\url{https://doi.org/10.1093/mnras/stu2643}}.

\bibitem[{Treves} and {Colpi}(1991)]{1991A&A...241..107T}
{Treves}, A.; {Colpi}, M.
\newblock {The observability of old isolated neutron stars.}
\newblock {\em \aap} {\bf 1991}, {\em 241},~107.

\bibitem[{Manning} \em{et~al.}(1996){Manning}, {Jeffries}, and
  {Willmore}]{1996MNRAS.278..577M}
{Manning}, R.A.; {Jeffries}, R.D.; {Willmore}, A.P.
\newblock {Are there any isolated old neutron stars in the ROSAT Wide Field
  Camera survey?}
\newblock {\em \mnras} {\bf 1996}, {\em 278},~577--585.
\newblock {\url{https://doi.org/10.1093/mnras/278.2.577}}.

\bibitem[{Treves} \em{et~al.}(2000){Treves}, {Turolla}, {Zane}, and
  {Colpi}]{2000PASP..112..297T}
{Treves}, A.; {Turolla}, R.; {Zane}, S.; {Colpi}, M.
\newblock {Isolated Neutron Stars: Accretors and Coolers}.
\newblock {\em \pasp} {\bf 2000}, {\em 112},~297--314,
  \href{http://xxx.lanl.gov/abs/astro-ph/9911430}{{\normalfont
  [arXiv:astro-ph/astro-ph/9911430]}}.
\newblock {\url{https://doi.org/10.1086/316529}}.

\bibitem[{Popov} \em{et~al.}(2000){Popov}, {Colpi}, {Prokhorov}, {Treves}, and
  {Turolla}]{2000ApJ...544L..53P}
{Popov}, S.B.; {Colpi}, M.; {Prokhorov}, M.E.; {Treves}, A.; {Turolla}, R.
\newblock {The LOG N-LOG S Distributions of Accreting and Cooling Isolated
  Neutron Stars}.
\newblock {\em \apjl} {\bf 2000}, {\em 544},~L53--L56,
  \href{http://xxx.lanl.gov/abs/astro-ph/0009225}{{\normalfont
  [arXiv:astro-ph/astro-ph/0009225]}}.
\newblock {\url{https://doi.org/10.1086/317295}}.

\bibitem[{Turner} \em{et~al.}(2010){Turner}, {Rutledge}, {Letcavage},
  {Shevchuk}, and {Fox}]{2010ApJ...714.1424T}
{Turner}, M.L.; {Rutledge}, R.E.; {Letcavage}, R.; {Shevchuk}, A.S.H.; {Fox},
  D.B.
\newblock {A Limit on the Number of Isolated Neutron Stars Detected in the
  ROSAT All-Sky-Survey Bright Source Catalog}.
\newblock {\em \apj} {\bf 2010}, {\em 714},~1424--1440,
  \href{http://xxx.lanl.gov/abs/1003.3955}{{\normalfont
  [arXiv:astro-ph.HE/1003.3955]}}.
\newblock {\url{https://doi.org/10.1088/0004-637X/714/2/1424}}.

\bibitem[{Sana} \em{et~al.}(2012){Sana}, {de Mink}, {de Koter}, {Langer},
  {Evans}, {Gieles}, {Gosset}, {Izzard}, {Le Bouquin}, and
  {Schneider}]{2012Sci...337..444S}
{Sana}, H.; {de Mink}, S.E.; {de Koter}, A.; {Langer}, N.; {Evans}, C.J.;
  {Gieles}, M.; {Gosset}, E.; {Izzard}, R.G.; {Le Bouquin}, J.B.; {Schneider},
  F.R.N.
\newblock {Binary Interaction Dominates the Evolution of Massive Stars}.
\newblock {\em Science} {\bf 2012}, {\em 337},~444,
  \href{http://xxx.lanl.gov/abs/1207.6397}{{\normalfont
  [arXiv:astro-ph.SR/1207.6397]}}.
\newblock {\url{https://doi.org/10.1126/science.1223344}}.

\bibitem[{Manchester}(2017)]{2017JApA...38...42M}
{Manchester}, R.N.
\newblock {Millisecond Pulsars, their Evolution and Applications}.
\newblock {\em Journal of Astrophysics and Astronomy} {\bf 2017}, {\em 38},~42,
   \href{http://xxx.lanl.gov/abs/1709.09434}{{\normalfont
  [arXiv:astro-ph.HE/1709.09434]}}.
\newblock {\url{https://doi.org/10.1007/s12036-017-9469-2}}.

\bibitem[{Archibald} \em{et~al.}(2009){Archibald}, {Stairs}, {Ransom}, {Kaspi},
  {Kondratiev}, {Lorimer}, {McLaughlin}, {Boyles}, {Hessels}, {Lynch}, {van
  Leeuwen}, {Roberts}, {Jenet}, {Champion}, {Rosen}, {Barlow}, {Dunlap}, and
  {Remillard}]{2009Sci...324.1411A}
{Archibald}, A.M.; {Stairs}, I.H.; {Ransom}, S.M.; {Kaspi}, V.M.; {Kondratiev},
  V.I.; {Lorimer}, D.R.; {McLaughlin}, M.A.; {Boyles}, J.; {Hessels}, J.W.T.;
  {Lynch}, R.;  et~al.
\newblock {A Radio Pulsar/X-ray Binary Link}.
\newblock {\em Science} {\bf 2009}, {\em 324},~1411,
  \href{http://xxx.lanl.gov/abs/0905.3397}{{\normalfont
  [arXiv:astro-ph.HE/0905.3397]}}.
\newblock {\url{https://doi.org/10.1126/science.1172740}}.

\bibitem[{Bahramian} and {Degenaar}(2023)]{2023hxga.book..120B}
{Bahramian}, A.; {Degenaar}, N.
\newblock {Low-Mass X-ray Binaries}. In {\em Handbook of X-ray and Gamma-ray
  Astrophysics. Edited by Cosimo Bambi and Andrea Santangelo}; Springer,  2023;
  p. 120.
\newblock {\url{https://doi.org/10.1007/978-981-16-4544-0_94-1}}.

\bibitem[{Strohmayer} \em{et~al.}(1996){Strohmayer}, {Zhang}, {Swank}, {Smale},
  {Titarchuk}, {Day}, and {Lee}]{1996ApJ...469L...9S}
{Strohmayer}, T.E.; {Zhang}, W.; {Swank}, J.H.; {Smale}, A.; {Titarchuk}, L.;
  {Day}, C.; {Lee}, U.
\newblock {Millisecond X-Ray Variability from an Accreting Neutron Star
  System}.
\newblock {\em \apjl} {\bf 1996}, {\em 469},~L9.
\newblock {\url{https://doi.org/10.1086/310261}}.

\bibitem[{Hasinger} and {van der Klis}(1989)]{1989A&A...225...79H}
{Hasinger}, G.; {van der Klis}, M.
\newblock {Two patterns of correlated X-ray timing and spectral behaviour in
  low-mass X-ray binaries.}
\newblock {\em \aap} {\bf 1989}, {\em 225},~79--96.

\bibitem[{M{\'e}ndez} and {Belloni}(2007)]{2007MNRAS.381..790M}
{M{\'e}ndez}, M.; {Belloni}, T.
\newblock {Is there a link between the neutron-star spin and the frequency of
  the kilohertz quasi-periodic oscillations?}
\newblock {\em \mnras} {\bf 2007}, {\em 381},~790--796,
  \href{http://xxx.lanl.gov/abs/0708.0015}{{\normalfont
  [arXiv:astro-ph/0708.0015]}}.
\newblock {\url{https://doi.org/10.1111/j.1365-2966.2007.12306.x}}.

\bibitem[{Malacaria} \em{et~al.}(2020){Malacaria}, {Jenke}, {Roberts},
  {Wilson-Hodge}, {Cleveland}, {Mailyan}, and {GBM Accreting Pulsars Program
  Team}]{2020ApJ...896...90M}
{Malacaria}, C.; {Jenke}, P.; {Roberts}, O.J.; {Wilson-Hodge}, C.A.;
  {Cleveland}, W.H.; {Mailyan}, B.; {GBM Accreting Pulsars Program Team}.
\newblock {The Ups and Downs of Accreting X-Ray Pulsars: Decade-long
  Observations with the Fermi Gamma-Ray Burst Monitor}.
\newblock {\em \apj} {\bf 2020}, {\em 896},~90,
  \href{http://xxx.lanl.gov/abs/2004.00051}{{\normalfont
  [arXiv:astro-ph.HE/2004.00051]}}.
\newblock {\url{https://doi.org/10.3847/1538-4357/ab855c}}.

\bibitem[{Raguzova} and {Popov}(2005)]{2005A&AT...24..151R}
{Raguzova}, N.V.; {Popov}, S.B.
\newblock {Be X-ray binaries and candidates}.
\newblock {\em Astronomical and Astrophysical Transactions} {\bf 2005}, {\em
  24},~151--185,  \href{http://xxx.lanl.gov/abs/astro-ph/0505275}{{\normalfont
  [arXiv:astro-ph/astro-ph/0505275]}}.
\newblock {\url{https://doi.org/10.1080/10556790500497311}}.

\bibitem[{Reig}(2011)]{2011Ap&SS.332....1R}
{Reig}, P.
\newblock {Be/X-ray binaries}.
\newblock {\em \apss} {\bf 2011}, {\em 332},~1--29,
  \href{http://xxx.lanl.gov/abs/1101.5036}{{\normalfont
  [arXiv:astro-ph.HE/1101.5036]}}.
\newblock {\url{https://doi.org/10.1007/s10509-010-0575-8}}.

\bibitem[{Corbet}(1984)]{1984A&A...141...91C}
{Corbet}, R.H.D.
\newblock {Be/neutron star binaries : a relationship between orbital period and
  neutron star spin period.}
\newblock {\em \aap} {\bf 1984}, {\em 141},~91--93.

\bibitem[{Kretschmar} \em{et~al.}(2021){Kretschmar}, {El Mellah},
  {Mart{\'\i}nez-N{\'u}{\~n}ez}, {F{\"u}rst}, {Grinberg}, {Sander}, {van den
  Eijnden}, {Degenaar}, {Ma{\'\i}z Apell{\'a}niz}, {Jim{\'e}nez Esteban},
  {Ramos-Lerate}, and {Utrilla}]{2021A&A...652A..95K}
{Kretschmar}, P.; {El Mellah}, I.; {Mart{\'\i}nez-N{\'u}{\~n}ez}, S.;
  {F{\"u}rst}, F.; {Grinberg}, V.; {Sander}, A.A.C.; {van den Eijnden}, J.;
  {Degenaar}, N.; {Ma{\'\i}z Apell{\'a}niz}, J.; {Jim{\'e}nez Esteban}, F.;
  et~al.
\newblock {Revisiting the archetypical wind accretor Vela X-1 in depth. Case
  study of a well-known X-ray binary and the limits of our knowledge}.
\newblock {\em \aap} {\bf 2021}, {\em 652},~A95,
  \href{http://xxx.lanl.gov/abs/2104.13148}{{\normalfont
  [arXiv:astro-ph.HE/2104.13148]}}.
\newblock {\url{https://doi.org/10.1051/0004-6361/202040272}}.

\bibitem[{Lai}(2014)]{2014EPJWC..6401001L}
{Lai}, D.
\newblock {Theory of Disk Accretion onto Magnetic Stars}.
\newblock In Proceedings of the European Physical Journal Web of Conferences,
  2014, Vol.~64, {\em European Physical Journal Web of Conferences}, p. 01001,
  \href{http://xxx.lanl.gov/abs/1402.1903}{{\normalfont
  [arXiv:astro-ph.SR/1402.1903]}}.
\newblock {\url{https://doi.org/10.1051/epjconf/20136401001}}.

\bibitem[{Ghosh} and {Lamb}(1979)]{1979ApJ...234..296G}
{Ghosh}, P.; {Lamb}, F.K.
\newblock {Accretion by rotating magnetic neutron stars. III. Accretion torques
  and period changes in pulsating X-ray sources.}
\newblock {\em \apj} {\bf 1979}, {\em 234},~296--316.
\newblock {\url{https://doi.org/10.1086/157498}}.

\bibitem[{Wang}(1995)]{1995ApJ...449L.153W}
{Wang}, Y.M.
\newblock {On the Torque Exerted by a Magnetically Threaded Accretion Disk}.
\newblock {\em \apjl} {\bf 1995}, {\em 449},~L153.
\newblock {\url{https://doi.org/10.1086/309649}}.

\bibitem[{Shakura} and {Sunyaev}(1973)]{1973A&A....24..337S}
{Shakura}, N.I.; {Sunyaev}, R.A.
\newblock {Black holes in binary systems. Observational appearance.}
\newblock {\em \aap} {\bf 1973}, {\em 24},~337--355.

\bibitem[{Shi} \em{et~al.}(2015){Shi}, {Zhang}, and {Li}]{2015ApJ...813...91S}
{Shi}, C.S.; {Zhang}, S.N.; {Li}, X.D.
\newblock {Super Strong Magnetic Fields of Neutron Stars in Be X-Ray Binaries
  Estimated with New Torque and Magnetosphere Models}.
\newblock {\em \apj} {\bf 2015}, {\em 813},~91,
  \href{http://xxx.lanl.gov/abs/1509.06126}{{\normalfont
  [arXiv:astro-ph.HE/1509.06126]}}.
\newblock {\url{https://doi.org/10.1088/0004-637X/813/2/91}}.

\bibitem[{Doroshenko} \em{et~al.}(2022){Doroshenko}, {Poutanen}, {Tsygankov},
  {Suleimanov}, {Bachetti}, {Caiazzo}, {Costa}, {Di Marco}, {Heyl}, {La
  Monaca}, {Muleri}, {Mushtukov}, {Pavlov}, {Ramsey}, {Rankin}, {Santangelo},
  {Soffitta}, {Staubert}, {Weisskopf}, {Zane}, {Agudo}, {Antonelli}, {Baldini},
  {Baumgartner}, {Bellazzini}, {Bianchi}, {Bongiorno}, {Bonino}, {Brez},
  {Bucciantini}, {Capitanio}, {Castellano}, {Cavazzuti}, {Ciprini}, {De Rosa},
  {Del Monte}, {Di Gesu}, {Di Lalla}, {Donnarumma}, {Dov{\v{c}}iak}, {Ehlert},
  {Enoto}, {Evangelista}, {Fabiani}, {Ferrazzoli}, {Garcia}, {Gunji},
  {Hayashida}, {Iwakiri}, {Jorstad}, {Karas}, {Kitaguchi}, {Kolodziejczak},
  {Krawczynski}, {Latronico}, {Liodakis}, {Maldera}, {Manfreda}, {Marin},
  {Marinucci}, {Marscher}, {Marshall}, {Matt}, {Mitsuishi}, {Mizuno}, {Ng},
  {O'Dell}, {Omodei}, {Oppedisano}, {Papitto}, {Peirson}, {Perri},
  {Pesce-Rollins}, {Pilia}, {Possenti}, {Puccetti}, {Ratheesh}, {Romani},
  {Sgr{\`o}}, {Slane}, {Spandre}, {Sunyaev}, {Tamagawa}, {Tavecchio},
  {Taverna}, {Tawara}, {Tennant}, {Thomas}, {Tombesi}, {Trois}, {Turolla},
  {Vink}, {Wu}, and {Xie}]{2022NatAs...6.1433D}
{Doroshenko}, V.; {Poutanen}, J.; {Tsygankov}, S.S.; {Suleimanov}, V.F.;
  {Bachetti}, M.; {Caiazzo}, I.; {Costa}, E.; {Di Marco}, A.; {Heyl}, J.; {La
  Monaca}, F.;  et~al.
\newblock {Determination of X-ray pulsar geometry with IXPE polarimetry}.
\newblock {\em Nature Astronomy} {\bf 2022}, {\em 6},~1433--1443,
  \href{http://xxx.lanl.gov/abs/2206.07138}{{\normalfont
  [arXiv:astro-ph.HE/2206.07138]}}.
\newblock {\url{https://doi.org/10.1038/s41550-022-01799-5}}.

\bibitem[{Johnson} \em{et~al.}(2014){Johnson}, {Venter}, {Harding},
  {Guillemot}, {Smith}, {Kramer}, {{\c{C}}elik}, {den Hartog}, {Ferrara},
  {Hou}, {Lande}, and {Ray}]{2014ApJS..213....6J}
{Johnson}, T.J.; {Venter}, C.; {Harding}, A.K.; {Guillemot}, L.; {Smith}, D.A.;
  {Kramer}, M.; {{\c{C}}elik}, {\"O}.; {den Hartog}, P.R.; {Ferrara}, E.C.;
  {Hou}, X.;  et~al.
\newblock {Constraints on the Emission Geometries and Spin Evolution of
  Gamma-Ray Millisecond Pulsars}.
\newblock {\em \apjs} {\bf 2014}, {\em 213},~6,
  \href{http://xxx.lanl.gov/abs/1404.2264}{{\normalfont
  [arXiv:astro-ph.HE/1404.2264]}}.
\newblock {\url{https://doi.org/10.1088/0067-0049/213/1/6}}.

\bibitem[{Bozzo} \em{et~al.}(2018){Bozzo}, {Ascenzi}, {Ducci}, {Papitto},
  {Burderi}, and {Stella}]{2018A&A...617A.126B}
{Bozzo}, E.; {Ascenzi}, S.; {Ducci}, L.; {Papitto}, A.; {Burderi}, L.;
  {Stella}, L.
\newblock {Magnetospheric radius of an inclined rotator in the magnetically
  threaded disk model}.
\newblock {\em \aap} {\bf 2018}, {\em 617},~A126,
  \href{http://xxx.lanl.gov/abs/1806.11516}{{\normalfont
  [arXiv:astro-ph.HE/1806.11516]}}.
\newblock {\url{https://doi.org/10.1051/0004-6361/201732004}}.

\bibitem[{Wang}(1997)]{1997ApJ...475L.135W}
{Wang}, Y.M.
\newblock {Torque Exerted on an Oblique Rotator by a Magnetically Threaded
  Accretion Disk}.
\newblock {\em \apjl} {\bf 1997}, {\em 475},~L135--L137.
\newblock {\url{https://doi.org/10.1086/310481}}.

\bibitem[{Aly} and {Kuijpers}(1990)]{1990A&A...227..473A}
{Aly}, J.J.; {Kuijpers}, J.
\newblock {Flaring interactions between accretion disk and neutron star
  magnetosphere.}
\newblock {\em \aap} {\bf 1990}, {\em 227},~473--482.

\bibitem[{Matt} and {Pudritz}(2005)]{2005ApJ...632L.135M}
{Matt}, S.; {Pudritz}, R.E.
\newblock {Accretion-powered Stellar Winds as a Solution to the Stellar Angular
  Momentum Problem}.
\newblock {\em \apjl} {\bf 2005}, {\em 632},~L135--L138,
  \href{http://xxx.lanl.gov/abs/astro-ph/0510060}{{\normalfont
  [arXiv:astro-ph/astro-ph/0510060]}}.
\newblock {\url{https://doi.org/10.1086/498066}}.

\bibitem[{Romanova} \em{et~al.}(2013){Romanova}, {Ustyugova}, {Koldoba}, and
  {Lovelace}]{2013MNRAS.430..699R}
{Romanova}, M.M.; {Ustyugova}, G.V.; {Koldoba}, A.V.; {Lovelace}, R.V.E.
\newblock {Warps, bending and density waves excited by rotating magnetized
  stars: results of global 3D MHD simulations}.
\newblock {\em \mnras} {\bf 2013}, {\em 430},~699--724,
  \href{http://xxx.lanl.gov/abs/1209.1161}{{\normalfont
  [arXiv:astro-ph.SR/1209.1161]}}.
\newblock {\url{https://doi.org/10.1093/mnras/sts670}}.

\bibitem[{Zanni} and {Ferreira}(2013)]{2013A&A...550A..99Z}
{Zanni}, C.; {Ferreira}, J.
\newblock {MHD simulations of accretion onto a dipolar magnetosphere. II.
  Magnetospheric ejections and stellar spin-down}.
\newblock {\em \aap} {\bf 2013}, {\em 550},~A99,
  \href{http://xxx.lanl.gov/abs/1211.4844}{{\normalfont
  [arXiv:astro-ph.SR/1211.4844]}}.
\newblock {\url{https://doi.org/10.1051/0004-6361/201220168}}.

\bibitem[{Ireland} \em{et~al.}(2022){Ireland}, {Matt}, and
  {Zanni}]{2022ApJ...929...65I}
{Ireland}, L.G.; {Matt}, S.P.; {Zanni}, C.
\newblock {Magnetic Braking of Accreting T Tauri Stars II: Torque Formulation
  Spanning Spin-up and Spin-down Regimes}.
\newblock {\em \apj} {\bf 2022}, {\em 929},~65,
  \href{http://xxx.lanl.gov/abs/2203.00326}{{\normalfont
  [arXiv:astro-ph.SR/2203.00326]}}.
\newblock {\url{https://doi.org/10.3847/1538-4357/ac59b2}}.

\bibitem[{Parfrey} and {Tchekhovskoy}(2023)]{2023arXiv231104291P}
{Parfrey}, K.; {Tchekhovskoy}, A.
\newblock {Accreting Neutron Stars in 3D GRMHD Simulations: Jets, Magnetic
  Polarity, and the Interchange Slingshot}.
\newblock {\em arXiv e-prints} {\bf 2023}, p. arXiv:2311.04291,
  \href{http://xxx.lanl.gov/abs/2311.04291}{{\normalfont
  [arXiv:astro-ph.HE/2311.04291]}}.
\newblock {\url{https://doi.org/10.48550/arXiv.2311.04291}}.

\bibitem[{Das} and {Porth}(2023)]{2023arXiv231105301D}
{Das}, P.; {Porth}, O.
\newblock {Three-dimensional GRMHD simulations of neutron star jets}.
\newblock {\em arXiv e-prints} {\bf 2023}, p. arXiv:2311.05301,
  \href{http://xxx.lanl.gov/abs/2311.05301}{{\normalfont
  [arXiv:astro-ph.HE/2311.05301]}}.
\newblock {\url{https://doi.org/10.48550/arXiv.2311.05301}}.

\bibitem[{Chashkina} and {Popov}(2012)]{2012NewA...17..594C}
{Chashkina}, A.; {Popov}, S.B.
\newblock {Magnetic field estimates for accreting neutron stars in massive
  binary systems and models of magnetic field decay}.
\newblock {\em NewA} {\bf 2012}, {\em 17},~594--602,
  \href{http://xxx.lanl.gov/abs/1112.1123}{{\normalfont
  [arXiv:astro-ph.HE/1112.1123]}}.
\newblock {\url{https://doi.org/10.1016/j.newast.2012.01.004}}.

\bibitem[{Vasilopoulos} \em{et~al.}(2019){Vasilopoulos}, {Petropoulou},
  {Koliopanos}, {Ray}, {Bailyn}, {Haberl}, and {Gendreau}]{2019MNRAS.488.5225V}
{Vasilopoulos}, G.; {Petropoulou}, M.; {Koliopanos}, F.; {Ray}, P.S.; {Bailyn},
  C.B.; {Haberl}, F.; {Gendreau}, K.
\newblock {NGC 300 ULX1: spin evolution, super-Eddington accretion, and
  outflows}.
\newblock {\em \mnras} {\bf 2019}, {\em 488},~5225--5231,
  \href{http://xxx.lanl.gov/abs/1905.03740}{{\normalfont
  [arXiv:astro-ph.HE/1905.03740]}}.
\newblock {\url{https://doi.org/10.1093/mnras/stz2045}}.

\bibitem[{Bildsten} \em{et~al.}(1997){Bildsten}, {Chakrabarty}, {Chiu},
  {Finger}, {Koh}, {Nelson}, {Prince}, {Rubin}, {Scott}, {Stollberg},
  {Vaughan}, {Wilson}, and {Wilson}]{1997ApJS..113..367B}
{Bildsten}, L.; {Chakrabarty}, D.; {Chiu}, J.; {Finger}, M.H.; {Koh}, D.T.;
  {Nelson}, R.W.; {Prince}, T.A.; {Rubin}, B.C.; {Scott}, D.M.; {Stollberg},
  M.;  et~al.
\newblock {Observations of Accreting Pulsars}.
\newblock {\em \apjs} {\bf 1997}, {\em 113},~367--408,
  \href{http://xxx.lanl.gov/abs/astro-ph/9707125}{{\normalfont
  [arXiv:astro-ph/astro-ph/9707125]}}.
\newblock {\url{https://doi.org/10.1086/313060}}.

\bibitem[{Parmar} \em{et~al.}(1989){Parmar}, {White}, {Stella}, {Izzo}, and
  {Ferri}]{1989ApJ...338..359P}
{Parmar}, A.N.; {White}, N.E.; {Stella}, L.; {Izzo}, C.; {Ferri}, P.
\newblock {The Transient 42 Second X-Ray Pulsar EXO 2030+375. I. The Discovery
  and the Luminosity Dependence of the Pulse Period Variations}.
\newblock {\em \apj} {\bf 1989}, {\em 338},~359.
\newblock {\url{https://doi.org/10.1086/167204}}.

\bibitem[{Klu{\'z}niak} and {Rappaport}(2007)]{2007ApJ...671.1990K}
{Klu{\'z}niak}, W.; {Rappaport}, S.
\newblock {Magnetically Torqued Thin Accretion Disks}.
\newblock {\em \apj} {\bf 2007}, {\em 671},~1990--2005,
  \href{http://xxx.lanl.gov/abs/0709.2361}{{\normalfont
  [arXiv:astro-ph/0709.2361]}}.
\newblock {\url{https://doi.org/10.1086/522954}}.

\bibitem[{Wang} \em{et~al.}(2006){Wang}, {Lai}, and {Han}]{2006ApJ...639.1007W}
{Wang}, C.; {Lai}, D.; {Han}, J.L.
\newblock {Neutron Star Kicks in Isolated and Binary Pulsars: Observational
  Constraints and Implications for Kick Mechanisms}.
\newblock {\em \apj} {\bf 2006}, {\em 639},~1007--1017,
  \href{http://xxx.lanl.gov/abs/astro-ph/0509484}{{\normalfont
  [arXiv:astro-ph/astro-ph/0509484]}}.
\newblock {\url{https://doi.org/10.1086/499397}}.

\bibitem[{Wang}(1981)]{1981A&A...102...36W}
{Wang}, Y.M.
\newblock {Spin-reversed accretion as the cause of intermittent spindown in
  slowX-ray pulsars.}
\newblock {\em \aap} {\bf 1981}, {\em 102},~36--44.

\bibitem[{El Mellah} \em{et~al.}(2018){El Mellah}, {Sundqvist}, and
  {Keppens}]{2018MNRAS.475.3240E}
{El Mellah}, I.; {Sundqvist}, J.O.; {Keppens}, R.
\newblock {Accretion from a clumpy massive-star wind in supergiant X-ray
  binaries}.
\newblock {\em \mnras} {\bf 2018}, {\em 475},~3240--3252,
  \href{http://xxx.lanl.gov/abs/1711.08709}{{\normalfont
  [arXiv:astro-ph.HE/1711.08709]}}.
\newblock {\url{https://doi.org/10.1093/mnras/stx3211}}.

\bibitem[{Biryukov} and {Abolmasov}(2021)]{2021MNRAS.505.1775B}
{Biryukov}, A.; {Abolmasov}, P.
\newblock {Magnetic angle evolution in accreting neutron stars}.
\newblock {\em \mnras} {\bf 2021}, {\em 505},~1775--1786,
  \href{http://xxx.lanl.gov/abs/2105.00754}{{\normalfont
  [arXiv:astro-ph.HE/2105.00754]}}.
\newblock {\url{https://doi.org/10.1093/mnras/stab1378}}.

\bibitem[{Kato} and {Yoshizawa}(1997)]{1997PASJ...49..213K}
{Kato}, S.; {Yoshizawa}, A.
\newblock {A Steady Hydrodynamical Turbulence in Differentially Rotating
  Disks}.
\newblock {\em \pasj} {\bf 1997}, {\em 49},~213--220.
\newblock {\url{https://doi.org/10.1093/pasj/49.2.213}}.

\bibitem[{Gonz{\'a}lez-Gal{\'a}n} \em{et~al.}(2012){Gonz{\'a}lez-Gal{\'a}n},
  {Kuulkers}, {Kretschmar}, {Larsson}, {Postnov}, {Kochetkova}, and
  {Finger}]{2012A&A...537A..66G}
{Gonz{\'a}lez-Gal{\'a}n}, A.; {Kuulkers}, E.; {Kretschmar}, P.; {Larsson}, S.;
  {Postnov}, K.; {Kochetkova}, A.; {Finger}, M.H.
\newblock {Spin period evolution of GX 1+4}.
\newblock {\em \aap} {\bf 2012}, {\em 537},~A66,
  \href{http://xxx.lanl.gov/abs/1111.6791}{{\normalfont
  [arXiv:astro-ph.HE/1111.6791]}}.
\newblock {\url{https://doi.org/10.1051/0004-6361/201117893}}.

\bibitem[{Hessels} \em{et~al.}(2006){Hessels}, {Ransom}, {Stairs}, {Freire},
  {Kaspi}, and {Camilo}]{2006Sci...311.1901H}
{Hessels}, J.W.T.; {Ransom}, S.M.; {Stairs}, I.H.; {Freire}, P.C.C.; {Kaspi},
  V.M.; {Camilo}, F.
\newblock {A Radio Pulsar Spinning at 716 Hz}.
\newblock {\em Science} {\bf 2006}, {\em 311},~1901--1904,
  \href{http://xxx.lanl.gov/abs/astro-ph/0601337}{{\normalfont
  [arXiv:astro-ph/astro-ph/0601337]}}.
\newblock {\url{https://doi.org/10.1126/science.1123430}}.

\bibitem[{Patruno} \em{et~al.}(2017){Patruno}, {Haskell}, and
  {Andersson}]{2017ApJ...850..106P}
{Patruno}, A.; {Haskell}, B.; {Andersson}, N.
\newblock {The Spin Distribution of Fast-spinning Neutron Stars in Low-mass
  X-Ray Binaries: Evidence for Two Subpopulations}.
\newblock {\em \apj} {\bf 2017}, {\em 850},~106,
  \href{http://xxx.lanl.gov/abs/1705.07669}{{\normalfont
  [arXiv:astro-ph.HE/1705.07669]}}.
\newblock {\url{https://doi.org/10.3847/1538-4357/aa927a}}.

\bibitem[{Bildsten}(1998)]{1998ApJ...501L..89B}
{Bildsten}, L.
\newblock {Gravitational Radiation and Rotation of Accreting Neutron Stars}.
\newblock {\em \apjl} {\bf 1998}, {\em 501},~L89--L93,
  \href{http://xxx.lanl.gov/abs/astro-ph/9804325}{{\normalfont
  [arXiv:astro-ph/astro-ph/9804325]}}.
\newblock {\url{https://doi.org/10.1086/311440}}.

\bibitem[{Owen} \em{et~al.}(1998){Owen}, {Lindblom}, {Cutler}, {Schutz},
  {Vecchio}, and {Andersson}]{1998PhRvD..58h4020O}
{Owen}, B.J.; {Lindblom}, L.; {Cutler}, C.; {Schutz}, B.F.; {Vecchio}, A.;
  {Andersson}, N.
\newblock {Gravitational waves from hot young rapidly rotating neutron stars}.
\newblock {\em \prd} {\bf 1998}, {\em 58},~084020,
  \href{http://xxx.lanl.gov/abs/gr-qc/9804044}{{\normalfont
  [arXiv:gr-qc/gr-qc/9804044]}}.
\newblock {\url{https://doi.org/10.1103/PhysRevD.58.084020}}.

\bibitem[{Patruno} \em{et~al.}(2012){Patruno}, {Haskell}, and
  {D'Angelo}]{2012ApJ...746....9P}
{Patruno}, A.; {Haskell}, B.; {D'Angelo}, C.
\newblock {Gravitational Waves and the Maximum Spin Frequency of Neutron
  Stars}.
\newblock {\em \apj} {\bf 2012}, {\em 746},~9,
  \href{http://xxx.lanl.gov/abs/1109.0536}{{\normalfont
  [arXiv:astro-ph.HE/1109.0536]}}.
\newblock {\url{https://doi.org/10.1088/0004-637X/746/1/9}}.

\bibitem[{Tauris}(2012)]{2012Sci...335..561T}
{Tauris}, T.M.
\newblock {Spin-Down of Radio Millisecond Pulsars at Genesis}.
\newblock {\em Science} {\bf 2012}, {\em 335},~561,
  \href{http://xxx.lanl.gov/abs/1202.0551}{{\normalfont
  [arXiv:astro-ph.SR/1202.0551]}}.
\newblock {\url{https://doi.org/10.1126/science.1216355}}.

\bibitem[{Patruno} and {Watts}(2021)]{2021ASSL..461..143P}
{Patruno}, A.; {Watts}, A.L.
\newblock {Accreting Millisecond X-ray Pulsars}.
\newblock In Proceedings of the Timing Neutron Stars: Pulsations, Oscillations
  and Explosions; {Belloni}, T.M.; {M{\'e}ndez}, M.; {Zhang}, C., Eds.,  2021,
  Vol. 461, {\em Astrophysics and Space Science Library}, pp. 143--208,
  \href{http://xxx.lanl.gov/abs/1206.2727}{{\normalfont
  [arXiv:astro-ph.HE/1206.2727]}}.
\newblock {\url{https://doi.org/10.1007/978-3-662-62110-3_4}}.

\bibitem[{Papitto} and {de Martino}(2022)]{2022ASSL..465..157P}
{Papitto}, A.; {de Martino}, D.
\newblock {Transitional Millisecond Pulsars}.
\newblock In Proceedings of the Astrophysics and Space Science Library;
  {Bhattacharyya}, S.; {Papitto}, A.; {Bhattacharya}, D., Eds.,  2022, Vol.
  465, {\em Astrophysics and Space Science Library}, pp. 157--200,
  \href{http://xxx.lanl.gov/abs/2010.09060}{{\normalfont
  [arXiv:astro-ph.HE/2010.09060]}}.
\newblock {\url{https://doi.org/10.1007/978-3-030-85198-9_6}}.

\bibitem[{Archibald} \em{et~al.}(2015){Archibald}, {Bogdanov}, {Patruno},
  {Hessels}, {Deller}, {Bassa}, {Janssen}, {Kaspi}, {Lyne}, {Stappers},
  {Tendulkar}, {D'Angelo}, and {Wijnands}]{2015ApJ...807...62A}
{Archibald}, A.M.; {Bogdanov}, S.; {Patruno}, A.; {Hessels}, J.W.T.; {Deller},
  A.T.; {Bassa}, C.; {Janssen}, G.H.; {Kaspi}, V.M.; {Lyne}, A.G.; {Stappers},
  B.W.;  et~al.
\newblock {Accretion-powered Pulsations in an Apparently Quiescent Neutron Star
  Binary}.
\newblock {\em \apj} {\bf 2015}, {\em 807},~62,
  \href{http://xxx.lanl.gov/abs/1412.1306}{{\normalfont
  [arXiv:astro-ph.HE/1412.1306]}}.
\newblock {\url{https://doi.org/10.1088/0004-637X/807/1/62}}.

\bibitem[{Veledina} \em{et~al.}(2019){Veledina}, {N{\"a}ttil{\"a}}, and
  {Beloborodov}]{2019ApJ...884..144V}
{Veledina}, A.; {N{\"a}ttil{\"a}}, J.; {Beloborodov}, A.M.
\newblock {Pulsar Wind-heated Accretion Disk and the Origin of Modes in
  Transitional Millisecond Pulsar PSR J1023+0038}.
\newblock {\em \apj} {\bf 2019}, {\em 884},~144,
  \href{http://xxx.lanl.gov/abs/1906.02519}{{\normalfont
  [arXiv:astro-ph.HE/1906.02519]}}.
\newblock {\url{https://doi.org/10.3847/1538-4357/ab44c6}}.

\bibitem[{Bisnovatyi-Kogan} and {Komberg}(1974)]{1974SvA....18..217B}
{Bisnovatyi-Kogan}, G.S.; {Komberg}, B.V.
\newblock {Pulsars and close binary systems}.
\newblock {\em \sovast} {\bf 1974}, {\em 18},~217.

\bibitem[{Melatos} and {Phinney}(2001)]{2001PASA...18..421M}
{Melatos}, A.; {Phinney}, E.S.
\newblock {Hydromagnetic Structure of a Neutron Star Accreting at Its Polar
  Caps}.
\newblock {\em \pasa} {\bf 2001}, {\em 18},~421--430.
\newblock {\url{https://doi.org/10.1071/AS01056}}.

\bibitem[{Payne} and {Melatos}(2004)]{2004MNRAS.351..569P}
{Payne}, D.J.B.; {Melatos}, A.
\newblock {Burial of the polar magnetic field of an accreting neutron star - I.
  Self-consistent analytic and numerical equilibria}.
\newblock {\em \mnras} {\bf 2004}, {\em 351},~569--584,
  \href{http://xxx.lanl.gov/abs/astro-ph/0403173}{{\normalfont
  [arXiv:astro-ph/astro-ph/0403173]}}.
\newblock {\url{https://doi.org/10.1111/j.1365-2966.2004.07798.x}}.

\bibitem[{Konar} and {Bhattacharya}(1999)]{1999MNRAS.303..588K}
{Konar}, S.; {Bhattacharya}, D.
\newblock {Magnetic field evolution of accreting neutron stars - II}.
\newblock {\em \mnras} {\bf 1999}, {\em 303},~588--594,
  \href{http://xxx.lanl.gov/abs/astro-ph/9808119}{{\normalfont
  [arXiv:astro-ph/astro-ph/9808119]}}.
\newblock {\url{https://doi.org/10.1046/j.1365-8711.1999.02287.x}}.

\bibitem[{Litwin} \em{et~al.}(2001){Litwin}, {Brown}, and
  {Rosner}]{2001ApJ...553..788L}
{Litwin}, C.; {Brown}, E.F.; {Rosner}, R.
\newblock {Ballooning Instability in Polar Caps of Accreting Neutron Stars}.
\newblock {\em \apj} {\bf 2001}, {\em 553},~788--795,
  \href{http://xxx.lanl.gov/abs/astro-ph/0101168}{{\normalfont
  [arXiv:astro-ph/astro-ph/0101168]}}.
\newblock {\url{https://doi.org/10.1086/320952}}.

\bibitem[{Kulsrud} and {Sunyaev}(2020)]{2020JPlPh..86f9002K}
{Kulsrud}, R.M.; {Sunyaev}, R.
\newblock {Anomalous diffusion across a tera-Gauss magnetic field in accreting
  neutron stars}.
\newblock {\em Journal of Plasma Physics} {\bf 2020}, {\em 86},~905860602.
\newblock {\url{https://doi.org/10.1017/S0022377820001026}}.

\bibitem[{Payne} and {Melatos}(2007)]{2007MNRAS.376..609P}
{Payne}, D.J.B.; {Melatos}, A.
\newblock {Burial of the polar magnetic field of an accreting neutron star -
  II. Hydromagnetic stability of axisymmetric equilibria}.
\newblock {\em \mnras} {\bf 2007}, {\em 376},~609--624,
  \href{http://xxx.lanl.gov/abs/astro-ph/0703203}{{\normalfont
  [arXiv:astro-ph/astro-ph/0703203]}}.
\newblock {\url{https://doi.org/10.1111/j.1365-2966.2007.11451.x}}.

\bibitem[{Lipunov} and {Prokhorov}(1984)]{1984Ap&SS..98..221L}
{Lipunov}, V.M.; {Prokhorov}, M.E.
\newblock {Ejection from Pulsars in Binary Systems}.
\newblock {\em \apss} {\bf 1984}, {\em 98},~221--236.
\newblock {\url{https://doi.org/10.1007/BF00651401}}.

\bibitem[{Popov}(2008)]{2008arXiv0812.4587P}
{Popov}, S.B.
\newblock {Scenarios for GCRT J1745-3009}.
\newblock {\em arXiv e-prints} {\bf 2008}, p. arXiv:0812.4587,
  \href{http://xxx.lanl.gov/abs/0812.4587}{{\normalfont
  [arXiv:astro-ph/0812.4587]}}.
\newblock {\url{https://doi.org/10.48550/arXiv.0812.4587}}.

\bibitem[{Sumiyoshi} \em{et~al.}(2022){Sumiyoshi}, {Kojo}, and
  {Furusawa}]{2022arXiv220700033S}
{Sumiyoshi}, K.; {Kojo}, T.; {Furusawa}, S.
\newblock {Equation of state in neutron stars and supernovae}.
\newblock {\em arXiv e-prints} {\bf 2022}, p. arXiv:2207.00033,
  \href{http://xxx.lanl.gov/abs/2207.00033}{{\normalfont
  [arXiv:nucl-th/2207.00033]}}.
\newblock {\url{https://doi.org/10.48550/arXiv.2207.00033}}.

\bibitem[{Gourgouliatos} \em{et~al.}(2022){Gourgouliatos}, {De Grandis}, and
  {Igoshev}]{2022Symm...14..130G}
{Gourgouliatos}, K.N.; {De Grandis}, D.; {Igoshev}, A.
\newblock {Magnetic Field Evolution in Neutron Star Crusts: Beyond the Hall
  Effect}.
\newblock {\em Symmetry} {\bf 2022}, {\em 14},~130,
  \href{http://xxx.lanl.gov/abs/2201.08345}{{\normalfont
  [arXiv:astro-ph.HE/2201.08345]}}.
\newblock {\url{https://doi.org/10.3390/sym14010130}}.

\bibitem[{Wang}(2016)]{2016AdAst2016E...3W}
{Wang}, J.
\newblock {Physical Environment of Accreting Neutron Stars}.
\newblock {\em Advances in Astronomy} {\bf 2016}, {\em 2016},~3434565,
  \href{http://xxx.lanl.gov/abs/1909.01057}{{\normalfont
  [arXiv:astro-ph.HE/1909.01057]}}.
\newblock {\url{https://doi.org/10.1155/2016/3424565}}.

\bibitem[{Armitage}(2022)]{2022arXiv220107262A}
{Armitage}, P.J.
\newblock {Lecture notes on accretion disk physics}.
\newblock {\em arXiv e-prints} {\bf 2022}, p. arXiv:2201.07262,
  \href{http://xxx.lanl.gov/abs/2201.07262}{{\normalfont
  [arXiv:astro-ph.HE/2201.07262]}}.
\newblock {\url{https://doi.org/10.48550/arXiv.2201.07262}}.

\end{thebibliography}

\PublishersNote{}
\end{adjustwidth}
\end{document}